\documentclass{aa}

\usepackage{graphicx}
\usepackage{txfonts}
\usepackage{lipsum}
\usepackage{subcaption}         
\usepackage{lscape}             
\usepackage{placeins}           
\usepackage{xcolor}
\usepackage[colorlinks,allcolors=blue]{hyperref}
\usepackage{breakurl}
\usepackage{stackengine}
\usepackage{amsmath}

\def \xmm {\mbox{\emph{XMM-Newton}}}

\newcommand{\rxte}{RXTE}

\newcommand{\nicer}{NICER}

\newcommand {\beq}{\begin {eqnarray}}
\newcommand {\eeq}{\end {eqnarray}}

\newcommand{\src}{A0538$-$66}

\begin{document}

   \title{A0538-66: A transient ultraluminous X-ray source with a relativistic outflow, 69\,ms pulsations, and anomalous optical activity}
\titlerunning{A0538-66: a transient ULX with relativistic outflow,  pulsations, and anomalous optical activity}

%

  \author{Lorenzo~Ducci\inst{\ref{tue}}\fnmsep\thanks{Corresponding author: ducci@astro.uni-tuebingen.de}
        \and Sandro~Mereghetti\inst{\ref{iasf-mi}}
        \and Ciro~Pinto\inst{\ref{iasf-pa}}
        \and Matteo~Bachetti\inst{\ref{inaf-ca}}
        \and Motoki~Nakajima\inst{\ref{nihon}}
        \and Rahul~Sharma\inst{\ref{iucaa}}
        \and Stefano~Covino\inst{\ref{oab}}
        \and Michela~Rigoselli\inst{\ref{iasf-mi}}
        \and Caterina~Tresoldi\inst{\ref{oab},\ref{como}}
        \and Tatehiro~Mihara\inst{\ref{riken}}
        \and Mutsumi~Sugizaki\inst{\ref{kanazawa-u}}
        }

   \institute{Institut f\"ur Astronomie und Astrophysik, Kepler Center for Astro and Particle Physics, University of Tuebingen, Sand 1, 72076 T\"ubingen, Germany \label{tue}
   \and
   INAF, Istituto di Astrofisica Spaziale e Fisica Cosmica di Milano, via Corti 12, I-20133 Milano, Italy \label{iasf-mi}   
   \and 
   INAF--IASF Palermo, Via U. La Malfa 153, 90146 Palermo, Italy \label{iasf-pa}
   \and 
   INAF -- Osservatorio Astronomico di Cagliari, Via della Scienza 5, 09047 Selargius (CA), Italy \label{inaf-ca}
   \and 
   School of Dentistry at Matsudo, Nihon University, 2-870-1 Sakaecho-nishi, Matsudo, Chiba 271-8587, Japan \label{nihon}
   \and
   Inter-University Centre for Astronomy and Astrophysics (IUCAA), Ganeshkhind, Pune 411007, India \label{iucaa}
   \and
   Osservatorio Astronomico di Brera, Via E.\ Bianchi 46, I-23807, Merate, Italy \label{oab}
   \and
   RIKEN Pioneering Research Institute (PRI), 2-1 Hirosawa, Wako, Saitama 351-0198, Japan \label{riken}
   \and
   Advanced Research Center for Space Science and Technology, Kanazawa University, Kakuma, Kanazawa, Ishikawa, 920-1192, Japan \label{kanazawa-u}  
   \and
   Como Lake Center for Astrophysics (CLAP), DiSAT, Università degli Studi dell'Insubria, via Valleggio 11, 22100 Como, Italy \label{como}
   }

   \date{}

  \abstract
   {A0538$-$66 is a peculiar Be/X-ray binary in the Large Magellanic Cloud, hosting the fastest spinning neutron star
   of its class ($\sim69$\,ms) and showing the brightest optical bursts of any system of this type.
   We report four \xmm\ observations obtained near periastron in 2025, together with MAXI and REM monitoring, and a
   reanalysis of archival RXTE data. 
   On 30 September 2025 the source reached $L_{\rm X} \approx 1.5\times 10^{39}$\,erg\,s$^{-1}$ (0.2$-$10\,keV),
   the highest level since 1980, and in February, May, and August 2026, MAXI detected  similarly bright outbursts.
   During the \xmm\ observation taken on 30 September 2025, we detected the pulsations, measuring $P_{\rm spin}=69.25878(2)$\,ms and $\dot{P}_{\rm spin}=6.58(9)\times10^{-10}$\,s\,s$^{-1}$; the pulsed fraction was low ($\approx3\%$ in 1$-$10\,keV), 
   increased with energy, and the soft and hard pulse profiles were shifted by half a cycle.
   A new detection in RXTE data of 3 December 1998 gives a faster spin period of $P=68.88813(28)$\,ms. 
   The \xmm/RGS spectra show emission lines from photoionised plasma at rest and blueshifted absorption lines that imply
   an outflow at $\sim 0.12c$. This indicates a highly super-Eddington accretion rate and 
   an intrinsic $L_{\rm X} \approx 1.5\times 10^{39}$\,erg\,s$^{-1}$,
   and it makes \src\ a transient ultra-luminous X-ray pulsar, a class of which only a few objects are known so far. 
   Since 2023 the optical bursts observed by REM have become longer and brighter, resembling those of the 1980s,
   but most of them lack a luminous X-ray counterpart ($L_{\rm X}>10^{37}$\,erg\,s$^{-1}$).
   We interpret these results in terms of a radiation pressure dominated accretion disc whose geometrically thick 
   inner regions intermittently screen the innermost emitting regions.}


   \maketitle
\nolinenumbers

\section{Introduction}

Be/neutron star (NS) systems are the most numerous class of accreting high-mass X-ray binaries 
(HMXBs; for a review, see, for example, \citealt{Reig26}).
They consist of a rapidly rotating Be star surrounded by a circumstellar disc and a NS that interacts with this disc as it moves along the orbit. As a class, their properties are relatively well understood. Their spin and orbital periods cover a wide range of values (tens of milliseconds to a few hours),
the orbits are often eccentric, and the systems show type I outbursts ($L_{\rm X-ray,\,peak}\approx 10^{37}$~erg~s$^{-1}$) close to periastron  and, in some cases, brighter (typically, $L_{\rm X-ray,\,peak}\approx 10^{38}$~erg~s$^{-1}$ and, in a very few cases, $L_{\rm X-ray}\approx 10^{39}$~erg~s$^{-1}$) less predictable and longer type II outbursts, probably
linked to major structural changes in the circumstellar disc \citep{Okazaki13,Martin14}. Several of these systems 
display long time intervals (several years) with no X-ray outbursts, that are connected to the depletion of the circumstellar disc, and the class as a whole spans many orders of magnitude in X-ray luminosity between quiescence and outburst (up to $\sim 6$).

A few Be/NS systems, however, do not fit well into this general picture and are particularly interesting since they probe the limits of the standard picture of accretion (e.g. Cir~X$-$1, \citealt{Schulz20}) and may represent transitional objects, linking Be/X-ray binaries to other classes of systems, such as gamma-ray binaries with Be stars paired with radio pulsars 
(e.g. PSR~B1259$-$63, \citealt{Johnston99}).

\src,  discovered in 1977 in the Large Magellanic Cloud \citep{White78}, belongs to this small group of outliers. Its NS has the shortest spin period known among Be/X-ray binaries (69~ms; \citealt{Skinner82}), spinning considerably faster than the next fastest pulsars in the class (a few seconds).
The system has an orbital period of $\sim 16.6$ days \citep{Johnston79} and an eccentricity of $e=0.72$ \citep{Rajoelimanana17}.
Its optical counterpart is equally unusual, showing bright and fast  periodic flares in phase
with the X-ray outbursts, together with a superorbital period of 
$\sim 421$ days that was previously well defined but has recently changed and then disappeared \citep{Alcock01,Ducci22}.
The peculiar optical flares in combination with the long-term optical variability  make \src\ unlike any other Be/X-ray binary. 
The optical flares recur every $\sim16.6$~d (the orbital period), in phase with the X-ray outbursts, and are the
brightest observed in any Be/X-ray binary, reaching $L_{\rm opt}\approx3\times10^{38}$~erg~s$^{-1}$
and $\Delta V\approx2.3$~mag during the 1981 events \citep{Skinner80,Charles83}. They are
unusual for the class, in which the optical emission is normally dominated by the donor star
and its circumstellar disc, and the corresponding maximum ratio $L_{\rm opt}/L_{\rm X}\approx0.3$ is
far higher than in any other system of this type \citep{Maraschi83}. The optical flares have been interpreted as reprocessing of the X-ray emission of the NS in a dense envelope surrounding the
binary, formed from material tidally displaced from the outer layers of the donor star over
many orbits \citep{Charles83,Densham83,Maraschi83,Apparao88}. Photoionisation modelling of the outburst
of 29 April 1981 has indicated a roughly spherical cloud with a shallow density profile and a size
of $\approx 10^{12}$~cm \citep{Ducci19b}. After 1981 the flares became shorter ($t\lesssim1$~d, against $t\approx3$~d)
and fainter ($\Delta V\approx0.5-1.5$), and there was also a decrease in the X-ray luminosities measured over
the same period \citep{Corbet97,Alcock01,Ducci16,Rajoelimanana17,Ducci22}.

In the first years after its discovery, \src\  displayed a series of X-ray outbursts that reached super-Eddington luminosities and in one case nearly $\sim 10^{39}$~erg~s$^{-1}$, when the 69~ms pulsations were first observed \citep{Skinner82}.
All the X-ray observations carried out since then, however, found \src\ at much lower luminosity levels \citep[][]{Mavromatakis93,Corbet97,Campana97,Campana02,Rigoselli25}.
As a result, a detailed study of its spectral and timing properties outside the early bright outbursts remained out of reach for decades.

It was only in 2018 that \src\ showed signs of renewed bright X-ray activity \citep{Ducci19},
and it took until 2023 for pulsations to be detected again for the first time since the discovery outburst, again at about 69~ms, when the source had an X-ray luminosity of  $\sim 10^{37}$ erg\,s$^{-1}$ in NICER data \citep{Ducci25}.

\src\ has also recently been detected in the radio band, first with the Australian Square
Kilometre Array Pathfinder (ASKAP) and subsequently in a weekly monitoring campaign with the
MeerKAT radio telescope, which showed it to be one of the most radio-luminous NS
X-ray binaries, reaching $\approx3\times10^{22}$~erg~s$^{-1}$~Hz$^{-1}$ at 1.28~GHz, with 
a radio emission that appears to be orbitally modulated and that peaks shortly after the optical
maximum, close to periastron \citep{Crook-Mansour26}. The origin of this emission is not yet
established, with discrete ejecta, an outflow powered by a propeller mechanism, and a shock
between a pulsar wind and the circumstellar material all remaining viable \citep{Crook-Mansour26}.

In September 2025, \src\ reached a peak X-ray luminosity of about $10^{39}$~erg~s$^{-1}$ \citep{Ducci25b}. 
This outburst, the first to reach such a brightness in the past 45 years, 
gives us  the opportunity to study the spectral and timing behaviour of \src\ 
in the super-Eddington  regime with the sensitivity and time resolution of modern X-ray instruments.

In this paper we report on four \xmm\ observations of \src\ performed near periastron
in 2025, complemented by MAXI and REM monitoring covering the 2023-2026 period
and by an analysis of archival RXTE data.
The observations and data reduction are described in Sect. \ref{sect.data reduction},
and the timing and spectral results are presented in Sects.~\ref{sect:timing results} and \ref{sect:spectral results}.
We discuss our findings in Sect.~\ref{sect:discussion}
and summarise our conclusions in Sect.~\ref{sect:conclusions}.

In the following we assume for \src\ a distance of $d=50$~kpc \citep{Alves04}, and unless noted differently, in all the equations we use an NS mass of $M=1.4$\,M$_\odot$ and a radius of 12\,km. We quote all uncertainties at a 1$\sigma$ confidence level.

\begin{table*}
\begin{center}
\caption{Log of the \xmm\ observations. Orbital phases obtained using $P_{\rm orb}=16.64002$~d and $T_0=55673.71$~MJD corresponding to the first optical peak \citep{Ducci22}.}
\label{Log of the observations}
\resizebox{\columnwidth+\columnwidth}{!}{
\begin{tabular}{lcccccccccc}
\hline
\hline
Obs.      &     ObsID     &        t$_{\rm start}$       &        t$_{\rm stop}$           & $\phi_{\rm start}$ & $\phi_{\rm stop}$ &   \multicolumn{5}{c}{effective exposure time}    \\
          &               &         (UTC)       &         (UTC)          &                    &                   &    \multicolumn{5}{c}{(ks)}             \\
          &               &                     &                        &                    &                   &    timing    & \multicolumn{4}{c}{spectra}  \\
          &               &                     &                        &                    &                   & \emph{pn} & \emph{pn} & MOS1 & MOS2 & RGS   \\
\hline
obs.\,1   &   0964320201  & 2025-08-11 15:08:21 &  2025-08-11 20:29:01   &         0.997      &     0.011         &   19.2    &   13.0    & 18.2 & 18.2 & 19.7  \\
obs.\,2   &   0964320301  & 2025-08-28 06:46:46 &  2025-08-28 10:59:05   &         0.998      &     0.009         &   15.1    &   10.1    & 13.9 & 13.8 & 15.1  \\
obs.\,3   &   0964320401  & 2025-09-30 13:58:29 &  2025-09-30 17:54:56   &         0.999      &     0.009         &   14.2    &    5.2    &      &      & 16.3  \\
obs.\,4   &   0973790101  & 2025-11-03 21:52:18 &  2025-11-04 00:38:58   &         0.062      &     0.069         & 10.5; 0.3 & 10.5; 0.3 &      &      &        \\
\hline
\end{tabular}
}
Notes: For obs.\,4, the two exposures refer to the timing mode and burst mode, respectively. The latter has a live time of $\sim3$\% ($\sim 99.5$\% for the timing mode).
\end{center}
\end{table*}

\section{Observations and data reduction} 
\label{sect.data reduction}

\subsection{XMM-Newton} \label{xmm data analysis}

We observed \src\ with \xmm\ 
\citep{Jansen01} in four separate exposures.
We processed the data from the European Photon Imaging Camera (EPIC;
\citealt{Strueder01,Turner01}) and the Reflection Grating Spectrometer (RGS; \citealt{denHerder01})
with the Science Analysis System (SAS) version 22.1.0 and the calibration files available in the \xmm\ calibration database at the time of reduction.
Calibrated event lists for the EPIC and RGS instruments were produced from the raw observation data files using the SAS tasks {\tt epproc} for \emph{pn}, {\tt emproc} for MOS1 and MOS2, and {\tt rgsproc} for RGS1 and RGS2.
Time intervals affected by high particle background were identified and excluded from the spectral analysis using standard criteria.\footnote{\url{https://www.cosmos.esa.int/web/xmm-newton/sas-threads}}
The resulting exposure times for timing and spectral analysis are listed in Table \ref{Log of the observations}.

For all the observations except obs.\,4, the \emph{pn} and MOS camera operated in small window mode.
For \emph{pn} we retained only single- and double-pixel events (PATTERN $\leq$ 4), while for MOS we included single- to quadruple-pixel events (PATTERN $\leq$ 12).
Source events were extracted from a circular region centred on the position of \src, and background events from source-free circular regions on the same CCD chip.
Response matrices and ancillary response files for each EPIC camera were generated with the SAS tasks {\tt rmfgen} and {\tt arfgen}.

The EPIC data of obs.\,3 were affected by pile-up due to the high  flux of \src. To mitigate this problem, for \emph{pn} we excised the 
core of the point-spread function where pile-up is most severe by extracting the source spectra from an annular region with an inner radius of $20^{\prime\prime}$ and an outer radius of $38.6^{\prime\prime}$.
For the MOS cameras, pile-up was even more severe, resulting in a distorted point-spread function with a depleted core. Despite attempts to correct the MOS data by excising the PSF centre, the residuals remained significant, and we therefore opted not to use the MOS data of this observation.
We also  corrected the \emph{pn} data for out-of-time events, which occur when photons are registered during the CCD readout and are assigned incorrect raw coordinates, leading to a broadening of spectral features. In small-window mode, out-of-time events account for 1.1\% of the total events. Following the procedure described in the SAS thread, we generated an out-of-time event list, scaled it by the appropriate factor, and subtracted it from the source spectrum.

For obs.\,3, the RGS source count rates exceeded the thresholds above which pile-up may become significant. No standard correction procedure is available for piled-up RGS spectra, and thus we assessed the impact of pile-up by comparing the fluxed first-order and second-order spectra, as recommended in the relevant SAS threads. We found that for RGS2, which is more susceptible to pile-up than RGS1 because of its longer CCD frame time,\footnote{See \xmm\ calibration note \url{https://xmmweb.esac.esa.int/docs/documents/CAL-TN-0030.pdf}} the first- and second-order spectra do not agree, owing to the presence of spurious lines. 
We therefore limited our analysis to RGS1 for this observation.

For obs.\,4, only the \emph{pn} camera was available: the first part of the observation was performed in burst mode, and the second one in timing mode.
In these modes spatial information is retained only in one dimension, along the   RAWX axis of CCD4.
We therefore extracted the source events from the region RAWX 34$-$40 and the background from RAWX 48$-$59.

The data from the Optical Monitor (OM, \citealt{Mason01}) were reduced using the standard SAS pipeline ({\tt omichain}), which performs source detection and aperture photometry in each of the filters available for a given observation.
We used this pipeline to extract the fluxes of the source in each filter, following the standard procedure described in the SAS analysis threads.

\subsection{RXTE}

\rxte\ has observed \src\ nine times in 1998. 
We downloaded all available data in Good Xenon or Science Event format from HEASARC.\footnote{https://heasarc.gsfc.nasa.gov/} These data can be considered ready for analysis, as the latest version of the \rxte\ pipeline was used to produce them when the mission was completed.

We used \texttt{stingray} \citep{Huppenkothen19,Bachetti24} to calculate rough values of the photon energy from the PHA values contained in the FITS files using the official PHA-energy conversion table.\footnote{\url{https://heasarc.gsfc.nasa.gov/docs/xte/e-c_table.html}}

\subsection{MAXI}

The sky region of \src\ has been regularly scanned every $\sim$92 minutes
by MAXI/GSC for 16 years since the start of the MAXI mission in 2009 \citep{Matsuoka09}.
So far, MAXI/GSC \citep{Mihara11} has clearly detected four bright X-ray outbursts,
in September 2025, February 2026, May 2026, and August 2026.
In addition, there is a possible fainter event in May 2025.
Since \src\ lies at only 0.6 degrees  from the   bright source LMC\,X$-$4,
the standard analysis procedure used for the public MAXI 
light curves\footnote{See the MAXI mission page: \url{https://maxi.riken.jp/top/index.html}} 
cannot disentangle its flux
from that of the contaminating source, LMC\,X$-$4.
To investigate the light curve of \src\ after MJD 60000 (2023 February 25),
we employed  for each scan the 'PSF-fit' analysis method introduced by \citet{Morii16}. 
The PSF-fit analysis is applicable to three energy bands: 2$-$4, 4$-$10, and 10$-$20\,keV.
In our analysis, we used the well-calibrated 4$-$10\,keV band and binned the light curve at 6-hour intervals.
Each time bin has an exposure of 40$-$160 seconds.
To check the validity of the PSF-fit and evaluate potential contamination,
we made light curves for both \src\ and LMC\,X$-$4.
We confirmed that
the respective light curves were not mutually affected by outbursts or flares.
Although the 30-day super-orbital modulation of LMC\,X$-$4,  remains in the \src\ light curve,
this flux level, with a maximum  of $\sim$6\,mCrab, is too small to affect the outburst profiles of \src.

\subsection{REM}
\label{sec:rem}
 
We obtained the photometric light curves of \src\ in the $g$- and $i$-bands,   from the daily monitoring carried out with the Rapid Eye Mount (REM) telescope \citep{Covino04, Zerbi04} from 2023 April 2 to 2026 July 11.
The data were reduced  following the same procedures described in \citet{Ducci16} and \citet{Ducci22}, 
where the previous REM monitoring campaigns of this source were presented.

   \begin{figure}
     \centering
     \includegraphics[width=\columnwidth]{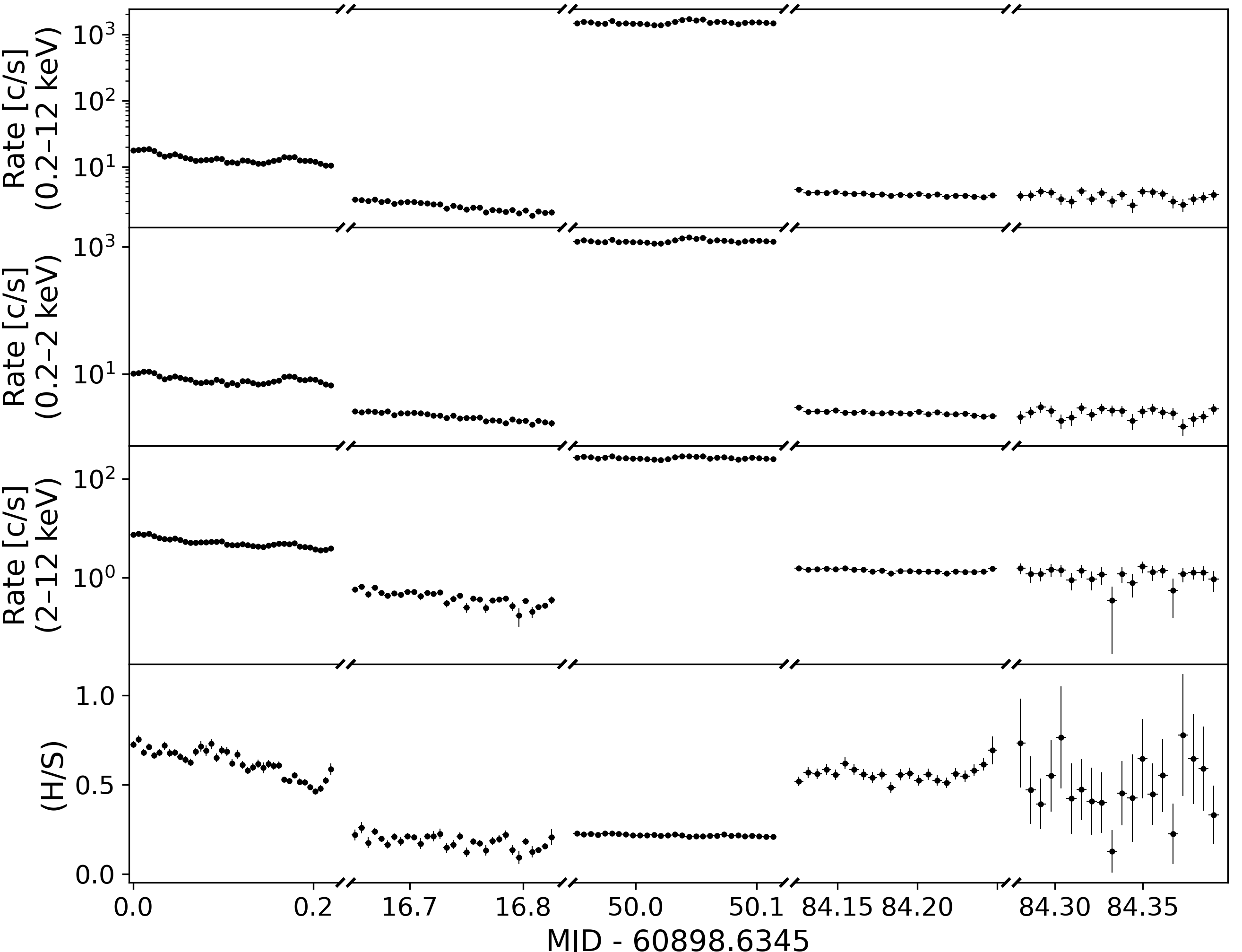}
   \caption{\emph{pn} light curves of \src\ in three energy bands, binned at 500~s, for all the \xmm\  observations. The top three panels show the light curve in the 0.2$-$12~keV (total),
     0.2$-$2~keV (soft), and 2$-$12~keV (hard) bands. The bottom panel shows the hardness ratio between the soft and hard bands. The last two parts of the light curve belong to the obs.\,4, timing and burst mode.}
   \label{fig. xmm lcr and HRs}
   \end{figure}

\section{Timing results} \label{sect:timing results}

In Fig. \ref{fig. xmm lcr and HRs} we plot the  \xmm\ light curves binned at  500~s in different energy ranges ($0.2-12$, $0.2-2$, and $2-12$~keV) as well as  the corresponding hardness ratio.
A bin time of 500\,s is adequate here, because none of the four observations shows the fast
flaring activity reported in \citet{Ducci19} and \citet{Ducci25},
with flux variations of up to three orders of magnitudes on timescales of a few seconds.
Figure \ref{fig. xmm lcr and HRs} clearly shows the much brighter source luminosity during obs.\,3.

We searched for pulsations in all the \xmm\  and \rxte\ observations  using their whole exposures.
For \rxte\ data, characterised by many short pointings,
we combined observations on a day-to-day and later on a week-to-week basis to gain signal to noise.
For all the data, the photon arrival times were corrected to the Solar System barycenter adopting  the source position from \citet{2020yCat.1350....0G} and the JPL DE405 ephemeris.
For \xmm, we used the SAS task {\tt barycen}, while for \rxte\  we used the \texttt{barycorr} FTOOL,

The search was performed using the Rayleigh $Z^2_{\rm n}$ test (e.g. \citealt{Buccheri83}), with $n=1$ harmonics. Given the known spin period of the source,
we restricted the search to the range 14.2$-$14.6\,Hz.

\begin{figure*}
    \centering
    \includegraphics[width=\textwidth]{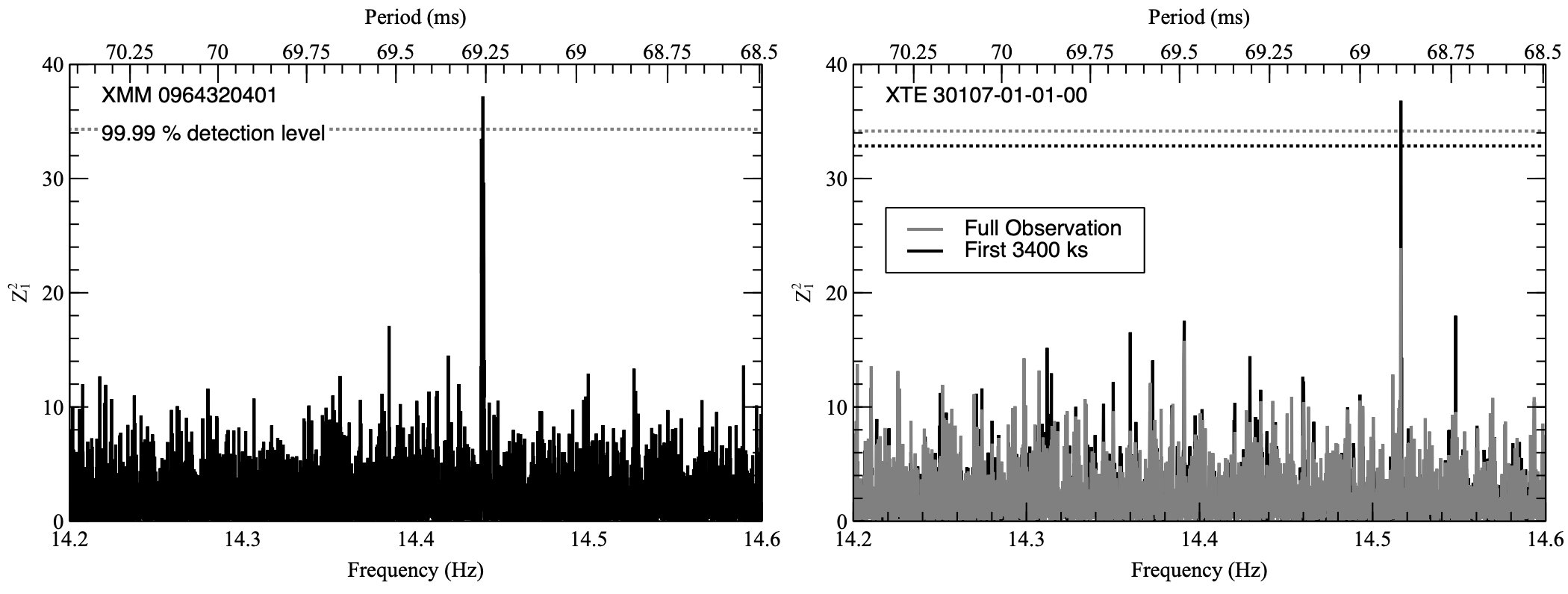}
    \caption{Discovery plots for (left) the \xmm\ data obs. 3 in the energy range 0.2-12 keV and (right) the \rxte\ data at MJD 51150 (obsid 30107-01-01-00) in the energy range 2-15 keV. For \rxte\, we plot the $Z^2_1$ search conducted on the whole data set (grey) and the first 3400 s of data (black). The detection levels are calculated as the 99.99\% inverse survival function of the $\chi^2_2$ distribution. They slightly differ because the number of trials, for a given frequency interval, increases at increasing observation lengths. The peak in \rxte\ corresponds to an amplitude of $p_{\rm f} = (3.0 \pm 0.5)$\%.}
   \label{fig Z^2}
\end{figure*}

In the \emph{pn} data of \xmm\ obs.\,3, we detected a significant pulsation in the energy band $0.2-12$~keV.
Figure \ref{fig Z^2}, left panel, shows the $Z^2$ test statistic as a function of the trial period.
A clear peak appears at a period of approximately $\approx 69.25$~ms, with a test statistic value $Z^2=37.76$.  
After accounting for the number of statistically independent trial periods, the overall probability that this peak arises from a random fluctuation is 0.036\%. To determine the period more accurately, we performed a phase fitting procedure. The resulting period is  $P = 69.25878(2)$~ms,
the period derivative is $\dot{P} = 6.58(9) \times 10^{-10}$\,s\,s$^{-1}$, and the epoch is $T_{\rm 0}=60948.66530440475$~MJD.
The $\sim 69$\,ms periodicity was not accessible to the MOS cameras, operated in small window mode with a time resolution of $\sim 0.3$\,s.
No pulsations were detected in all the other \xmm\  observations, in which the source flux was lower. 
The derived $3\sigma$ upper-limits on the
pulsed fraction\footnote{The pulsed fraction is defined as $p_{\rm f} = (F_{\rm max} - F_{\rm min})/(F_{\rm max} + F_{\rm min})$,
where $F_{\rm max}$ and $F_{\rm min}$ are the maximum and minimum flux of the folded light curve.} are reported in Table \ref{Table p_f}.

\begin{figure}
     \centering
   \includegraphics[width=\columnwidth]{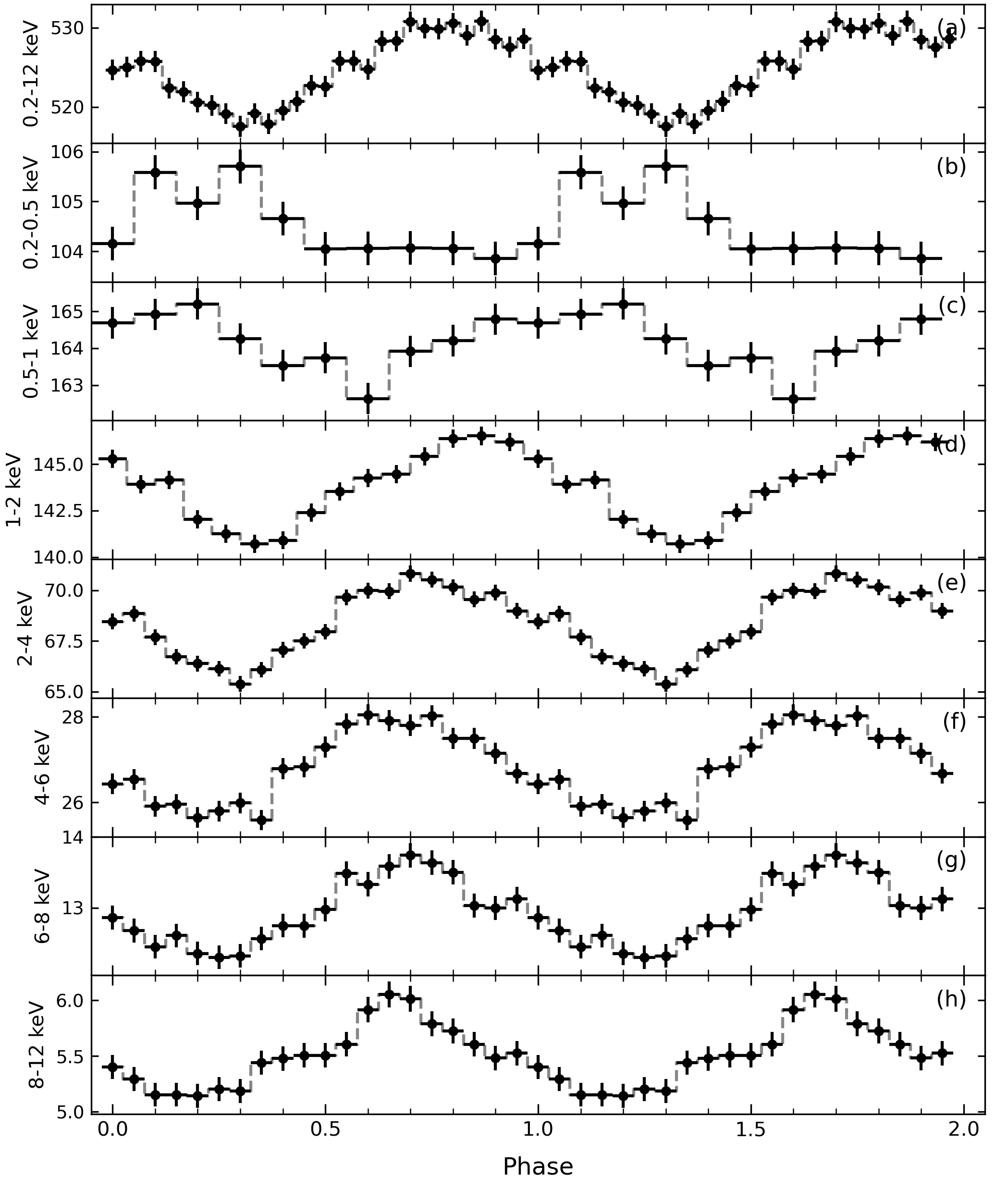}
   \caption{Pulse profiles of \src\ for the full energy range $0.2-12$~keV (top panel) and for seven sub-energy bands (lower panels), from obs.\,3.
All profiles have been folded with the same reference time.}
   \label{multi pp}
   \end{figure}

\begin{figure}
     \centering
   \includegraphics[width=\columnwidth]{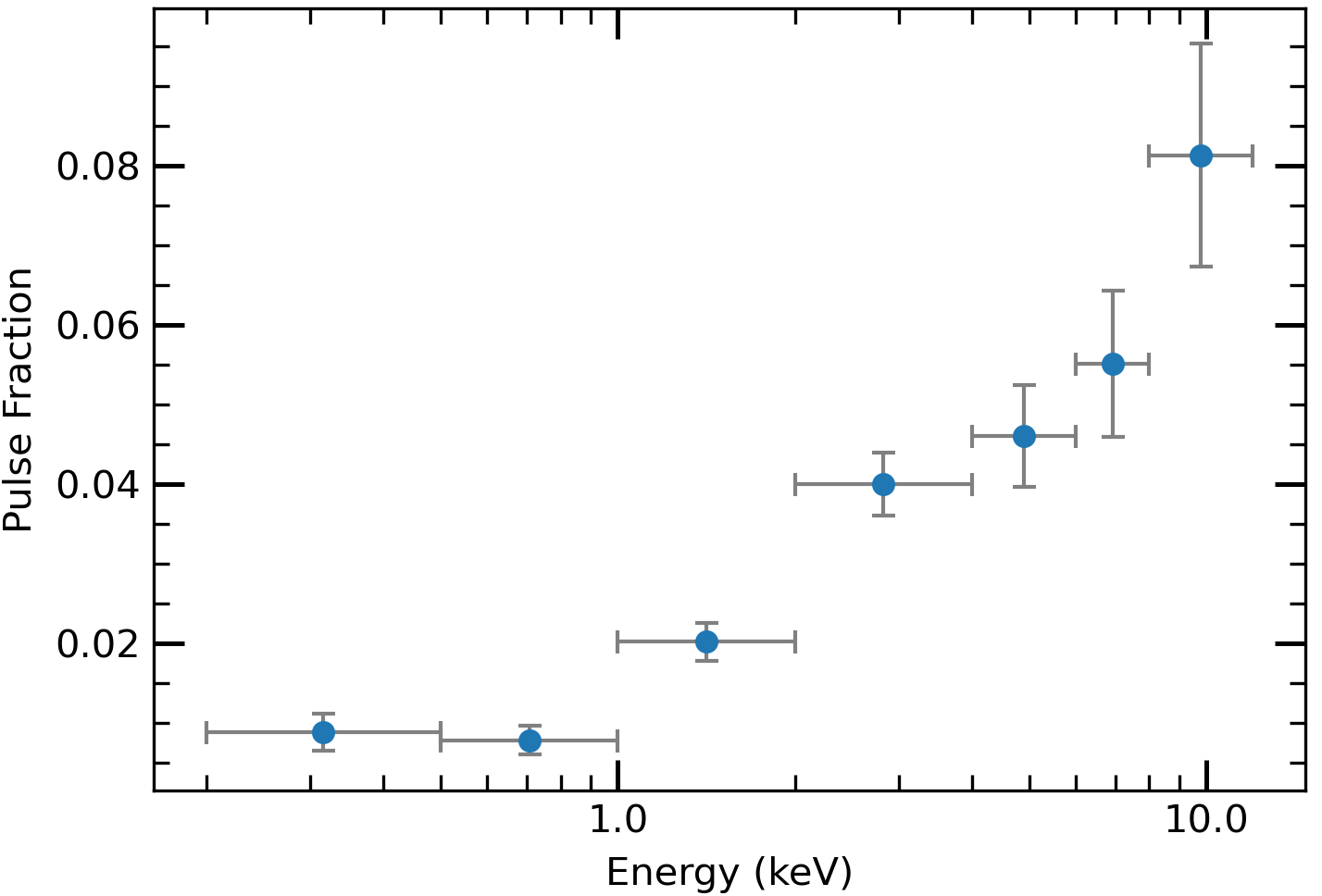}
   \caption{Pulsed fraction as a function of energy for \src\ in the same seven energy bands shown in Fig.~\ref{multi pp}, from obs.\,3. Error bars represent $1\sigma$ uncertainties.}
   \label{pf vs E}
   \end{figure}

Figure \ref{multi pp} displays the background subtracted pulse profiles observed in obs.\,3 in different energy ranges.
A clear phase shift appears between low energy ($\lesssim 1$~keV) and high energy profiles.
Energy-dependent pulse profiles are commonly observed in accreting X-ray pulsars. They are generally attributed to the energy-dependent beaming of the radiation emitted from the accretion column o polar caps, combined with spectral components produced in different regions of the system, such as soft ($\lesssim 1$\,keV) emission reprocessed by the inner accretion disc or the NS surface (see \citealt{Pottschmidt26} for a recent review on pulse profiles). In Sect.~\ref{sect:discussion}, we interpret the phase shift  observed in Fig.~\ref{multi pp} in the context of the scenario proposed for this source.
Figure \ref{pf vs E} shows that the pulsed fraction increases with energy,
a trend that has been observed in many other accreting pulsars (see e.g. \citealt{Lutovinov09}).

For the entire RXTE observations, we also used the \texttt{HENDRICS} \citep{2015ascl.soft02021B} tool \texttt{HENzsearch} with the \texttt{--fast} option that allows exploration of the full $\nu-\dot\nu$ plane, enabling for the detection of accelerated pulsations.
In addition,  we also ran a transient pulsation search using again the \texttt{HENDRICS} tool \texttt{HENzsearch}, this time with the \texttt{--transient} option. This method divides the observation into a power-of-two number intervals, calculates the $Z^2_n$ statistics on each interval, then combines the intervals in groups of two and re-calculates the statistics, and so on until a single interval is left.
We searched the frequency interval 14.2-14.6\,Hz with different combinations of energy bands: 
the full band, the soft band (E$<$10 keV) and the hard band (E$>$10 keV) when available.

With this search, we could reproduce the detection by \citet{Ducci25} in \nicer, which allowed us to verify the reliability of the method. Moreover, we found an additional detection in \rxte\ data (Fig.~\ref{fig Z^2}). Pulsations appear in the first 3400\,s of the observation at MJD 51150 (obsid 30107-01-01, PI Corbet).
We get a detection at more than 99.99\% level at a period of 68.88813(28)\,ms (see Fig.~\ref{fig Z^2}, right panel). 
The pulsed fraction is $(3\pm0.5)\%$ below 15 keV, and drops at higher energies.

\begin{table}
\begin{center}
\caption{Pulsed fractions and $3\sigma$ upper limits where the pulsed fraction is not significantly detected for the \xmm\ observations considered in this work.}
\label{Table p_f}
\begin{tabular}{lccc}
\hline
\hline
Obs.    & $0.2-1$~keV  &  $1-3$~keV  & $3-10$~keV        \\
\hline
obs.\,1 & $\leq2.9$\%   &  $\leq3.5$\%   &  $\leq3.4$\%   \\
obs.\,2 & $\leq4.7$\%   &  $\leq8.2$\%   &  $\leq10.4$\%  \\
obs.\,3 & $0.70\pm0.17$\% & $2.6\pm0.2$\% & $4.89\pm0.39$\%   \\
obs.\,4 & $\leq2.7$\%   &  $\leq6.7$\%   &  $\leq7.2$\%     \\
\hline
\end{tabular}
\end{center}
\end{table}

\section{Spectral results}
\label{sect:spectral results}

The spectra were modelled with the {\scriptsize{SPEX}} fitting package 3.08.01 \citep{kaastraspex}, which has a full suite of emitting / absorbing optically thin plasma. The spectra were grouped according to the optimal binning directly in {\scriptsize{SPEX}}, {which provides at least one count per bin and a bin size $\geq 1/3$ of the spectral resolution,} and fit by minimising the $C$-statistics \citep{Cash1979,Kaastra2017}.

All the emission components were corrected for absorption from the circumstellar and interstellar medium with the \texttt{hot} model (freezing the gas temperature to $10^{-6}$ keV, which provides a neutral gas in {\scriptsize{SPEX}}). To limit the computation time, we adopted the recommended solar abundances of \citet{Lodders2009}, which are the default in {\scriptsize{SPEX}}, for all emitting and absorbing plasma components.\footnote{This is generally a good approximation, although it might have an impact on the detection of the winds given that their abundances may differ from the solar pattern as  recently shown in Holmberg II (UL)X-1 \citep{Barra2024}.}
Each model is fitted simultaneously to the EPIC MOS 1, 2 / \emph{pn} and RGS first- and second-order spectra with a free multiplicative constant that accounts for the typical 1--5\,\% cross-calibration uncertainties. We fixed the ISM column density to $8\times10^{20}$\,cm$^{-2}$ \citep{HI4PI16} since it had an insignificant impact on the spectral modelling.

For all observations we adopted the following steps: continuum modelling of the high-statistics EPIC spectra, addition of RGS, implementation of line emission from optically thin plasma, photoionised grids, and best-fit physical model. 

\subsection{Continuum modelling}
\label{sect:continuum}

\begin{figure*}
   \centering
   \includegraphics[scale=0.32500]{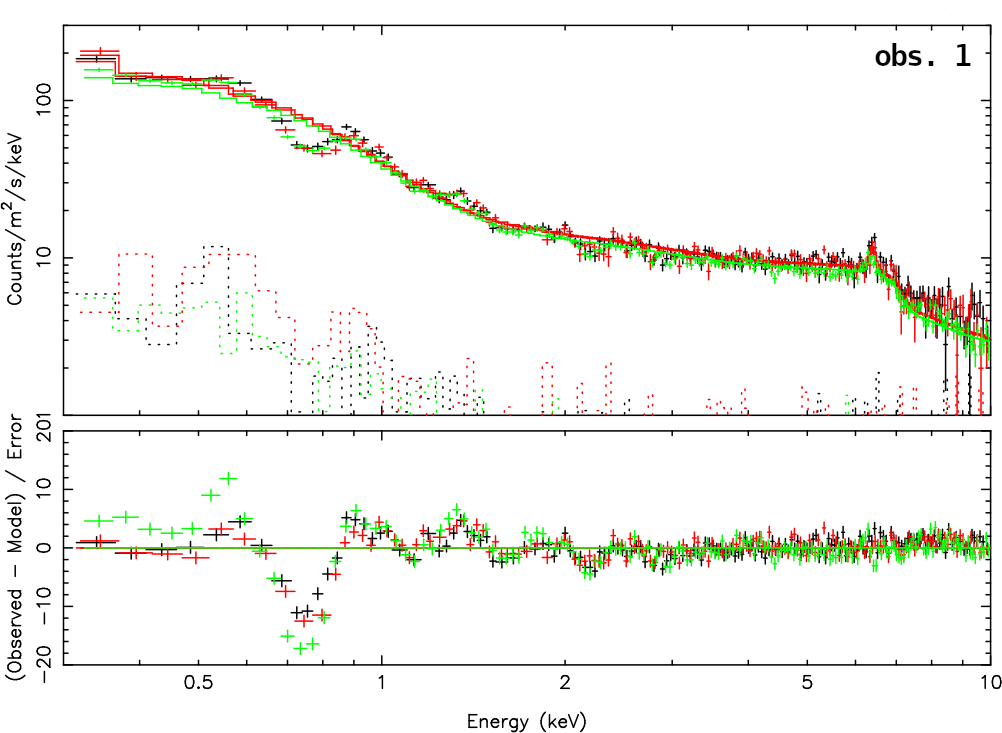} \hspace{0.cm}
   \includegraphics[scale=0.32500]{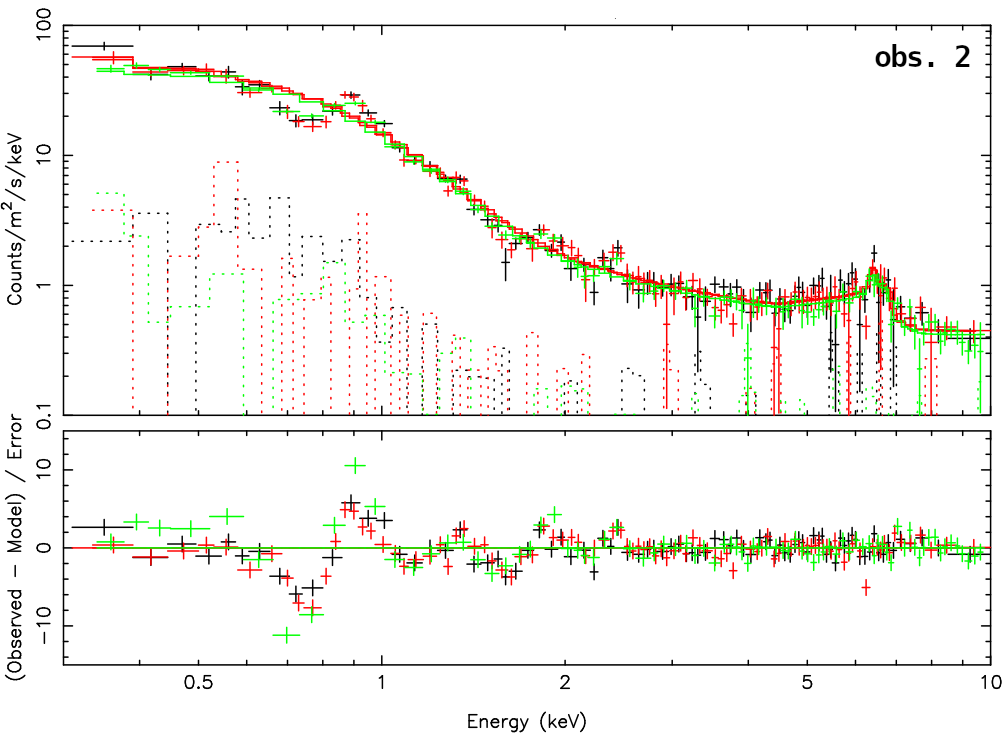}
   \includegraphics[scale=0.320500]{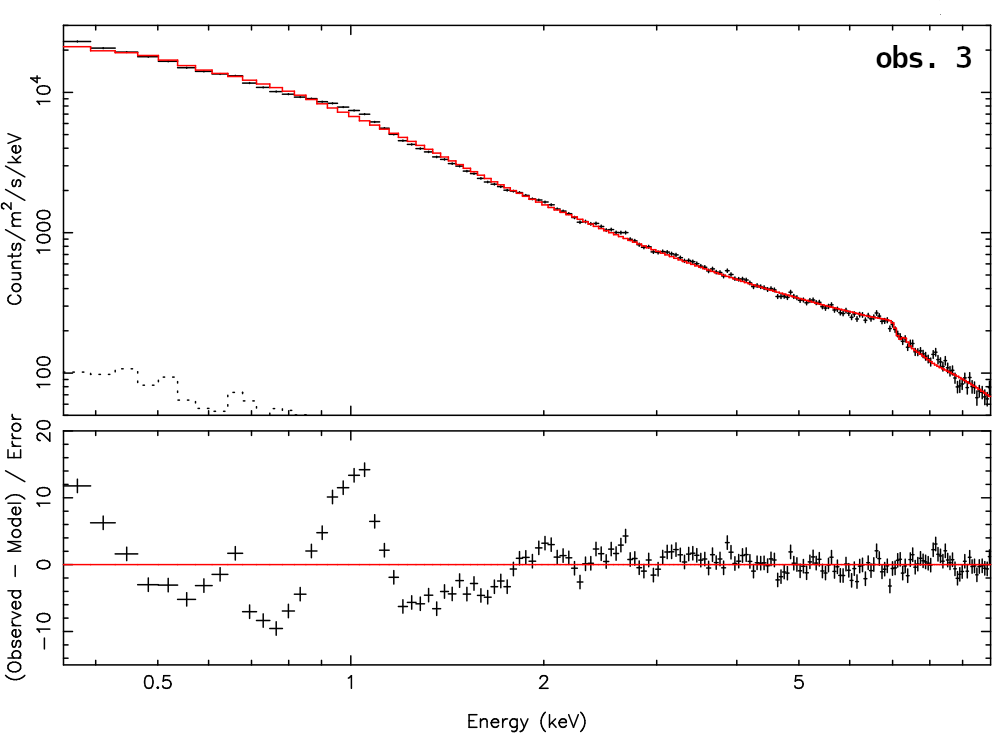} \hspace{0.cm}
   \includegraphics[scale=0.320500]{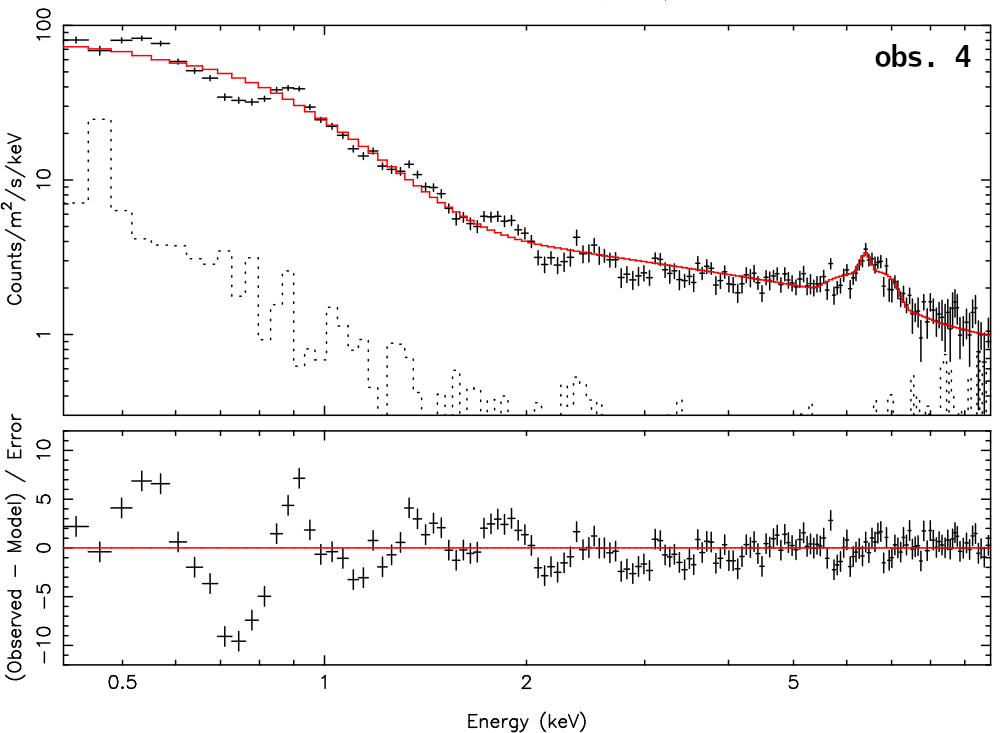}
   \caption{\xmm\ EPIC MOS 1 (black), 2 (red), and \emph{pn} (green) spectra and continuum modelling of the individual observations. 
   For the third and fourth observations, only EPIC-pn spectra were used.}
    \label{fig:spex_continuum}
\vspace{-0.1cm}
    \end{figure*}

We first model the broadband MOS and \emph{pn} spectra of each observation simultaneously. For obs.\,3 and obs.\,4 we could only use the \emph{pn} spectra. We used a three-component emission model  composed of a cool blackbody (about 0.15$-$0.18\,keV, \texttt{bb} component in {\scriptsize{SPEX}}), a cutoff power law with slope that significantly vary between the observations ($\Gamma \sim 0.4-1.8$) and a relativistic reflection component to model the broad Fe\,K line (\texttt{refl} component). The best-fit continuum model is shown in Fig.\,\ref{fig:spex_continuum}.
Sharp residuals can be seen around 1\,keV, similarly to those observed in ultra-luminous X-ray sources (ULXs; e.g. \citealt{Pinto2023}). Two cutoff (\texttt{etau} components in {\scriptsize{SPEX}}) were adopted at 0.1\,keV and 7\,keV. The first \texttt{etau} is a low-energy cutoff inserted to prevent the power law (especially in obs.\,3) to diverge at energy below 0.3\,keV, while the second \texttt{etau} mimics the curvature at 7\,keV that appears in the spectra, especially the ULX-like spectrum in obs.\,3 (such a 7\,keV curvature is present in all pulsating ULXs; see e.g. \citealt{Pintore2017}).

The best-fitting spectral parameters, obtained with the ionisation parameter of the reflection component fixed to zero (neutral reflector), are reported in Table~\ref{table:fit_params}. We verified that leaving this parameter free does not significantly improve the fit, owing to the limited spectral resolution around $\sim7$\,keV. Given the phenomenological nature of the reflection model, we therefore adopt the simpler neutral-reflector scenario throughout.

\subsection{Emission-line plasma modelling}
\label{sect:emission-line}

\begin{figure*}
   \centering
   \includegraphics[scale=0.32500]{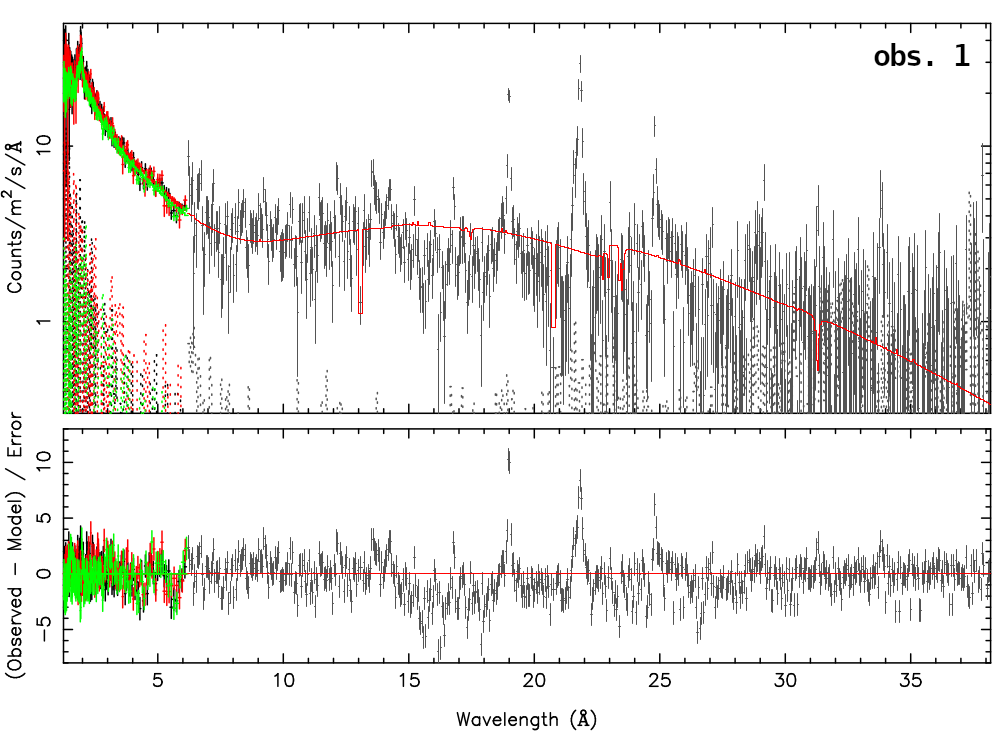} \hspace{0.cm}
   \includegraphics[scale=0.32500]{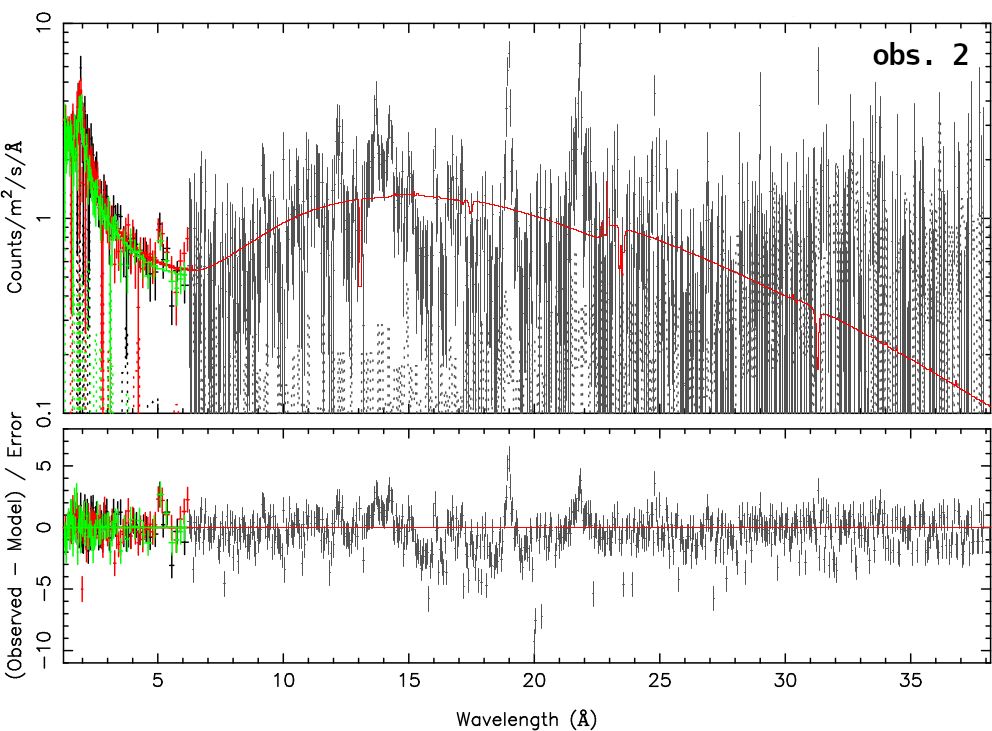}
   \includegraphics[scale=0.320500]{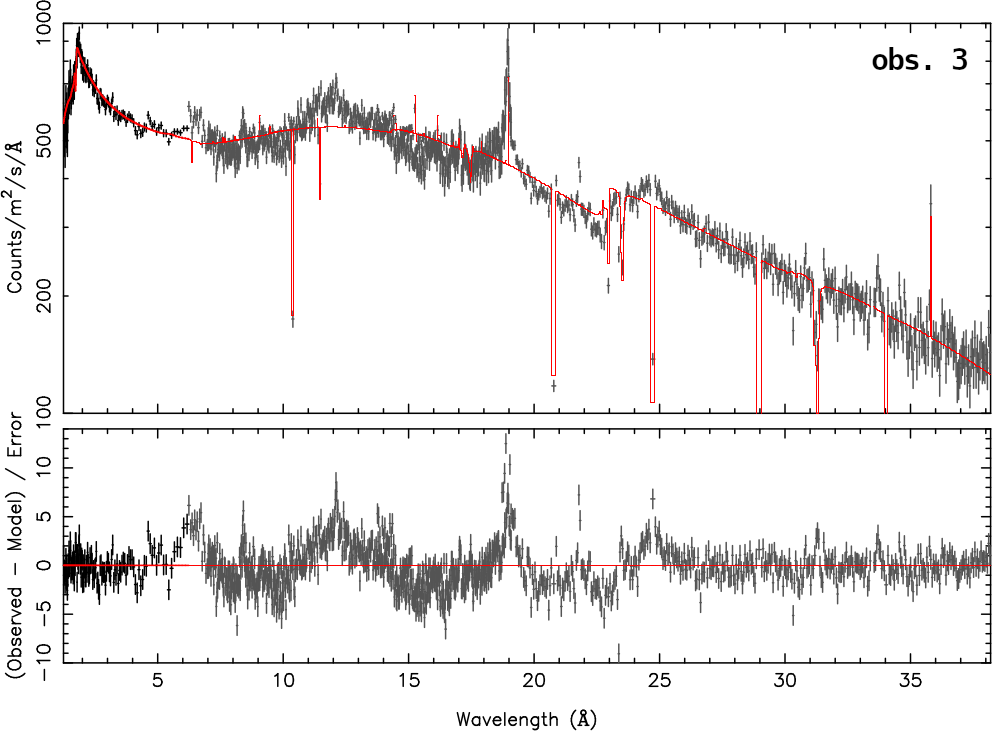} \hspace{0.cm}
   \includegraphics[scale=0.320500]{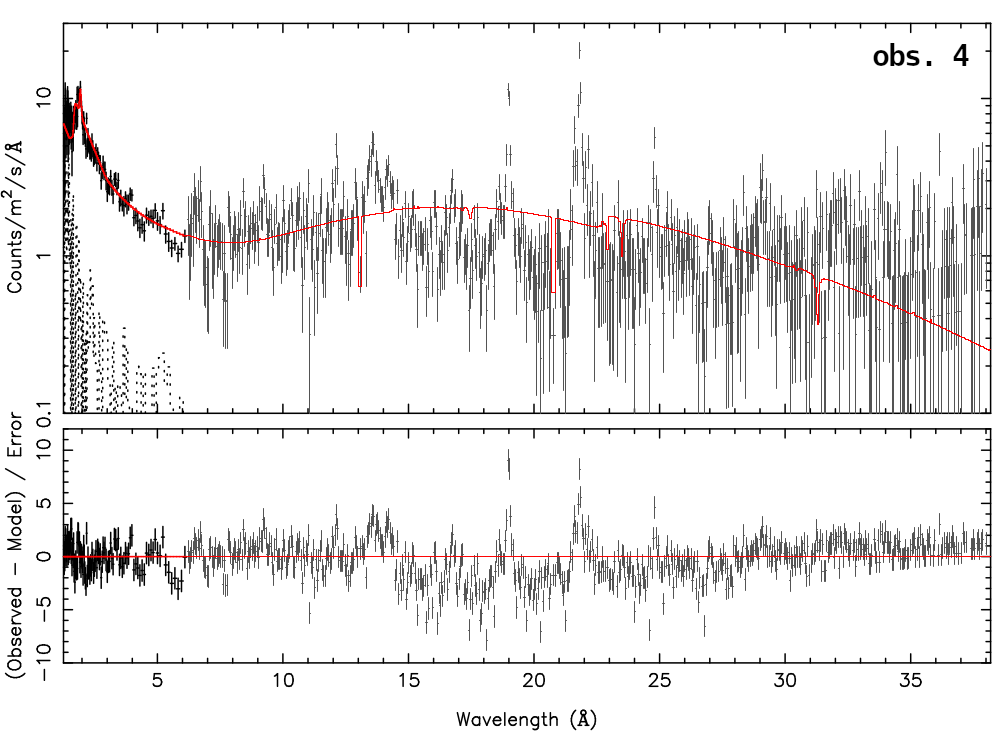}
   \caption{\xmm\ EPIC MOS 1 (black), 2 (red), \emph{pn} (green), and RGS (grey) spectra and continuum modelling of the individual observations. For the third and fourth observations, only EPIC-\emph{pn} and RGS were used. For the third observation only RGS 1 (first and second orders) are used due to RGS 2 pile-up.
   EPIC data were ignored below 2\,keV (i.e. above 6\,\AA) to take advantage of the higher spectral resolution of RGS.
   }
   \label{fig:spex_continuum_RGS}
   \vspace{-0.1cm}
   \end{figure*}

In order to fully exploit  the high spectral  resolution of the RGS and decrease the degeneracy between models, we ignored the EPIC data below 2\,keV, i.e. in the operating energy band of the RGS spectrometers. The RGS detector has an energy resolution that is constant in wavelengths, and therefore from this point on, we plot the spectra as a function of wavelength (see Fig.\,\ref{fig:spex_continuum_RGS}). By visual inspection, it is easy to spot the well-known O\,{\small VIII} H-like emission line around 19\,{\AA}, the He-like O\,{\small VII} emission feature near 22\,{\AA}, and the 23\,{\AA} oxygen K-edge, which are routinely found in HMXBs and in most high-quality RGS spectra of ULXs (\citealt{Kosec2021}). Strong emission lines are also found in correspondence of the He-like triplets of neon and nitrogen at 14 and 29\,{\AA}, along with the Lyman $\alpha$ lines of Ne\,{\small X} and N\,{\small VII} at 12.1 and 24.8\,{\AA}, respectively.

\begin{figure*}
   \centering
   \includegraphics[scale=0.32500]{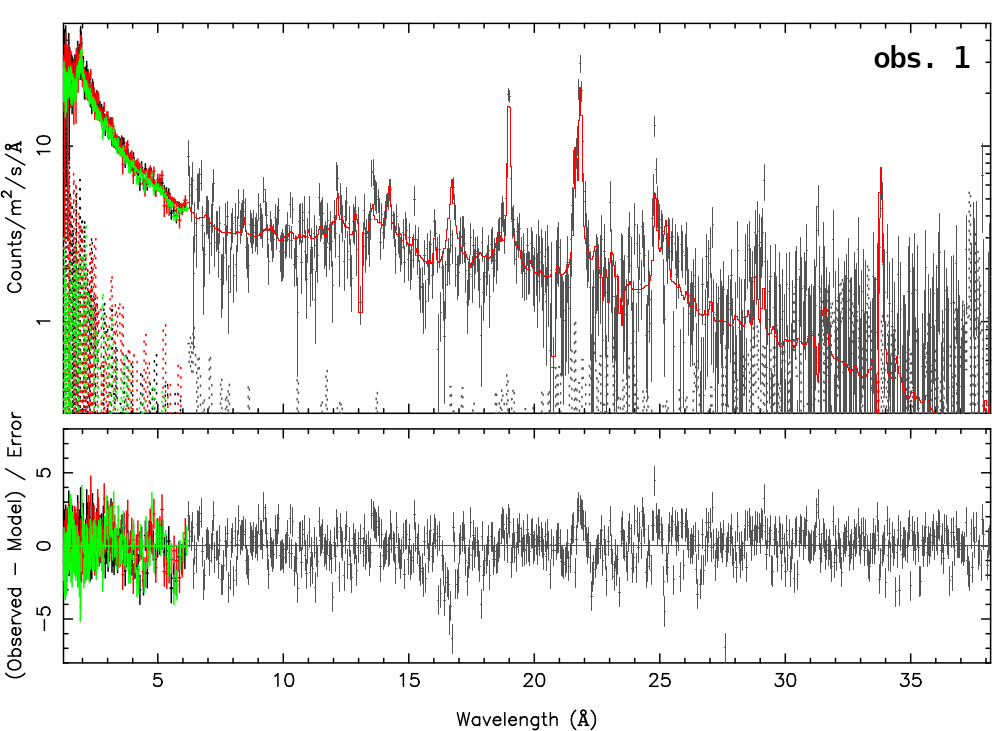} \hspace{0.cm}
   \includegraphics[scale=0.32500]{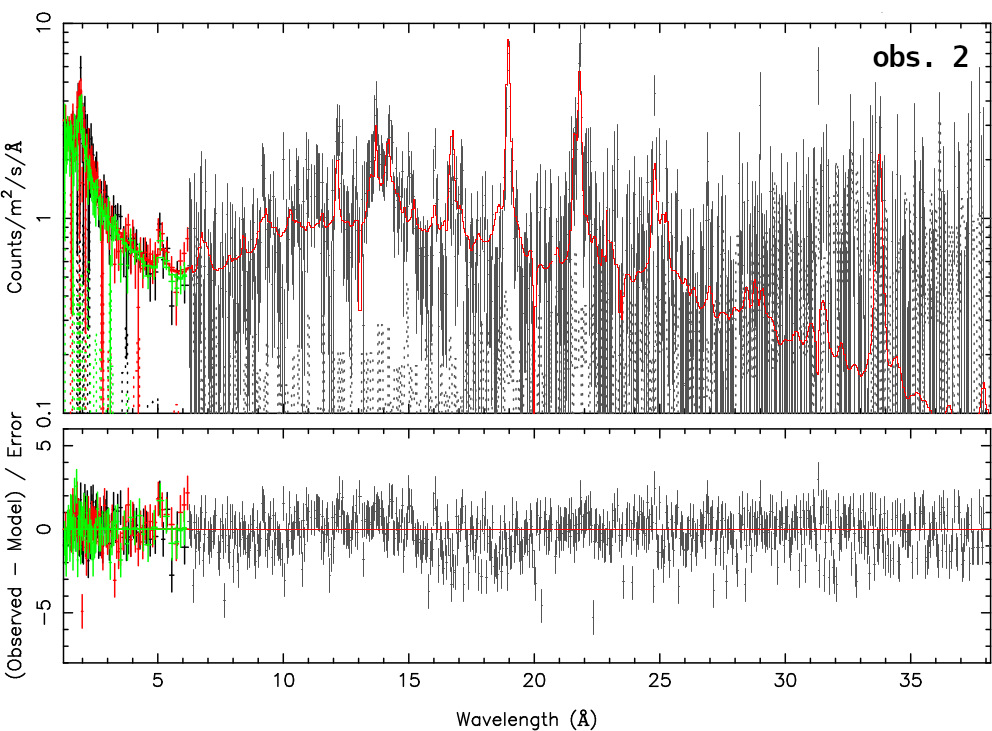}
   \includegraphics[scale=0.320500]{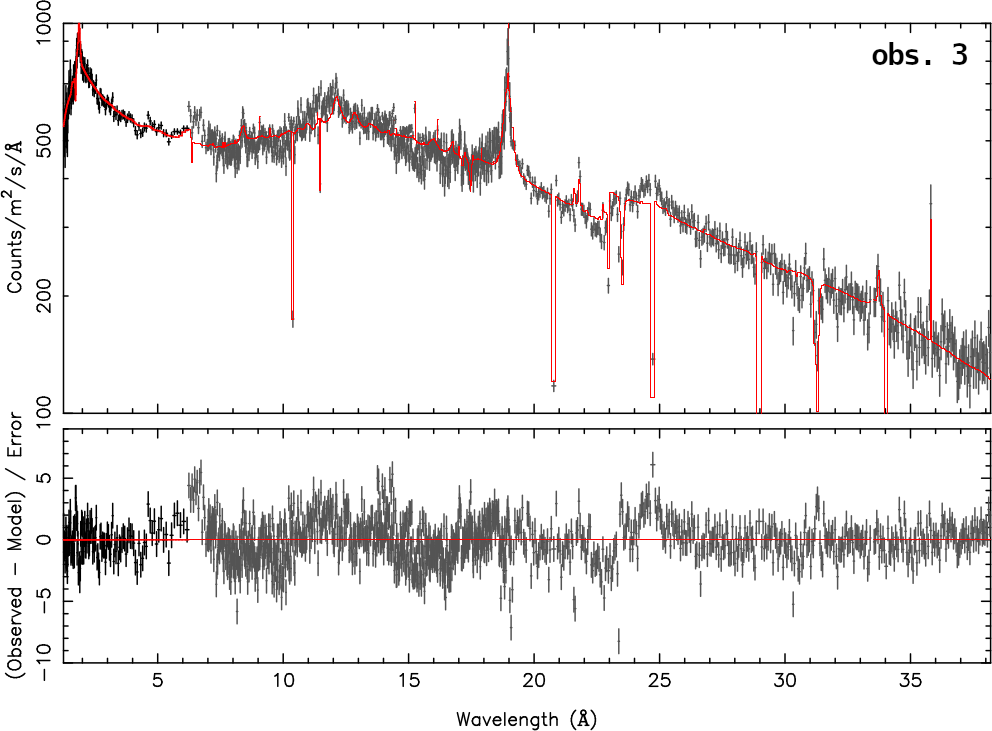} \hspace{0.cm}
   \includegraphics[scale=0.320500]{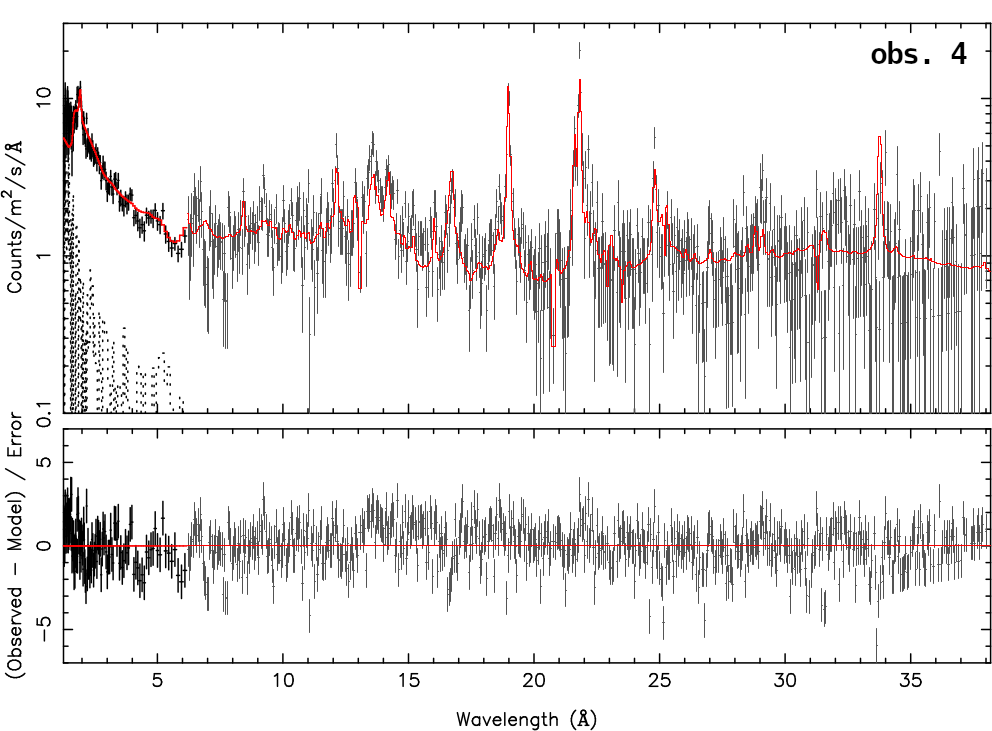}
   \caption{\xmm\ EPIC and RGS spectra and line emission modelling of the individual observations. Two photoionised emission components 
   (\texttt{pion}) have been employed for each spectrum.
   }
   \label{fig:spex_pion}
   \vspace{-0.1cm}
   \end{figure*}

We attempted to model these emission lines with a combination of collisionally-ionised and photoionised plasma components (\texttt{cie} and \texttt{pion} in {\scriptsize{SPEX}}). 
The \texttt{cie} models are easily computed starting from their electron temperature, while \texttt{pion} receives the current continuum model in input and instantaneously computes the ionisation balance, with the main parameter being the ionisation parameter ($\xi=L_{\rm ion}/n_{\rm H}R^2$, where $L_{\rm ion}$ is the ionising luminosity, $n_{\rm H}$ the density, and $R$ the distance of the plasma from the ionising source).
Collisional models provide worse fits than photoionisation due to the weak resonant lines, especially the O\,{\small VII} at 21.6\,{\AA}, which are four times weaker than the very strong intercombination line. The latter is also stronger (by a factor of five) than the forbidden line in all observations, resulting in a high density for the emission-line plasma (from a few $\times 10^{11}$\,cm$^{-3}$ to a few $\times 10^{13}$\,cm$^{-3}$). The broad range of ionisation states (from N\,{\small VI} to Ne\,{\small X}) cannot be fully described with a single component, and therefore we implemented a two-\texttt{pion} model, which provides a satisfactory description of the emission lines in obs.\,1, 2 and 4 (see Fig.\,\ref{fig:spex_pion}). The ionisation parameters of these two components do not vary significantly between the four observations (log $\xi_1 \, \rm [erg/s\,cm] \sim 1.2-1.4$ and log $\xi_2 \rm [erg/s\,cm] \sim 2.2-2.8$ with typical uncertainties of 0.1 and 0.3, respectively). The emission components are consistent with being at rest, i.e. with Doppler shifts of a few 100\,km\,s$^{-1}$, compatible with the spectral resolution and comparable broadening. There is excess in the data (with respect to the model) in the correspondence of the nitrogen VI and VII lines, suggesting a very likely nitrogen over-abundance with respect to oxygen and other species that can occur in Be-like XRBs and has also been observed in ULXs \citep{Barra2024}. 

Obs.\,3, where \src\ shows a ULX-like luminosity, is the exception here. Not only are two emission line components not sufficient to describe all the features, but the lines are characterised by an asymmetric and broader shape (which sticks out strongly for O\,{\small VIII} at 19\,{\AA}). Indeed, a satisfactory fit can be obtained by including a third \texttt{pion} component with a broadening and blueshift of about 6000\,km\,s$^{-1}$ each.

\subsection{Absorption-line plasma modelling}
\label{sect:absorption-line}

The emission-line modelling leaves some residuals in the spectra. In some cases are positive, suggesting  a non-perfect modelling of the emission lines at rest (see e.g. N\,{\small VII} at 24.8\,{\AA} and Ne\,{\small IX} at 13.4\,{\AA}),  likely due to the simplistic models adopted (e.g. the solar abundances). Some residuals are instead and frequently found between 15.5-16.5\,{\AA} where no major lines are expected. This means that there might be absorption by moving plasmas, which is expected at accretion rates around or above the Eddington limit. 

Modelling of the absorption lines is also performed with another \texttt{pion} component, with exception that here the solid angle is assumed equal to zero and the covering fraction equal to one (i.e. pure absorbing plasma, the opposite of the assumptions for the line emission). 
In order to avoid getting stuck into local minima, before performing direct fits, we performed a deep exploration of the parameter space for the two most relevant parameters: the Doppler shift or velocity projected along the line of sight ($v_{\rm LOS}$) and the plasma state parameter, $\xi$.

\begin{figure*}
   \centering
   \includegraphics[scale=0.500]{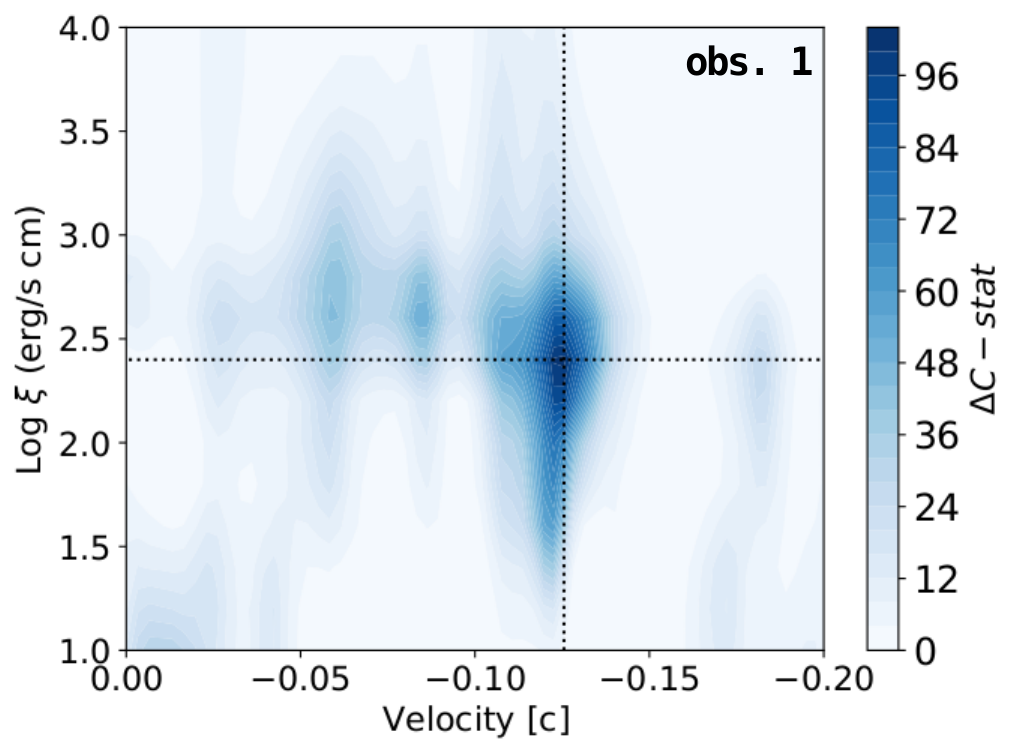} \hspace{0.cm}
   \includegraphics[scale=0.500]{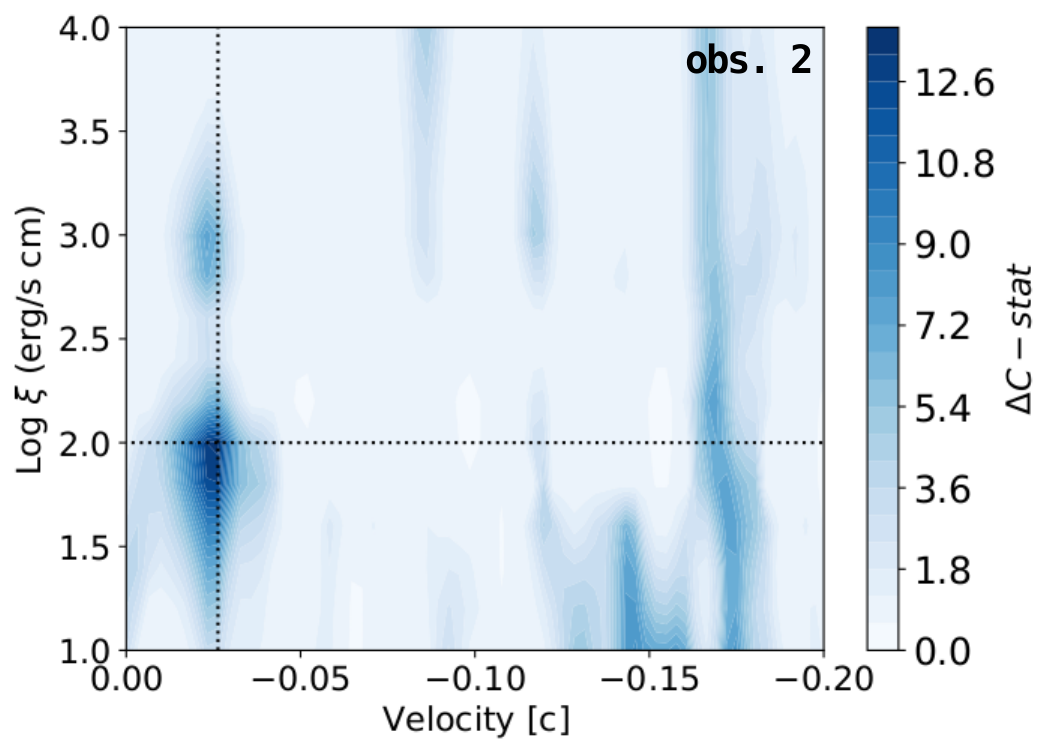}
   \includegraphics[scale=0.5000]{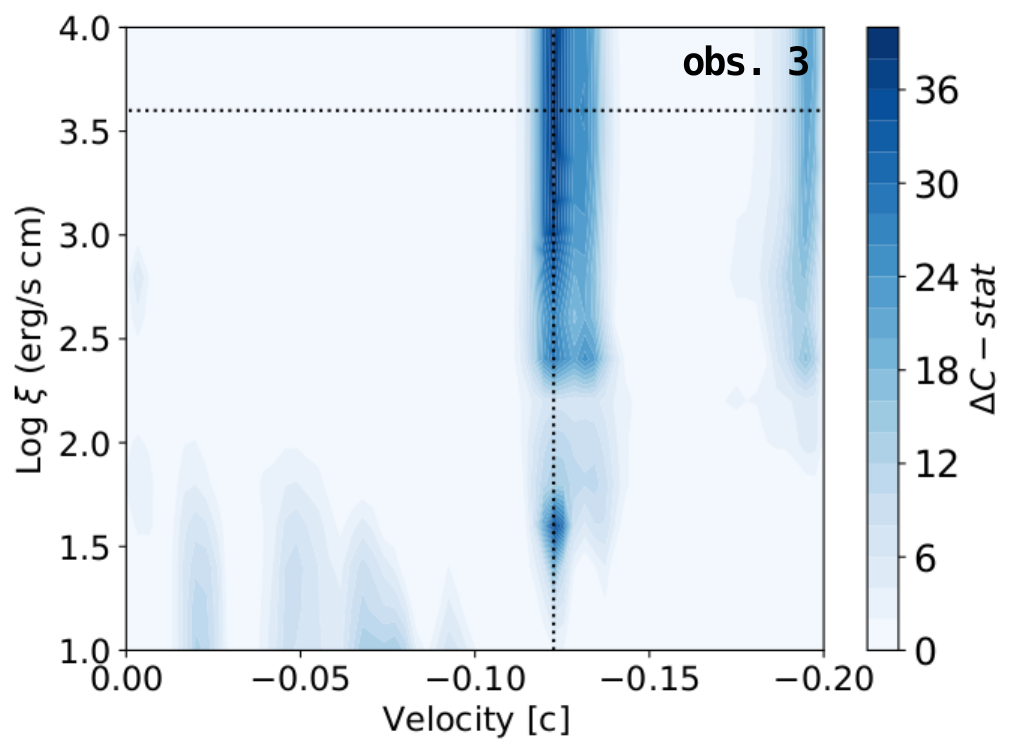} \hspace{0.cm}
   \includegraphics[scale=0.5000]{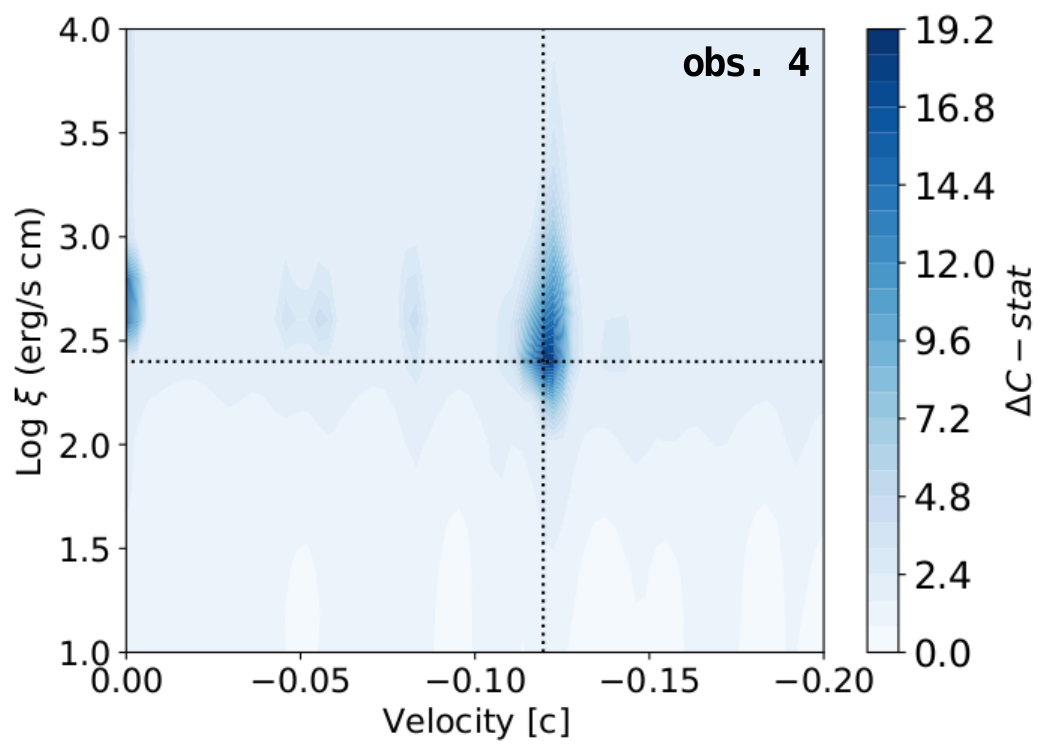}
   \caption{\xmm\ physical model grids of the individual observations: photoionised plasma in absorption. The results show the improvement in the spectral fits provided by the absorbing plasma on top of a model consisting of continuum plus two line-emission components. The dotted black lines identify the best-fitting solution.
   }
   \label{fig:spex_scan_xabs}
   \vspace{-0.1cm}
   \end{figure*}

We adopted a log $\xi$ grid between 1 and 4 (31 points), which provides the range where most He- and H-like transitions from the ions mentioned above are expected and a linear grid of line-of-sight velocities, $v_{\rm LOS}$,
between zero and $-0.3c$ (blueshifts or outflowing plasma as in typical XRB and ULX winds; see e.g. review by \citealt{Pinto2023}). We tested a velocity dispersion $v_{\sigma}$ of 1000 with LOS velocity steps of 1000\,km\,s$^{-1}$, respectively, to speed up the computation, as was done in previous work (see e.g. \citealt{Kosec2018b}). The results are shown in Fig.\,\ref{fig:spex_scan_xabs}. Aside from some small degeneracy between multiple solutions, it is obvious that the absorbing plasma flows out with a relativistic speed of $-0.12c$ for three out of four observations.
The very large $\Delta C$-stat, in two cases above 40, suggests a very high confidence level, most likely above $5\,\sigma$. In these cases there is no need to perform expensive Monte Carlo simulations (see \citealt{Pinto2026a,Pinto2026b}). In obs.\,2 there is no detection most likely due to the low, noisy continuum in the RGS spectrum, while in obs.\,4 the detection is weak but encouragingly at the same location.  

\begin{figure*}
   \centering
   \includegraphics[scale=0.32500]{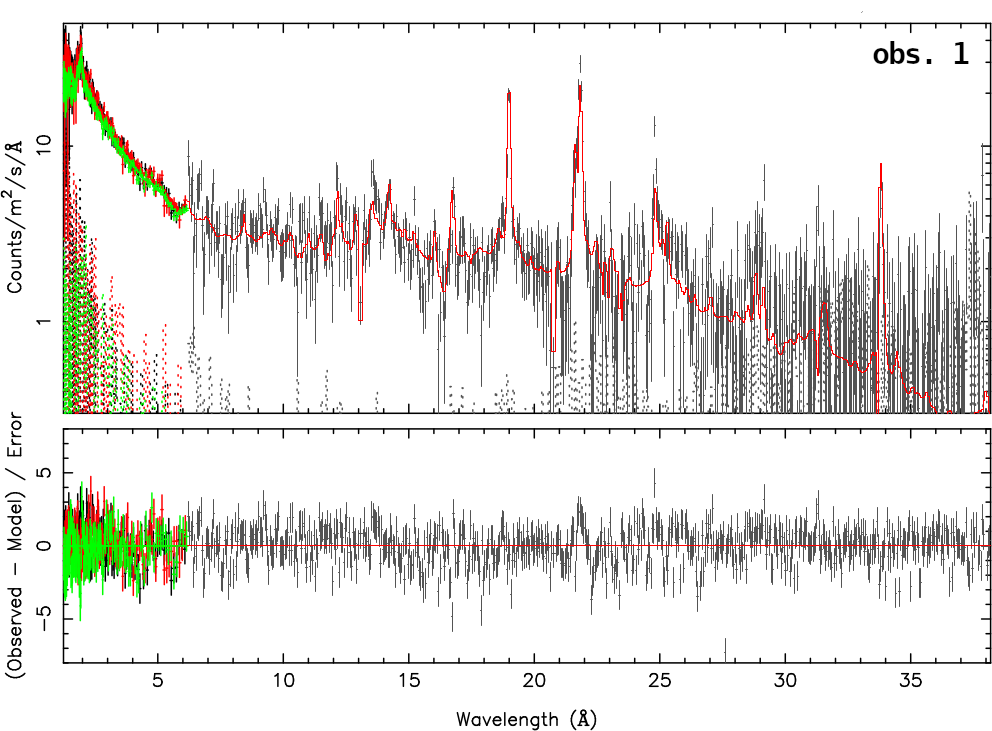} \hspace{0.cm}
   \includegraphics[scale=0.32500]{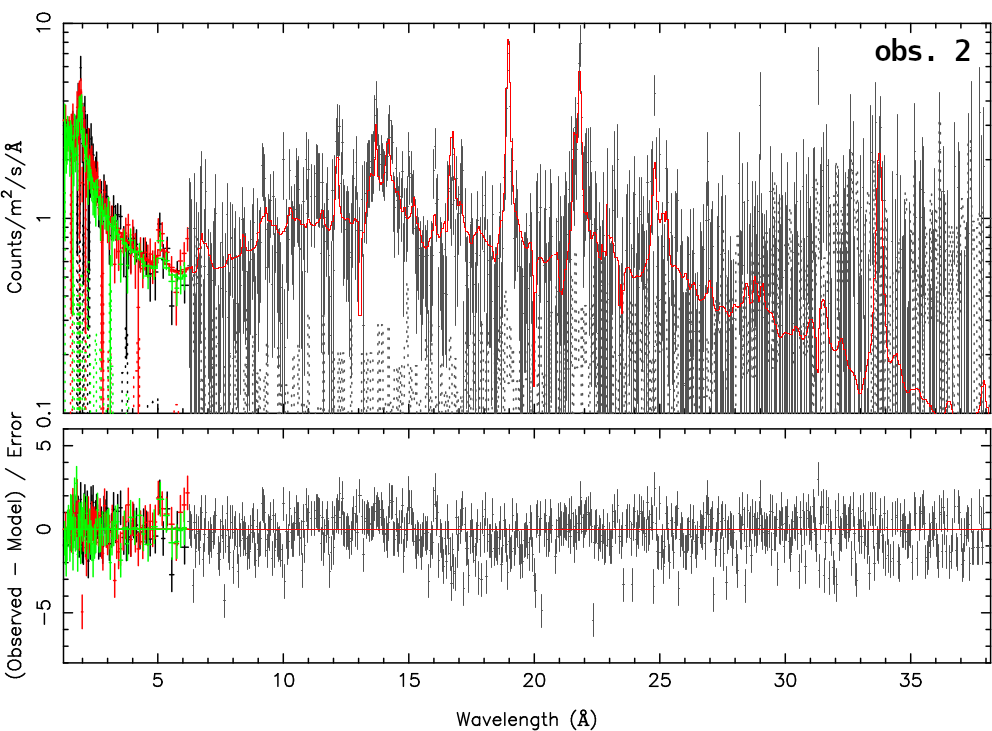}
   \includegraphics[scale=0.320500]{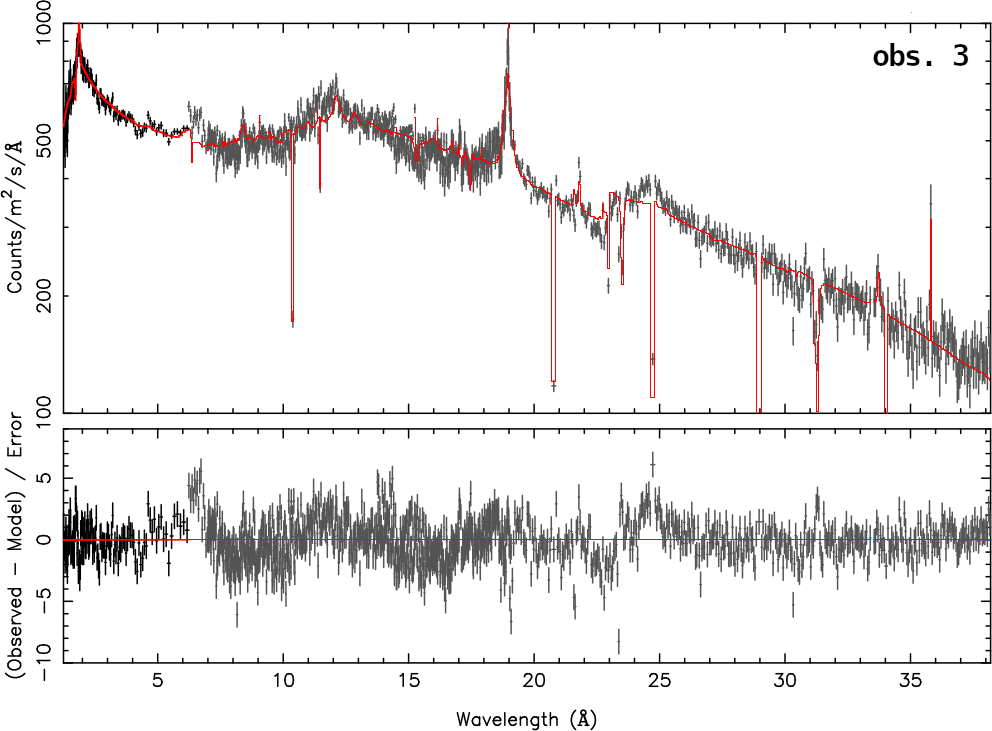} \hspace{0.cm}
   \includegraphics[scale=0.320500]{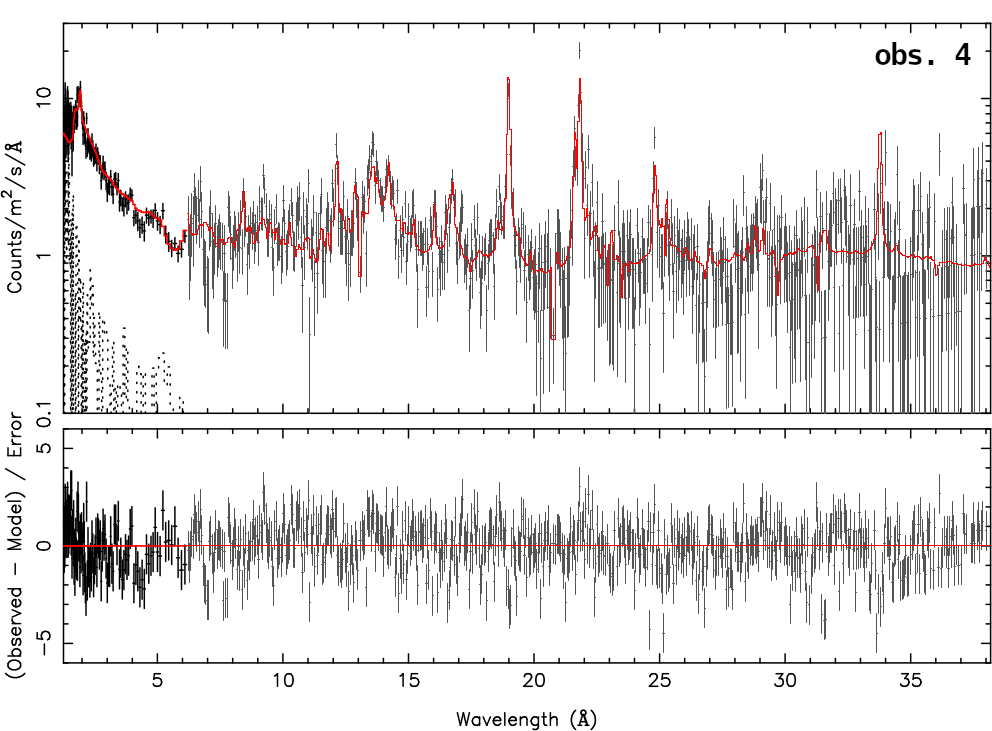}
   \caption{\xmm\ EPIC and RGS spectra and line emission and absorption modelling of the individual observations. Two photoionised emission components plus a photoionised absorption component have been employed for each spectrum.
   }
   \label{fig:spex_pion_xabs}
   \vspace{-0.1cm}
   \end{figure*}

Starting from these solutions we refit each spectrum with the complete model consisting of continuum plus three \texttt{pion} components (two in emission and one in absorption). This provides the current best-fit models as shown in Fig.\,\ref{fig:spex_pion_xabs}. As it can be noticed, spectral residuals are now generally small, with the exception of obs.\,3 for which a third, broader and blueshifted \texttt{pion} component was needed (see above). The photoionised absorption components improve the fits in correspondence  of the Ne\,{\small IX-X}, O\,{\small VIII} and some Fe L lines that are blueshifted by $0.12c$ at 10.5-11.5 and 15.5-16.5\,{\AA}.

\begin{table*}
\begin{center}
\caption{Best-fit parameters for the four \xmm\ observations. \label{table:fit_params}}
\begin{tabular}{l c c c c c}
\hline
\noalign{\smallskip}
Parameter & Units & obs.\,1 & obs.\,2 & obs.\,3 & obs.\,4 \\
\noalign{\smallskip}
\hline
\noalign{\smallskip}
BB norm
& $10^{16}$\,m$^2$
& $8.91{+0.05\atop-0.04}\times10^{-5}$
& $1.44{+0.04\atop-0.04}\times10^{-5}$
& $1.50{+0.05\atop-0.08}\times10^{-3}$
& $2.61{+0.12\atop-0.11}\times10^{-5}$ \\
\noalign{\smallskip}
BB $kT$
& keV
& $0.13683{+0.00014\atop-0.00013}$
& $0.1742{+0.0009\atop-0.0009}$
& $0.1748{+0.0012\atop-0.0008}$
& $0.1645{+0.0017\atop-0.0017}$ \\
\noalign{\smallskip}
Power law norm
& $10^{44}$\,ph\,s$^{-1}$\,keV$^{-1}$ 
& $10.09{+0.03\atop-0.03}$
& $1.05{+0.07\atop-0.09}$
& $2444{+10\atop-15}$
& $3.08{+0.20\atop-0.15}$ \\
\noalign{\smallskip}
Photon index
&
& $0.7809{+0.0020\atop-0.0017}$
& $0.84{+0.09\atop-0.11}$
& $2.097{+0.010\atop-0.005}$
& $1.00{+0.11\atop-0.08}$ \\
\noalign{\smallskip}
Reflection norm
& $10^{44}$\,ph\,s$^{-1}$\,keV$^{-1}$ 
& $65.96{+0.48\atop-0.52}$
& $7.6{+2.7\atop-1.6}$
& $9.85{+0.24\atop-0.40}\times10^{3}$
& $34.3{+10.5\atop-7.6}$ \\
\noalign{\smallskip}
$\cos i$
&
& $0.7575{+0.0057\atop-0.0014}$
& $0.69{+0.03\atop-0.63}$
& $0.751{+0.010\atop-0.006}$
& $0.697{+0.015\atop-0.030}$ \\
\noalign{\smallskip}
$R_{\rm in}$
& $GM/c^2$
& $6.0014{+0.0252\atop-0.0004}$
& $12.8{+3.4\atop-0.9}$
& $6.00{+0.06\atop-0.00}$               
& $6.08{+0.24\atop-0.08}$ \\
\noalign{\smallskip}
Gaussian norm
& $10^{44}$\,ph\,s$^{-1}$
& $0.111{+0.013\atop-0.015}$
& $3.23{+0.59\atop-0.53}\times10^{-2}$
& $0.00{+0.09\atop-0.00}$               
& $0.081{+0.014\atop-0.014}$ \\
\noalign{\smallskip}
\hline
\end{tabular}
\end{center}
\noindent\emph{Notes.}\\
\noindent $N_{\rm H}$ was frozen at $8\times10^{20}$\,cm$^{-2}$.\\
\noindent The power law and reflection normalisations are given at 1\,keV.\\
\noindent Two cutoff components were used, with $\tau=1$.\\
\noindent High-energy cutoff: $E_{\rm cut}=7.0$\,keV.\\
\noindent Low-energy cutoff: $E_{\rm cut}=0.1$\,keV.\\
\noindent The reflector ionisation parameter was fixed to $\xi=0$ (neutral reflector).\\
\noindent The Gaussian line centroid was fixed to 6.4 keV (neutral reflector). 

\end{table*}

   \begin{figure*}
     \centering
     \includegraphics[width=1.9\columnwidth]{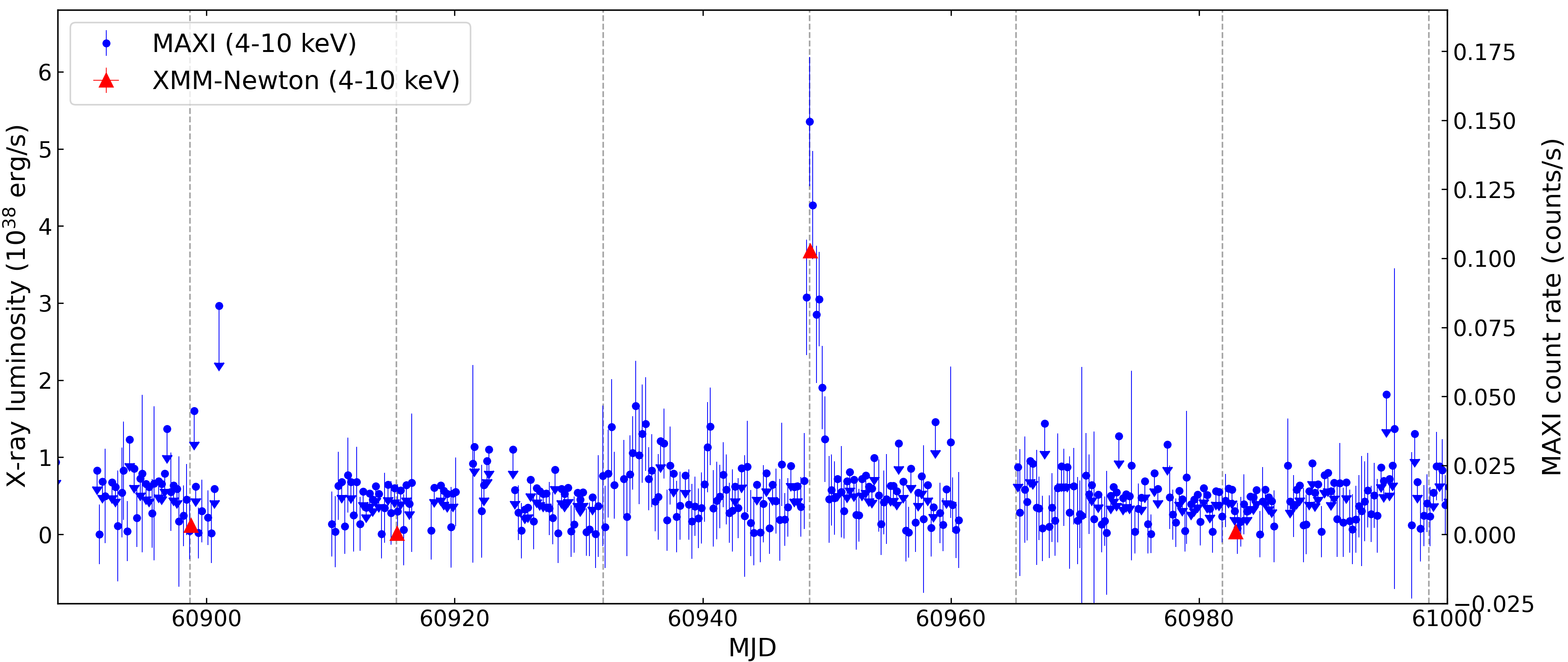}
   \caption{Long-term $4-10$~keV variability of \src\ from MAXI (blue circles) with the four \xmm\ observations (red triangles) overplotted. MAXI bins consistent with a count rate $<0$ are plotted as 2$\sigma$ upper-limits (downward arrows). Vertical dashed grey lines indicate the expected times of the optical bursts, calculated from the $T_0$ and orbital period from \citet{Ducci22}. 
   }
   \label{fig. maxi xmm lcr}
   \end{figure*}

\section{Discussion}
\label{sect:discussion}

\subsection{Return to $\sim 10^{39}$\,erg\,s$^{-1}$ outbursts after about 45 years.}
\label{sect:results 1}

\src\ was observed by \xmm\ four times near periastron (see orbital phases in Table \ref{Log of the observations}). 
In Fig.~\ref{fig. maxi xmm lcr} we put the luminosities measured with \xmm\ in the context of the long-term light curve obtained with  the MAXI monitoring in the  4$-$10~keV range.
In obs.\,3, the unabsorbed 0.2$-$10~keV flux reached $\sim 4.9\times 10^{-9}$~erg~cm$^{-2}$~s$^{-1}$, corresponding to a luminosity of $\sim 1.5\times 10^{39}$~erg~s$^{-1}$ (see Table \ref{tab:om vs X-ray}).
This was the highest luminosity level observed in this source since the \emph{Einstein} observations in 1980 \citep{Skinner82}. In the other three \xmm\ observations, the 0.2$-$10~keV luminosity was $(2-20)\times10^{36}$~erg~s$^{-1}$.

   \begin{figure}
     \centering
     \includegraphics[width=\columnwidth]{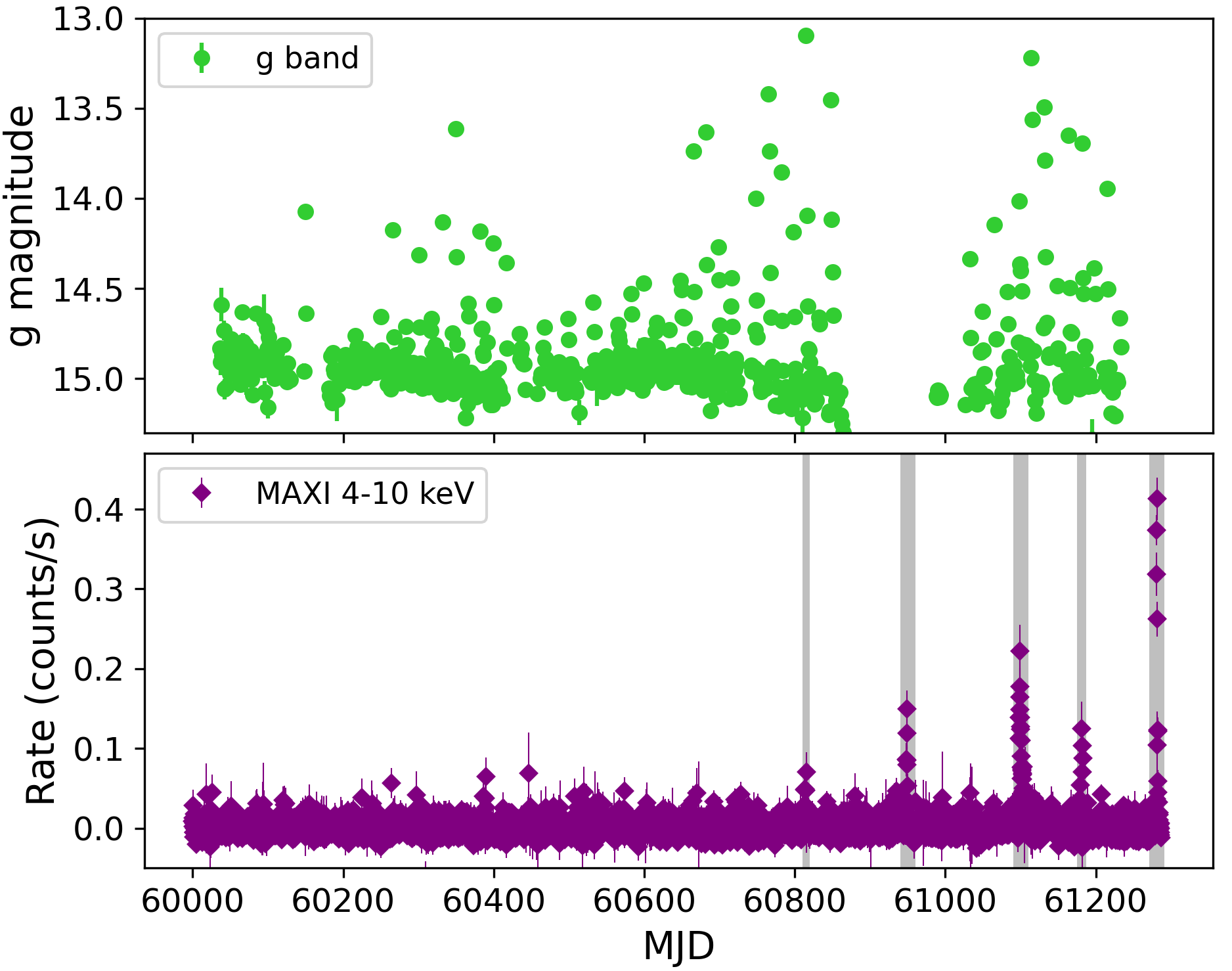}
  \caption{Long-term monitoring of \src\ from MJD~60000 to $\sim$61200. \textit{Top:} REM $g$-band light curve. \textit{Bottom:} MAXI/GSC light curve in the 4$-$10~keV band, binned at 6~hr.  Grey shaded regions mark the five X-ray outbursts detected by MAXI (see text).}
\label{fig:lc_maxi_rem}
   \end{figure}

Figure~\ref{fig:lc_maxi_rem} shows the simultaneous MAXI (4--10\,keV) and REM ($g$-band) light curves of \src\ from MJD~60000 (25 February 2023) to $\sim$MJD~61286 (3 September 2026). Apart from a few gaps in the REM data due to operational interruptions at the telescope, the two monitoring campaigns are nearly simultaneous over the whole interval. The MAXI light curve reveals five distinct X-ray outbursts,  highlighted by the grey-shaded regions in the figure. 
The last four events were already reported. They occurred 
on 30 September 2025 (MJD: 60948; $L_{\rm peak}\approx5\times 10^{38}$~erg~s$^{-1}$; \citealt{Nakajima25}; also detected in \xmm\ obs.\,3),
on 27 February 2026 (MJD: 61098; $L_{\rm peak}\approx 8\times 10^{38}$~erg~s$^{-1}$; \citealt{Nakajima26a}),
on 21 May 2026 (MJD: 61181; $L_{\rm peak}\approx 3\times 10^{38}$~erg~s$^{-1}$; \citealt{Nakajima26b}),
and on 28 August 2026 (MJD: 61280; $L_{\rm peak}\approx 1.1\times 10^{39}$~erg~s$^{-1}$; \citealt{Nakajima26c})
The first   outburst,   reported here for the first time, occurred on  20th May 2025 (MJD~60815) and reached $L_{\rm peak}\approx 1\times 10^{38}$~erg~s$^{-1}$ (all the quoted MAXI luminosities refer to the 4--10 keV range). 
For the August 2026 outburst, using the \xmm\ spectrum of obs.\,3 to extrapolate the MAXI luminosity to the 0.2$-$10~keV band, we obtain $L_{\rm x} \approx 3\times 10^{39}$~erg~s$^{-1}$. 
This is the brightest X-ray outburst detected so far from \src\ and implies, together with the other outbursts  at $10^{39}$~erg~s$^{-1}$ reported here, that the source belongs to the class of transient pulsating ULXs. This class comprises a few nearby pulsars that reach ULX-like luminosities during outbursts lasting about a month 
\citep[e.g. ][]{Wilson-Hodge18, Vasilopoulos20}.
\src\ stands out among them, since its outbursts have a much sharper profile and exceed 
$10^{39}$~erg~s$^{-1}$ for only a few hours.

   \begin{figure*}
     \centering
     \includegraphics[width=\columnwidth+\columnwidth]{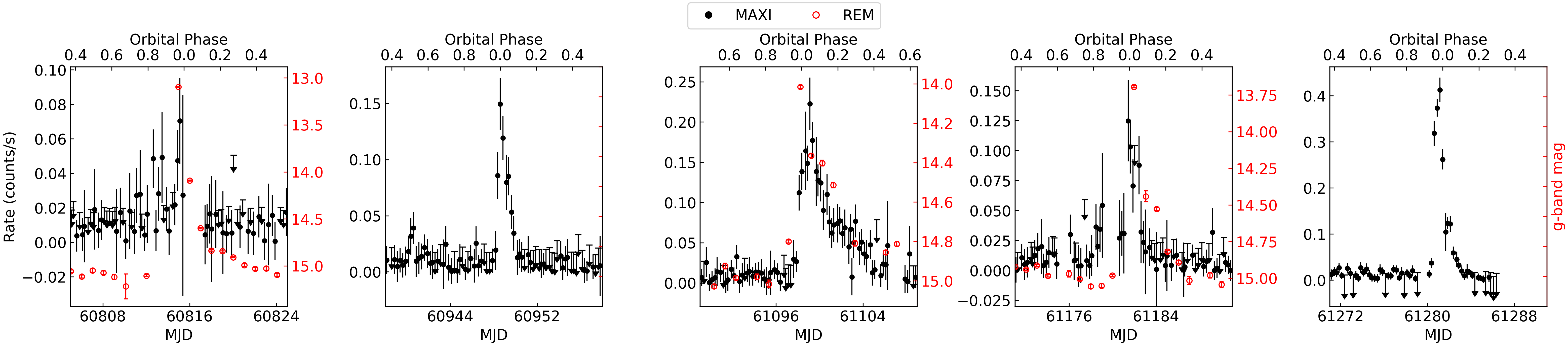}
   \caption{Five outbursts of \src\ detected by MAXI. Black filled circles: MAXI 4$-$10\,keV; red open circles: REM $g$-band magnitude.}
   \label{fig. maxi rem 4 bursts}
   \end{figure*}

A close-up of these five outbursts is shown in Fig.~\ref{fig. maxi rem 4 bursts}, together with the simultaneous REM photometry, where available. The figure also reports the corresponding orbital phase, computed adopting $P_{\rm orb}=16.64002$\,d and $T_0=55673.71$\,MJD, with the latter corresponding to the phase of the optical maximum derived by \citet{Ducci22} from archival REM, OGLE, and MACHO data ($<59000$\,MJD). For the three X-ray outbursts with simultaneous optical coverage, 
Fig.~\ref{fig. maxi rem 4 bursts} shows that
the X-ray and optical events occur nearly simultaneously, 
with only small shifts in phase. Interestingly, for the outburst at MJD\,$\sim$61098 the optical peak 
seems to slightly precede the X-ray peak.

To further characterise the optical burst properties, we focused on the recent REM monitoring campaign ($\geq 60000$\,MJD). We folded this subset of data on the orbital ephemeris given above and compared the resulting profile with the optical orbital profile previously derived from archival REM, OGLE, and MACHO data by \citet{Ducci22}. 
We show this comparison in Fig.~\ref{fig:fold_evolution}, where the grey shaded band indicates the (re-normalised) orbital profile from \citet{Ducci22}. 
To highlight the properties and temporal evolution of the recent optical variability, REM data were folded without binning or averaging, and were further divided into three time intervals of comparable duration, each shown in a different colour.
Figure~\ref{fig:fold_evolution} shows that the optical bursts is still observed to peak close to the previously reported orbital phase, although over a broader phase interval. In the most recent subset of data (after MJD~60806; blue colour in Fig.~\ref{fig:fold_evolution}), this broadening is more pronounced, and the burst peak seems to anticipate phase zero more significantly than in earlier epochs.

   \begin{figure}
     \centering
     \includegraphics[width=\columnwidth]{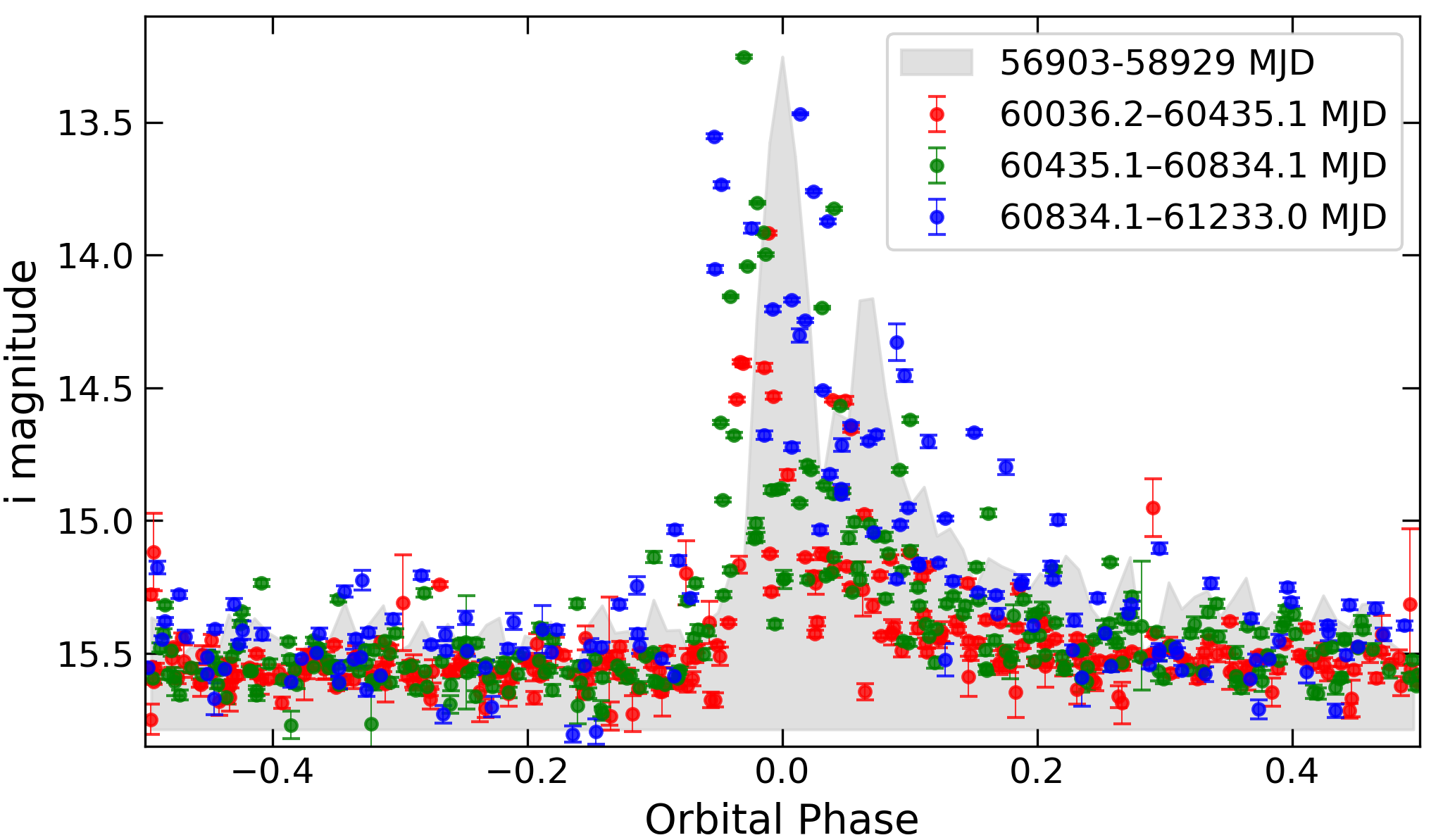}
   \caption{Orbital modulation of the optical emission from \src\ in the most recent REM monitoring campaign ($\geq 60000$~MJD), folded on the ephemeris $P_{\rm orb}=16.64002$~d and $T_0=55673.71$~MJD. Data are shown unbinned and divided into three time intervals of comparable duration to highlight the temporal evolution of the optical variability. The grey shaded band shows the (re-normalised) average profile obtained from archival REM, OGLE, and MACHO data  ($<59000$~MJD) by \citet{Ducci22}, for comparison.}
\label{fig:fold_evolution}
   \label{fig:fold_evolution}
   \end{figure}

   \begin{figure*}
     \centering
   \includegraphics[width=\columnwidth]{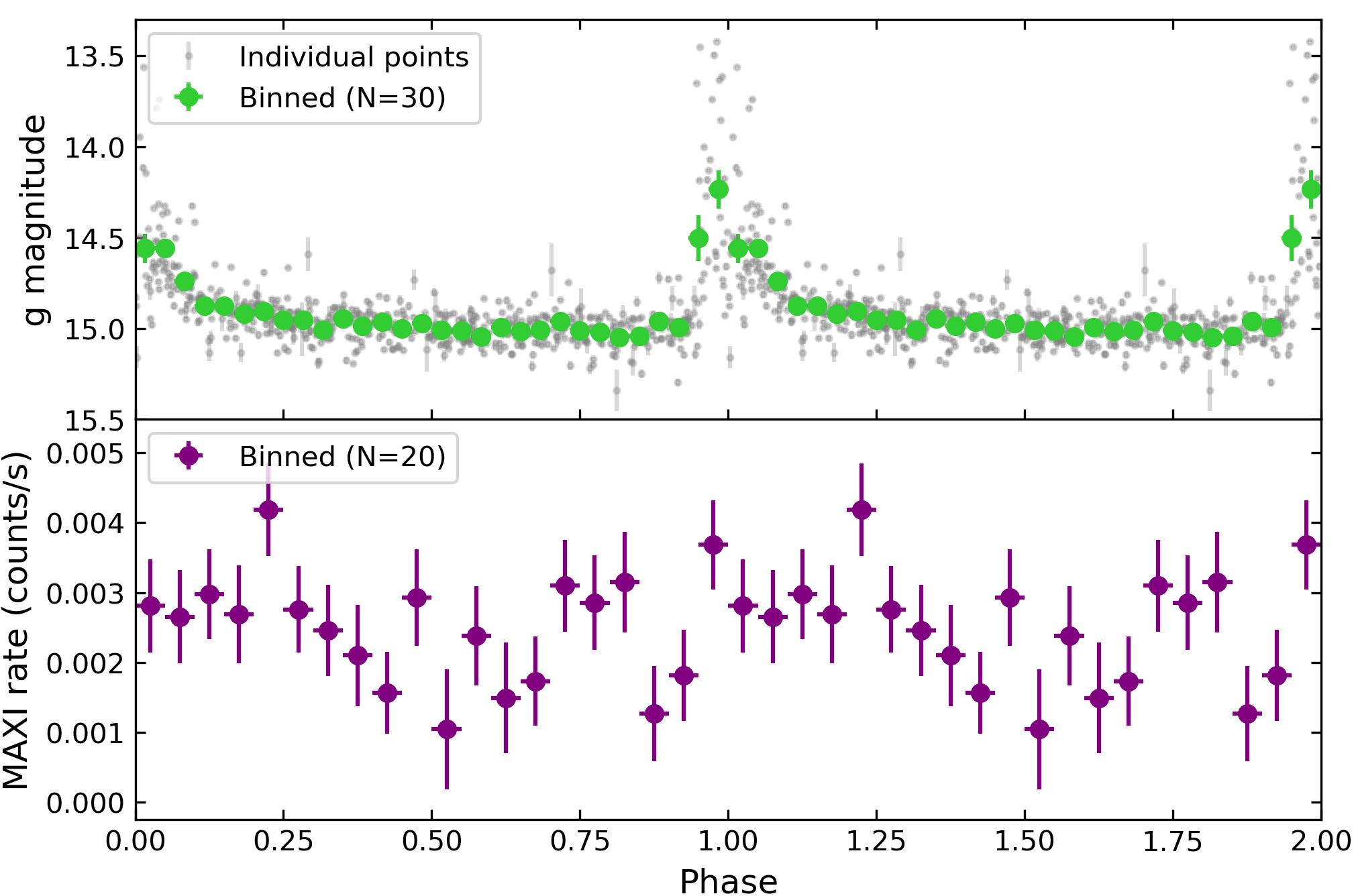}
   \includegraphics[width=\columnwidth]{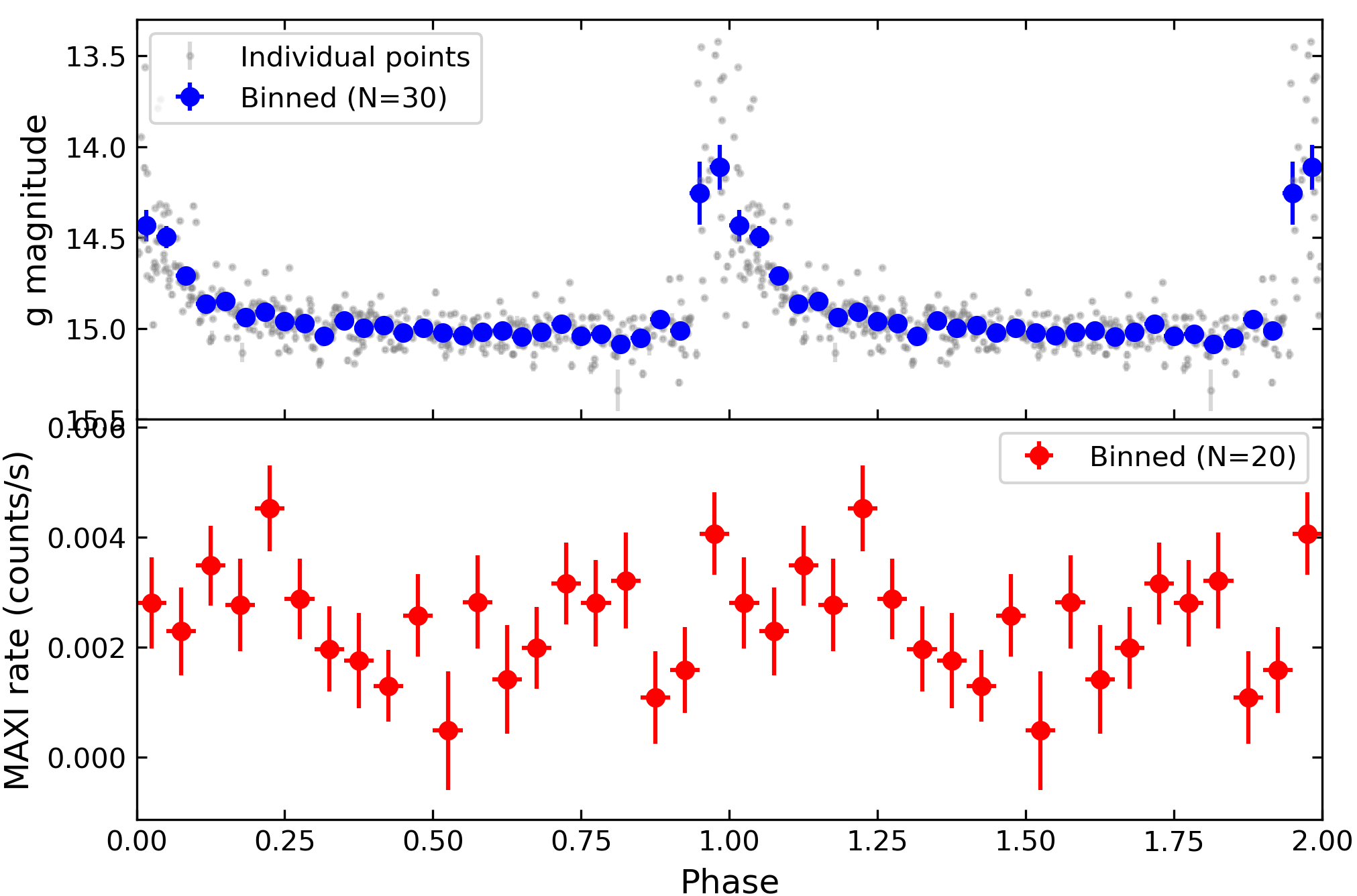}
   \caption{\emph{Left panel:} Orbital folding of the 
   REM $g$-band (\textit{top}) and MAXI/GSC $g$-band (\textit{bottom}) light curves after excluding the time intervals corresponding to the five bright X-ray outbursts (grey shaded regions in Fig.~\ref{fig:lc_maxi_rem}). The light curves were folded on the same ephemeris as in Fig.~\ref{fig:fold_evolution}.
\emph{Right panel:} Folding further restricted to the orbital cycles in which the REM light curve exhibited the optical bursts.}
   \label{fig:fold_test1}
   \end{figure*}

As it is apparent from Fig.~\ref{fig:lc_maxi_rem}, several bright optical bursts occur even when no corresponding X-ray outburst is detected by MAXI. 
To investigate whether a weaker X-ray modulation is nonetheless present, we excluded the time intervals corresponding to the five bright MAXI outbursts (grey-shaded regions in Fig.~\ref{fig:lc_maxi_rem}) and folded the remaining MAXI light curve on the orbital period. We applied the same filtering and folding procedure to the REM optical light curve. 
The result is shown in the left panel of Fig.~\ref{fig:fold_test1}.
As a further test, we repeated this analysis after excluding the bright  MAXI outbursts, but this time restricting the X-ray light curve to only those orbital cycles in which the optical light curve exhibited the recurrent periastron outburst. 
Folding both the X-ray and optical light curves under this stricter selection confirms the previous result 
(see the right panel of Fig.~\ref{fig:fold_test1}), indicating that the absence of X-ray modulation is not simply diluted by orbital cycles lacking optical activity.

We point out that the folded MAXI light curve shows a non-zero average count rate over the full orbital cycle ($\sim0.0025$\,c\,s$^{-1}$). Adopting a distance of $d=50$\,kpc and converting this count rate to a luminosity using PIMMS \citep{Mukai93}, this corresponds to a persistent X-ray luminosity of 
$\sim3\times10^{37}$\,erg\,s$^{-1}$. We caution, however, that the presence of such a persistent flux level in the folded light curve may be all or at least partly ascribable to contamination from the nearby source LMC\,X$-$4 ($\sim 34^\prime$ separation), given the limited angular resolution of MAXI.

\begin{table}
\caption{Simultaneous X-ray luminosities and OM $V$-band magnitudes (when available) of \src\ during the four
\xmm\ observations.}
\label{tab:om vs X-ray}
\centering
\resizebox{\columnwidth}{!}{
\begin{tabular}{llcc}
\hline\hline
\noalign{\smallskip}
Obs. & ObsID & $L_{\rm X}$     & $V$ \\
     &       & (erg\,s$^{-1}$) & (mag) \\
\noalign{\smallskip}
\hline
\noalign{\smallskip}
obs.\,1 & 0964320201  & $(1.89\pm 0.03)\times10^{37}$             & $13.784\pm0.002$   \\
\noalign{\smallskip}
obs.\,2 & 0964320301  & $(2.24\pm0.14)\times10^{36}$              & $14.096\pm0.002$   \\
\noalign{\smallskip}
obs.\,3 & 0964320401  & $(1.456{+0.003\atop-0.005})\times10^{39}$ & $13.645\pm0.002$   \\
\noalign{\smallskip}
obs.\,4 & 0973790101  & $(5.48{+0.05\atop-0.12})\times10^{36}$    & --$^a$             \\
\noalign{\smallskip}
\hline
\end{tabular}
}
\tablefoot{$V$ is the magnitude measured with the OM. $L_{\rm X}$ is the luminosity in the
0.2$-$10\,keV band. 
$^a$: No $V$-band photometry is available for obs.\,4,
for which the OM observed
in the UV filters: $UVW2 = 14.050 \pm 0.003$, $UVM2 = 14.033 \pm 0.005$, and
$UVW1 = 13.958 \pm 0.003$~mag.
}
\end{table}

\subsection{Optical bursts without X-ray counterparts: a possible interpretation}

We showed that \src\ has, since 2023, 
entered a renewed phase of activity characterised by longer and
brighter optical outbursts, closely resembling those recorded during the   1980-1981
discovery epoch (\citealt{Densham83} for the old outbursts and Figs. \ref{fig. maxi rem 4 bursts} and \ref{fig:fold_evolution} for the recent ones).

We showed that the MAXI and REM monitoring over this recent
interval reveals an unexpected mismatch. Several of the newly detected optical bursts reach
peak magnitudes comparable to, or brighter than, those of the three outbursts for which
quasi-simultaneous MAXI detections exist (Fig. \ref{fig:lc_maxi_rem}), yet no corresponding enhancement is seen in
the X-ray light curve for the majority of these events. 
We showed that this cannot be due to the MAXI sensitivity.
Using the folded MAXI light curves (Fig. \ref{fig:fold_test1}) and excluding the five
clearly identified bright outbursts, we can place a conservative upper limit on the luminosity
of any residual periastron-phase X-ray activity during the intervals of REM-detected bright
optical bursts, constraining it to $L_{\rm X} \lesssim 10^{37}$~erg\,s$^{-1}$, i.e. substantially
fainter than the outbursts associated with comparably bright optical events.

We outline below a qualitative scenario that reproduces the largest number of observables collected for \src.
It is based on the hypothesis that a radiation pressure dominated (RPD) accretion disc forms, and that its geometrically thick 
inner regions occasionally screen from the observer the emission produced close to the NS surface. 
This scenario can account for the decoupling between the optical and the X-ray light curves, for the elusive and transient
character of the 69\,ms pulsations,
for the  absorption lines detected in the \xmm\ spectra reported here, which indicate material moving at relativistic speed,
and for the absence of neutral X-ray absorption in excess of the 
interstellar value.

If the optical bursts of \src\ are produced by reprocessing of the X-ray emission in an
extended envelope, as the evidence collected over four decades indicates \citep{Charles83, Maraschi83, Apparao88, Densham83, Ducci19b}, then the optical
output depends on the total energy released by accretion, largely independent of the direction
along which that energy is emitted.
Under this hypothesis, the recent bright outbursts detected by REM suggest
that the energy release at periastron remains high in all these epochs, and that what varies is the
fraction of the X-ray emission that reaches the observer.
We therefore checked whether an RPD accretion disc can form at  the periastron passages in
which the brightest optical bursts are observed. 
The luminosity above which the  inner disc becomes RPD scales with the magnetic field of the NS as $B^{6/11}$
\citep[see e.g. ][]{Mushtukov23}.
Since the magnetic field of \src\ is constrained to $B\lesssim3\times10^{10}$\,G 
\citep{Ducci25}, this threshold is a few $\times10^{37}$\,erg\,s$^{-1}$,
well below the luminosity we measured in obs.\,3. 
A radiation driven outflow can form if the disc extends inside 
the spherisation radius $R_{\rm sp}$ ($R_{\rm m} < R_{\rm sp}$; see e.g. \citealt{Mushtukov19}). 
For the accretion rates considered here, 
$\dot{M}_{\rm acc} \approx 10^{39} {\rm erg\,s}^{-1} R_{\rm NS} / (G M_{\rm NS})\approx 6.4\times 10^{18}$\,g\,s$^{-1}$,
this condition is satisfied for $B \lesssim 3\times10^{10}$\,G
\citep[see equations 23, 25, and 57 in ][ assuming $L_{\rm Edd}=1.76\times 10^{38}$\,erg\,s$^{-1}$,  $\Lambda=0.5$, $\varepsilon_{\rm w}=0.5$]{Mushtukov23}.

Configurations of this kind are observed in other accreting compact objects. SS\,433, likely
hosting a black hole, is seen close to edge-on, so that the inner funnel is hidden and its
X-ray luminosity is several orders of magnitude below the bolometric luminosity released by accretion \citep{Middleton21}.
In Cyg\,X$-$3 (another HMXB with an accreting black hole), X-ray polarimetry has shown that the
central source is hidden and that the observed emission is scattered or reflected, with an
apparent luminosity more than an order of magnitude below the luminosity emitted along the
funnel \citep{Veledina24}.
Among extragalactic ULXs, obscuration by a precessing supercritical
flow has been invoked to explain some deep low states \citep[e.g. ][]{Vasilopoulos21,Fuerst21}.                  

In the case of \src, 
we assume that the line of sight is close to the rim of the cavity,
which corresponds to an inclination of the accretion disc with respect to the observer comparable
to the opening angle of the funnel. In this configuration, a small change in the height of the inner disc
is sufficient to bring the rim across the line of sight, and therefore to alternately 
hide and leave visible the innermost regions of the accretion flow.
The pulsating emission from the polar cap and the accretion column originates at the base of the funnel and is the first component 
to be screened, while the regions responsible for the reprocessed optical emission, of size of
$\sim10^{12}$\,cm, remain visible at all times.
The bright flux in obs.\,3 and the other MAXI
outbursts would then correspond to epochs in which the height of the inner disc was slightly lower, or the
viewing geometry more favourable, for instance because of disc precession or local warping,
so that part of the direct emission reached the observer. The three fainter \xmm\ observations,
and the REM bursts without MAXI counterparts, would correspond to epochs in which the base of
the funnel was partially screened.
The pulsed fraction expected in this geometry is low even when part of the direct emission is
observed, since the reflected component, produced by an extended surface illuminated by a
rotating beam, is averaged over pulse phase and therefore largely unpulsed. 
A similar explanation has been proposed for the accreting ULX pulsar M51 ULX$-$7 \citep{Vasilopoulos21}.
This is in agreement with the relatively low pulsed fraction measured in the \xmm\ obs.\,3 (Sect. \ref{sect:timing results}).

Figure~\ref{multi pp} shows that the peak of the pulse
profile shifts from $\phi \approx 0.75$ in the 1$-$12\,keV range to
$\phi \approx 1.25$ in the 0.2$-$1\,keV range, i.e. by half a rotational
cycle, and Fig.~\ref{pf vs E} shows that the pulsed fraction increases with energy. 
This behaviour fits the framework discussed above.
The hard maximum would be the direct emission
of the accretion column, seen when the beam sweeps across our line of sight.
The soft maximum would come from the reprocessing of the same beam by the
inner wall of the RPD disc: its illuminated surface faces us when the
beam points away, half a cycle later.
The increase of the pulsed fraction with energy is also expected in this geometry, 
since only a small fraction of the beam power is intercepted by the
wall and returned to the observer.
The observed half-cycle shift would be geometric in origin, since any propagation
delay involving the inner radius of the RPD disc is negligible in this
source. Accretion at the observed rate requires the magnetospheric radius to lie
within the corotation radius,\footnote{The corotation radius $R_{\rm co}=[GM_{\rm NS}P_{\rm spin}^2/(4\pi^2)]^{1/3}$ is the distance at which matter in a Keplerian orbit corotates with a NS of mass $M_{\rm NS}$ and spin period $P_{\rm spin}$.} which for a spin period of 69\,ms is only about
$\sim 2.8\times10^{7}$\,cm; the corresponding light-travel time is of the order of
one millisecond, too short to produce an appreciable phase lag.

The measured column density is consistent with the interstellar value in all four
\xmm\ observations, as expected in this framework, since the inner regions of an RPD disc are
highly ionised and electron scattering and geometrical screening are largely achromatic.
Our analysis shows that the observed X-ray emission is harder when the source is fainter.
A possible explanation is that the screening suppresses mainly the direct continuum,
so that faint states are dominated by the radiation reprocessed by the walls of the cavity.
Multiple scattering in a supercritical funnel can indeed produce a hard, curved emergent spectrum, 
as found in SS\,433 \citet{Medvedev10}.
The model of \citet{Karmakar25}, on the other hand, predicts the opposite behaviour.
In that model, the emission reaching an observer outside the opening angle of the cavity is the
component reflected and reprocessed by the walls, which are assumed to re-emit locally as blackbodies. 
The blackbody temperatures are higher close to the disc plane, where the irradiation is strongest, and decrease towards
the upper parts of the cavity.
Therefore, an  observer viewing the system along a direction close to the normal to the disc plane sees the hottest parts of the wall and measures a hard spectrum, whereas at larger inclinations
the visible portion  of the wall is confined to its cooler outer parts, and the observed spectrum becomes softer.
Since large inclinations are also those at which the direct emission from the NS is screened, 
and hence those of lowest apparent luminosity,
the model by \citet{Karmakar25} predicts that the spectrum softens as the apparent luminosity decreases.

\begin{table*}
\begin{center}
\caption{Period and period derivative measurements}
\label{tab:periods}
\begin{tabular}{lcllcll}
\hline
\hline
Satellite   & Date & MJD   & P (ms)   & $\dot P$ ($10^{-10}$  s s$^{-1}$) & $\phi_{orb}$ & Reference \\
\hline
Einstein  & 1980 Dec 16 &  44589.9792   & 69.21166(17)   & 5.01(86) &  0.911 &  \citet{Skinner82} \\
RossiXTE & 1998 Dec  3 &  51150.1452  & 68.88813(28)   & -- &  0.152   &       this work\\
NICER     &  	2023 Jan 9 & 59953.6772  & 69.3055(5)     & -- &  0.209 & \citet{Ducci25}\\
\xmm    & 2025 Sep 30 &   60948.6653  & 69.25878(2)    & 6.58(9) &  0.004 &  this work\\
\hline 
\end{tabular}
\end{center}
\end{table*}

\subsection{ULX-like outflow properties}
\label{sub:outflow}

The application of emission models of optically thin plasma 
to the strong lines in the RGS spectrum unveiled multiphase photoionised plasma with ionisation
parameters log $\xi_1 = 1.3 \pm 0.1$ and log $\xi_2 = 2.5 \pm 0.3$. This is consistent with the results obtained in ULXs, which are likely due to thermal instabilities found at log $\xi\sim2.0$ (see e.g. \citealt{Pinto2023}). As in most ULX RGS spectra, the emission components are consistent with being at rest with the exception of obs.\,3, where the lines appear broader (particularly for the strong O\,{\small VIII} line at 19\,{\AA}). The velocity broadening and displacement, each of about 6000\,km\,s$^{-1}$, is fully consistent with the findings in the bright-regime RGS spectra of ULX NGC\,1313 X$-$1 \citep{Pinto2020b} and imply that we are looking at an inner portion of the accretion disc. The underestimation of the nitrogen VI and VII lines, suggest a very likely nitrogen over-abundance with respect to oxygen and other species that has been observed in Be-like XRBs and transient ULXs \citep{Barra2024}. 

The plasma densities of $10^{11-13}$\,cm$^{-3}$ also match those observed in ULXs but the luminosities are significantly lower. Summing up the contribution from the two \texttt{pion} components we obtain, for the brighter obs.\,3, $L_{\rm X, \rm pion} \approx 4\times 10^{37}$\,erg\,s$^{-1}$ (0.2$-$10\,keV). Averaging the four observations, the emission lines contribute to about 5\% of the total 0.2--10\,keV luminosity, in full agreement with canonical ULXs. However, we note that the portion raises from 3\% in obs.\,1 to 9\% during obs.\,2, which is characterised by a lower bolometric luminosity and implies that the line-emitting region is broader than the region providing most of the continuum X-ray emission but the flux halving of the \texttt{pion} from obs.\,1 to obs.\,2 ($6 \rightarrow 3 \times 10^{35}$\,erg\,s$^{-1}$) agrees with the proposed scenario of partial obscuration. From the definition of ionisation parameter we can estimate an average distance of the O\,{\small VII} \texttt{pion} component (log $\xi=$ 1.3, for which the density was constrained thanks to the O\,{\small VII} line ratio) as $R = \sqrt{ L_{\rm ion} / ( n_{\rm H} \xi) }$, which yields $2\times 10^{12}$ cm (for $L_{\rm ion}=10^{39}$\,erg\,s$^{-1}$ as in obs.\,3). Assuming an ionising luminosity of $10^{37}$\,erg\,s$^{-1}$, i.e. closer to the other observations, it gives a size of about $2\times 10^{11}$\,cm, slightly smaller than the optical-emitting region but indeed still larger than that emitting the X-ray continuum.

Grids of absorption models of photoionised plasma unveiled the presence of an extreme outflow component blowing at more than 10\% of the speed of light with an ionisation parameter consistent with that of the hottest (and likely innermost) line-emitting component. Such outflows are also found in more than 40\% ULXs (particularly those associated with steep spectra; e.g. $\Gamma > 2$, or very high-quality RGS spectra with $\gtrsim$ 5,000 counts in total, see e.g. \citealt{Kosec2021,Pinto2023}). 
The very strong detection of fast outflows in obs.\,1 implies that the accretion rate is intrinsically, highly super-Eddington that would correspond to a real luminosity of $10^{39}$\,erg\,s$^{-1}$ or higher. We note that the plasma absorption-corrected 0.2$-$10\,keV luminosity of the source is of $1.5\times10^{39}$\,erg\,s$^{-1}$ during obs.\,3. The weaker of absorption features in obs.\,2 (and to some extent to obs.\,4) are likely due to a combination of a lower number of counts with respect to obs.\,1 and 3, wind variability and optically thick obscuration of the inner regions as previously mentioned.

   \begin{figure*}
     \centering
     \includegraphics[width=18cm]{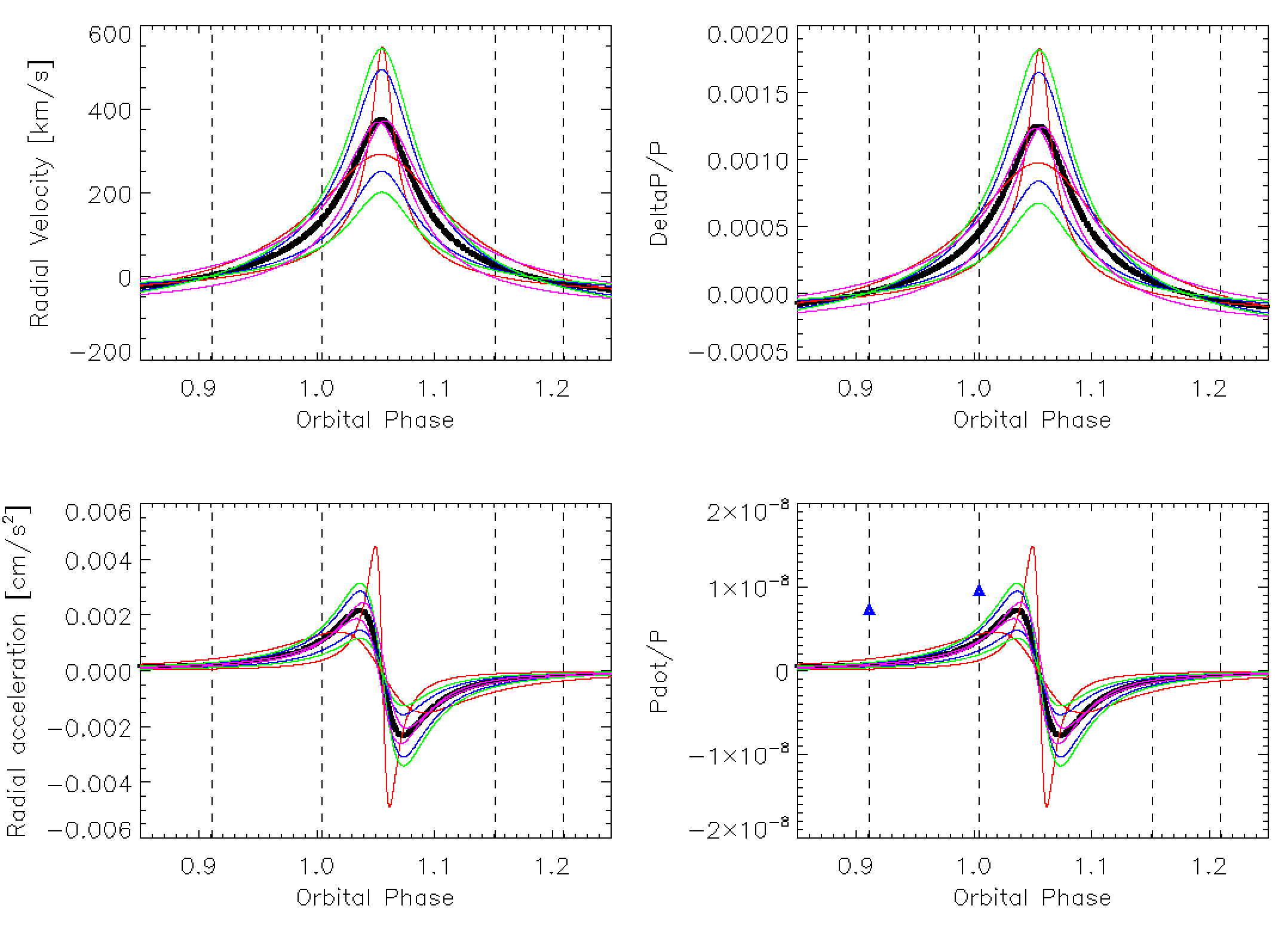}
   \caption{NS radial velocity (top left) and radial acceleration (bottom left) as a function of the orbital phase (periastron at $\phi_{\rm orb}$=0.056). The two right panels give the corresponding effect on the observed period and period derivative. 
  In all panels, the vertical dashed lines indicate the phases of the four period measurements (from the left: 1980 \emph{Einstein}, 2025 \xmm, 1998 RXTE, 2023 NICER).  The solid line corresponds to the best-fit orbital parameters. The thin lines were obtained by varying each single parameter at a time in the following intervals: eccentricity $e=$[0.58--0.86] (red),  mass ratio q=[3-18] (blue), $a_1 sini$=[4.36--11.78] $R_\odot$ (green), longitude of periastron $\omega =$[173.6--192.4]  deg (purple). 
  The two triangles in the bottom-right panel indicate the spin-down rates observed in the 1980 and 2025 observations. }
   \label{fig:orbit}
   \end{figure*}

   \begin{figure}
     \centering
     \includegraphics[width=\columnwidth]{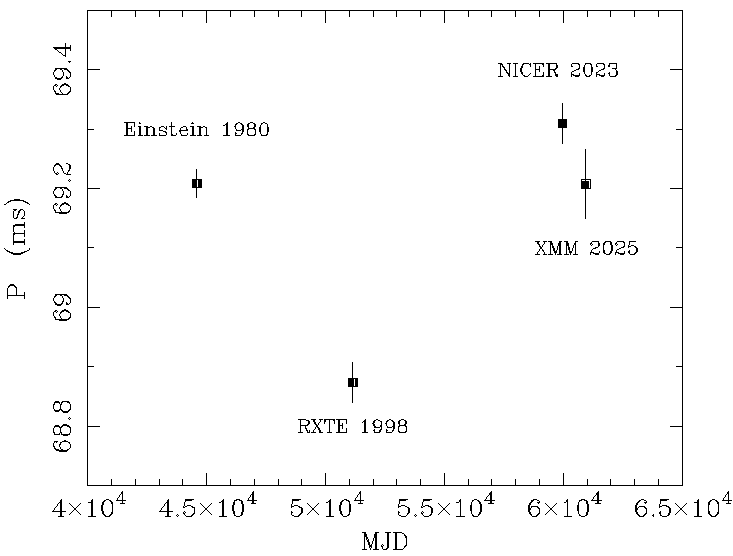}
   \caption{Spin period evolution of \src . The plotted values have been corrected for the orbital motion using the nominal values of the orbital solution from \citet{Rajoelimanana17}.}
   \label{fig:periods}
   \end{figure}

\subsection{Timing evolution}
\label{sub:timev}

All the spin period measurements of \src\ are summarised in Table~\ref{tab:periods}. Except for the 1998 \rxte\ period, they are all within 0.1 ms from the first discovered value, implying very small variations possibly explainable by orbital Doppler effects. Although the statistical significance of the 1998 detection is lower than those of the other period measurements ($\sim 3.7\sigma$), we believe it is still sufficiently high to consider this point in the following discussion. 

For a proper comparison, the measured spin periods must be corrected for the   motion of the NS in the binary system. We have done this  using the orbital parameters reported by \citet{Rajoelimanana17}, which however are affected by non negligible uncertainties. This is illustrated in Fig.~\ref{fig:orbit}, where we plot in the left panels the NS radial velocity and acceleration along the orbit and in the right panels the corresponding corrections to $P$ and $\dot P$.  The dashed vertical lines indicate the phases of the four period measurements. The thick solid lines are obtained using the  best fit orbital parameters, while the coloured thin lines indicate the effect of changing the   parameter values (see details in the figure caption). Clearly the measurements most affected by the orbit uncertainties are those obtained close to periastron ($\phi_{\rm orb}=0.056)$, where the NS radial velocity changes rapidly.

The long term spin evolution of \src\   is shown in Fig.~\ref{fig:periods}, where the $P$ values have been corrected assuming the best fit orbital parameters of  \citet{Rajoelimanana17}. We estimated the errors on $P$ with a Montecarlo simulation in which we simultaneously varied the orbital parameters based on the uncertainties reported by these authors and the mass ratios $q$ with a flat distribution in the conservatively wide interval [3,18]. 

The RXTE period measurement implies an average spin-up rate of $(-5.9\pm0.7)\times10^{-13}$ s s$^{-1}$ between 1980 and 1998, followed by a spin-down at an average rate of  $(5.1\pm0.6)\times10^{-13}$ s s$^{-1}$ in the following $\sim$24-27 years.  The last two measurements in 2023 and 2025 correspond to a spin-up of $(-1.2\pm0.8)\times10^{-12}$ s s$^{-1}$,  but the large uncertainties make this result non-significant.

If the spin-up between 1980 and 1998 was due to accretion torques, a simple estimate of the required accretion rate $\dot M$ can be obtained by the relation

\begin{equation}
\dot{\nu}\sim{\sqrt{GMR_{\rm m}}\over{2\pi 0.4 MR^2}} \dot{M}\sim {G^{3/7}\mu^{2/7}\dot{M}^{6/7}\over{2\pi 0.4 M^{4/7} R^2}} \mbox{\ ,}
\label{spinup_formula}
\end{equation}

\noindent 
where  $M$, $R$ and $\mu$ are the NS mass, radius and magnetic moment and $R_{\rm m}$ is the magnetospheric radius (see e.g. \citealt{2015SSRv..191..293R}). Note that this is an optimistic estimate since it assumes that accretion occurs through a disc and neglects possible spin-down torques. With this relation,
the observed spin-up rate between 1980 and 1998 ($1.2\times10^{-10}$ Hz s$^{-1}$), implies an accretion luminosity of  $L_X\sim1.4\times10^{39} \mu_{30}^{-1/3}$ erg s$^{-1}$. 
Similar spin-up rates have indeed been observed in highly super-Eddington pulsating ULXs.
Although bright outbursts were seen in the first years after the source discovery, there is no evidence that such a high luminosity persisted continuously until 1998 \citep{Rigoselli25}. If the spin up was not continuous, but occurred only during a few very bright outbursts even larger spin-up rates than the above average value would be required.

The spin-down after 1998 raises the question of what physical mechanism was responsible.
In principle, if mass accretion stops, the NS might turn-on as a rotation powered-radio pulsar subject to spin-down at $\dot P= 1.4\times10^{-14}~(B/10^{12}~{\rm G})^2~(0.069~{\rm s/P)}$ s s$^{-1}$ due to magnetic dipole radiation losses.  A dipole field of several 10$^{12}$ G would be required, but this is a lower limit because we cannot exclude that episodes of spin-up occurred in the 2018-2023 time period. Furthermore, no evidence of the presence of an active radio pulsar has ever been reported in \src.

An alternative is that the spin-down between 1998 and 2023 was produced by a propeller torque.
For most of this time interval, \src\ was not detected at 
luminosities above 
$\sim10^{37}$\,erg\,s$^{-1}$, with  the first re-detection occurring  only around 2018 \citep{Ducci19}. Therefore, it is reasonable to assume that 
the source spent this period with  $R_{\rm m} > R_{\rm co}$.
Under this assumption, the matter captured at the magnetosphere is expelled,
carrying away the specific angular momentum $\Omega R_{\rm m}^2$. In first  approximation, 
the resulting spin-down torque is 
$N=I\dot{\Omega}= -\dot{M} R_{\rm m}^2 \Omega$ 
(e.g. \citealt{Dai06} and references therein).
If we assume the definition of magnetospheric radius from \citet{2015SSRv..191..293R} (their eq.\,4), we obtain
\begin{eqnarray}
\dot{P} & = & 1.5\times 10^{-15}~\xi^2~(P/69\,\mbox{ms})~(\dot{M}/10^{16}\mbox{g\,s}^{-1})^{3/7} \times \nonumber \\
        &   & \times~(\mu/10^{28}\mbox{G\,cm}^3)^{8/7}~(I/10^{45}\mbox{g\,cm}^2)^{-1} \mbox{~s\,s}^{-1} \mbox{\ ,}
\end{eqnarray}
where $\xi=1$ for spherical inflow and $\xi=0.4$ for a Keplerian disc.
For $\mu=10^{28}$\,G\,cm$^3$ (corresponding to $B \approx 10^{10}$\,G)
the propeller condition  ($R_{\rm m} > R_{\rm co}$) requires
$\dot{M} \lesssim 6\times 10^{16}$\,g\,s$^{-1}$. Even at this maximum acretion rate, the predicted
spin-down ($\dot{P}\approx 3\times 10^{-15}$\,s\,s$^{-1}$, assuming $\xi=1$), is too small compared to the observed value.
At the given $\dot{M}$, the observed spin-down would require $\mu \approx 10^{30}$\,G\,cm$^3$ ($B\approx 10^{12}$\,G), which is in contradiction with
the weak magnetic field inferred until now for \src\ from the absence of pulsations at low X-ray luminosities ($\lesssim 10^{37}$\,erg\,s$^{-1}$) \citep[][ and references therein]{Ducci25}.
Using the torque prescription of \citet{Romanova03}, we obtain an estimate of the expected $\dot{P}$  that is comparable to the one derived above.

Finally, we consider the $\dot P$ values measured in the 1980 and 2025 observations, which are indicated by the triangles in the lower right panel of Fig.~\ref{fig:orbit}. \citet{Skinner82}  noticed that the high spin-down rate measured in their data could be naturally explained by a Doppler effect implying an eccentricity of at least 0.4 (corresponding to the most favourable relative geometry at the time of the observation). However, with the orbital solution later derived by \citet{Rajoelimanana17} and used here, it is not possible to explain the observed $\dot P$ value as an orbital Doppler shift (see lower right panel of Fig.~\ref{fig:orbit}).   The same applies to the spin-down of $\sim6.6\times10^{-10}$ s s$^{-1}$ we measured in the \xmm\ observation of 2025 September 30. 
In both cases, an intrinsic spin-down of the order of a few $\times10^{-10}$ s s$^{-1}$ is required.

\section{Conclusions}
\label{sect:conclusions}

We have presented the first observations of \src\ in the super-Eddington regime carried out with
modern X-ray instrumentation. The source returned, after about 45 years, to peak X-ray luminosities
of $\sim10^{39}$\,erg\,s$^{-1}$.  During the   \xmm\ observation in which \src\ reached the highest flux,  we detected the 69\,ms pulsations
and measured, for the second time only, their period derivative.
We also report a detection of pulsations from archival RXTE data taken in 1998.
The RGS spectra reveal a rich set of emission lines from photoionised plasma at rest,
together with blueshifted absorption lines indicating an outflow with a relativistic velocity of $\sim0.12c$ 
(which implies a highly super-Eddington accretion rate and makes \src\ a bonafide transient ULX pulsar, a class of which only a few objects are known so far).
The long-term MAXI and REM monitoring shows that the recent bright optical bursts are, in most cases, not accompanied by a comparably bright X-ray outburst.

We proposed a qualitative scenario based on an RPD disc whose geometrically thick inner regions intermittently
screen the emission produced close to the NS. This scenario can explain  the decoupling between the optical
and X-ray light curves,  the elusive character of the pulsations, their low pulsed fraction, the energy dependence of the pulse profile, and the outflow signatures seen in the RGS spectra.
Alternative or complementary scenarios remain to be explored.

In principle, the new measurements of the spin period and its time derivative offer an important diagnostic tool to understand the peculiarities of this source. The long term  evolution implied by the four available $P$ measurements (and in particular by the 1998 one), as well as the two  $\dot{P}_{\rm spin}$ values measured within the 1980 and 2025 observations, are not easily explained, if indeed the orbital parameters are close  to those of \citet{Rajoelimanana17}. It is also remarkable that the spin period in 2023-2025 is so similar to that of the first measurement, obtained more than four decades earlier. 
A dedicated spectroscopic study of the optical counterpart is clearly required to obtain a more precise orbital solution and disentangle the intrinsic torques acting on the NS from the orbital contribution. Although the NS pulsations are seen only sporadically, their presence makes this system equivalent to a double spectroscopic binary, with the possibility to derive the masses of the two components, which in turn is important to understand the system evolutionary history.

\begin{acknowledgements}
LD acknowledges funding from the Deutsche Forschungsgemeinschaft (DFG, German Research Foundation) - Projektnummer 549824807.
This work is based on observations obtained with XMM-Newton, an ESA science 
mission with instruments and contributions directly funded by ESA Member 
States and the USA (NASA).
This research has made use of MAXI data provided by RIKEN, JAXA and the MAXI team.
This research has made use of the NuSTAR Data Analysis Software (NuSTARDAS) jointly developed by the ASI Space Science Data Center (SSDC, Italy) and the California Institute of Technology (Caltech, USA).
This work made significant use of the Fornax Science Console, which is part of the NASA Astrophysics cloud-based Fornax Initiative jointly developed by Goddard Space Flight Center's Astrophysics Projects Division (GSFC/ApPD) and the astrophysics archives-the High Energy Astrophysics Science Archive Research Center (HEASARC), the Infrared Science Archive (IRSA), and the Mikulski Archive for Space Telescopes (MAST).
This research has made use of data and/or software provided by the High Energy Astrophysics Science Archive Research Center (HEASARC), which is a service of the Astrophysics Science Division at NASA/GSFC.
CP is supported by INAF Large Grant 2023 BLOSSOM O.F. 1.05.23.01.13.
\end{acknowledgements}

\bibliographystyle{aa} 
\bibliography{aa63088-26}

@ARTICLE{Rigoselli25,
       author = {{Rigoselli}, Michela and {Tresoldi}, Caterina and {Ducci}, Lorenzo and {Mereghetti}, Sandro},
        title = "{XMM-Newton Observations of the Peculiar Be X-Ray Binary A0538-66}",
      journal = {Astronomische Nachrichten},
         year = 2025,
        month = jan,
       volume = {346},
       number = {1},
          eid = {e20240098},
        pages = {e20240098},
          doi = {10.1002/asna.20240098},
archivePrefix = {arXiv},
       eprint = {2410.18606},
 primaryClass = {astro-ph.HE},
       adsurl = {https://ui.adsabs.harvard.edu/abs/2025AN....34640098R}
}

@ARTICLE{2015SSRv..191..293R,
       author = {{Revnivtsev}, Mikhail and {Mereghetti}, Sandro},
        title = "{Magnetic Fields of Neutron Stars in X-Ray Binaries}",
      journal = {\ssr},
         year = 2015,
        month = oct,
       volume = {191},
       number = {1-4},
        pages = {293-314},
          doi = {10.1007/s11214-014-0123-x},
archivePrefix = {arXiv},
       eprint = {1411.5843},
 primaryClass = {astro-ph.HE},
       adsurl = {https://ui.adsabs.harvard.edu/abs/2015SSRv..191..293R}
}

@ARTICLE{Jansen01,
       author = {{Jansen}, F. and {Lumb}, D. and {Altieri}, B. and {Clavel}, J. and {Ehle}, M. and {Erd}, C. and {Gabriel}, C. and {Guainazzi}, M. and {Gondoin}, P. and {Much}, R. and {Munoz}, R. and {Santos}, M. and {Schartel}, N. and {Texier}, D. and {Vacanti}, G.},
        title = "{XMM-Newton observatory. I. The spacecraft and operations}",
      journal = {\aap},
         year = 2001,
        month = jan,
       volume = {365},
        pages = {L1-L6},
          doi = {10.1051/0004-6361:20000036},
       adsurl = {https://ui.adsabs.harvard.edu/abs/2001A&A...365L...1J}
}

@ARTICLE{Strueder01,
       author = {{Str{\"u}der}, L. and {Briel}, U. and {Dennerl}, K. and {Hartmann}, R. and {Kendziorra}, E. and {Meidinger}, N. and {Pfeffermann}, E. and {Reppin}, C. and {Aschenbach}, B. and {Bornemann}, W. and {Br{\"a}uninger}, H. and {Burkert}, W. and {Elender}, M. and {Freyberg}, M. and {Haberl}, F. and {Hartner}, G. and {Heuschmann}, F. and {Hippmann}, H. and {Kastelic}, E. and {Kemmer}, S. and {Kettenring}, G. and {Kink}, W. and {Krause}, N. and {M{\"u}ller}, S. and {Oppitz}, A. and {Pietsch}, W. and {Popp}, M. and {Predehl}, P. and {Read}, A. and {Stephan}, K.~H. and {St{\"o}tter}, D. and {Tr{\"u}mper}, J. and {Holl}, P. and {Kemmer}, J. and {Soltau}, H. and {St{\"o}tter}, R. and {Weber}, U. and {Weichert}, U. and {von Zanthier}, C. and {Carathanassis}, D. and {Lutz}, G. and {Richter}, R.~H. and {Solc}, P. and {B{\"o}ttcher}, H. and {Kuster}, M. and {Staubert}, R. and {Abbey}, A. and {Holland}, A. and {Turner}, M. and {Balasini}, M. and {Bignami}, G.~F. and {La Palombara}, N. and {Villa}, G. and {Buttler}, W. and {Gianini}, F. and {Lain{\'e}}, R. and {Lumb}, D. and {Dhez}, P.},
        title = "{The European Photon Imaging Camera on XMM-Newton: The pn-CCD camera}",
      journal = {\aap},
         year = 2001,
        month = jan,
       volume = {365},
        pages = {L18-L26},
          doi = {10.1051/0004-6361:20000066},
       adsurl = {https://ui.adsabs.harvard.edu/abs/2001A&A...365L..18S}
}

@ARTICLE{Turner01,
       author = {{Turner}, M.~J.~L. and {Abbey}, A. and {Arnaud}, M. and {Balasini}, M. and {Barbera}, M. and {Belsole}, E. and {Bennie}, P.~J. and {Bernard}, J.~P. and {Bignami}, G.~F. and {Boer}, M. and {Briel}, U. and {Butler}, I. and {Cara}, C. and {Chabaud}, C. and {Cole}, R. and {Collura}, A. and {Conte}, M. and {Cros}, A. and {Denby}, M. and {Dhez}, P. and {Di Coco}, G. and {Dowson}, J. and {Ferrando}, P. and {Ghizzardi}, S. and {Gianotti}, F. and {Goodall}, C.~V. and {Gretton}, L. and {Griffiths}, R.~G. and {Hainaut}, O. and {Hochedez}, J.~F. and {Holland}, A.~D. and {Jourdain}, E. and {Kendziorra}, E. and {Lagostina}, A. and {Laine}, R. and {La Palombara}, N. and {Lortholary}, M. and {Lumb}, D. and {Marty}, P. and {Molendi}, S. and {Pigot}, C. and {Poindron}, E. and {Pounds}, K.~A. and {Reeves}, J.~N. and {Reppin}, C. and {Rothenflug}, R. and {Salvetat}, P. and {Sauvageot}, J.~L. and {Schmitt}, D. and {Sembay}, S. and {Short}, A.~D.~T. and {Spragg}, J. and {Stephen}, J. and {Str{\"u}der}, L. and {Tiengo}, A. and {Trifoglio}, M. and {Tr{\"u}mper}, J. and {Vercellone}, S. and {Vigroux}, L. and {Villa}, G. and {Ward}, M.~J. and {Whitehead}, S. and {Zonca}, E.},
        title = "{The European Photon Imaging Camera on XMM-Newton: The MOS cameras}",
      journal = {\aap},
         year = 2001,
        month = jan,
       volume = {365},
        pages = {L27-L35},
          doi = {10.1051/0004-6361:20000087},
archivePrefix = {arXiv},
       eprint = {astro-ph/0011498},
 primaryClass = {astro-ph},
       adsurl = {https://ui.adsabs.harvard.edu/abs/2001A&A...365L..27T}
}

@ARTICLE{denHerder01,
       author = {{den Herder}, J.~W. and {Brinkman}, A.~C. and {Kahn}, S.~M. and {Branduardi-Raymont}, G. and {Thomsen}, K. and {Aarts}, H. and {Audard}, M. and {Bixler}, J.~V. and {den Boggende}, A.~J. and {Cottam}, J. and {Decker}, T. and {Dubbeldam}, L. and {Erd}, C. and {Goulooze}, H. and {G{\"u}del}, M. and {Guttridge}, P. and {Hailey}, C.~J. and {Janabi}, K. Al and {Kaastra}, J.~S. and {de Korte}, P.~A.~J. and {van Leeuwen}, B.~J. and {Mauche}, C. and {McCalden}, A.~J. and {Mewe}, R. and {Naber}, A. and {Paerels}, F.~B. and {Peterson}, J.~R. and {Rasmussen}, A.~P. and {Rees}, K. and {Sakelliou}, I. and {Sako}, M. and {Spodek}, J. and {Stern}, M. and {Tamura}, T. and {Tandy}, J. and {de Vries}, C.~P. and {Welch}, S. and {Zehnder}, A.},
        title = "{The Reflection Grating Spectrometer on board XMM-Newton}",
      journal = {\aap},
         year = 2001,
        month = jan,
       volume = {365},
        pages = {L7-L17},
          doi = {10.1051/0004-6361:20000058},
       adsurl = {https://ui.adsabs.harvard.edu/abs/2001A&A...365L...7D}
}

@ARTICLE{Mason01,
       author = {{Mason}, K.~O. and {Breeveld}, A. and {Much}, R. and {Carter}, M. and {Cordova}, F.~A. and {Cropper}, M.~S. and {Fordham}, J. and {Huckle}, H. and {Ho}, C. and {Kawakami}, H. and {Kennea}, J. and {Kennedy}, T. and {Mittaz}, J. and {Pandel}, D. and {Priedhorsky}, W.~C. and {Sasseen}, T. and {Shirey}, R. and {Smith}, P. and {Vreux}, J.-M.},
        title = "{The XMM-Newton optical/UV monitor telescope}",
      journal = {\aap},
         year = 2001,
        month = jan,
       volume = {365},
        pages = {L36-L44},
          doi = {10.1051/0004-6361:20000044},
archivePrefix = {arXiv},
       eprint = {astro-ph/0011216},
 primaryClass = {astro-ph},
       adsurl = {https://ui.adsabs.harvard.edu/abs/2001A&A...365L..36M}
}

@ARTICLE{Alves04,
       author = {{Alves}, David R.},
        title = "{A review of the distance and structure of the Large Magellanic Cloud}",
      journal = {\nar},
         year = 2004,
        month = jul,
       volume = {48},
       number = {9},
        pages = {659-665},
          doi = {10.1016/j.newar.2004.03.001},
archivePrefix = {arXiv},
       eprint = {astro-ph/0310673},
 primaryClass = {astro-ph},
       adsurl = {https://ui.adsabs.harvard.edu/abs/2004NewAR..48..659A}
}

@ARTICLE{Lutovinov09,
       author = {{Lutovinov}, A.~A. and {Tsygankov}, S.~S.},
        title = "{Timing characteristics of the hard X-ray emission from bright X-ray pulsars based on INTEGRAL data}",
      journal = {Astronomy Letters},
         year = 2009,
        month = jul,
       volume = {35},
       number = {7},
        pages = {433-456},
          doi = {10.1134/S1063773709070019},
archivePrefix = {arXiv},
       eprint = {0907.4288},
 primaryClass = {astro-ph.HE},
       adsurl = {https://ui.adsabs.harvard.edu/abs/2009AstL...35..433L}
}

@ARTICLE{HI4PI16,
       author = {{HI4PI Collaboration} and {Ben Bekhti}, N. and {Fl{\"o}er}, L. and {Keller}, R. and {Kerp}, J. and {Lenz}, D. and {Winkel}, B. and {Bailin}, J. and {Calabretta}, M.~R. and {Dedes}, L. and {Ford}, H.~A. and {Gibson}, B.~K. and {Haud}, U. and {Janowiecki}, S. and {Kalberla}, P.~M.~W. and {Lockman}, F.~J. and {McClure-Griffiths}, N.~M. and {Murphy}, T. and {Nakanishi}, H. and {Pisano}, D.~J. and {Staveley-Smith}, L.},
        title = "{HI4PI: A full-sky H I survey based on EBHIS and GASS}",
      journal = {\aap},
         year = 2016,
        month = oct,
       volume = {594},
          eid = {A116},
        pages = {A116},
          doi = {10.1051/0004-6361/201629178},
archivePrefix = {arXiv},
       eprint = {1610.06175},
 primaryClass = {astro-ph.GA},
       adsurl = {https://ui.adsabs.harvard.edu/abs/2016A&A...594A.116H}
}

@ARTICLE{Ducci19,
       author = {{Ducci}, Lorenzo and {Mereghetti}, Sandro and {Santangelo}, Andrea},
        title = "{Awakening of the Fast-spinning Accreting Be/X-Ray Pulsar A0538-66}",
      journal = {\apjl},
         year = 2019,
        month = aug,
       volume = {881},
       number = {1},
          eid = {L17},
        pages = {L17},
          doi = {10.3847/2041-8213/ab32f0},
archivePrefix = {arXiv},
       eprint = {1907.08078},
 primaryClass = {astro-ph.HE},
       adsurl = {https://ui.adsabs.harvard.edu/abs/2019ApJ...881L..17D}
}

@ARTICLE{Johnston79,
       author = {{Johnston}, M.~D. and {Bradt}, H.~V. and {Doxsey}, R.~E. and {Griffiths}, R.~E. and {Schwartz}, D.~A. and {Schwarz}, J.},
        title = "{Location of the recurrent (LMC?) X-ray transient A0538-66 with the HEAO 1 scanning modulation collimator.}",
      journal = {\apjl},
         year = 1979,
        month = may,
       volume = {230},
        pages = {L11-L14},
          doi = {10.1086/182952},
       adsurl = {https://ui.adsabs.harvard.edu/abs/1979ApJ...230L..11J}
}

@ARTICLE{Romanova03,
       author = {{Romanova}, M.~M. and {Toropina}, O.~D. and {Toropin}, Yu. M. and {Lovelace}, R.~V.~E.},
        title = "{Magnetohydrodynamic Simulations of Accretion onto a Star in the ``Propeller'' Regime}",
      journal = {\apj},
         year = 2003,
        month = may,
       volume = {588},
       number = {1},
        pages = {400-407},
          doi = {10.1086/373990},
archivePrefix = {arXiv},
       eprint = {astro-ph/0209548},
 primaryClass = {astro-ph},
       adsurl = {https://ui.adsabs.harvard.edu/abs/2003ApJ...588..400R}
}

@ARTICLE{Apparao88,
       author = {{Apparao}, K.~M.~V. and {Tarafdar}, S.~P.},
        title = "{Optical and Ultraviolet Line and Continuum Emission from the Large Magellanic Cloud X-Ray Be Star A0538-66}",
      journal = {\apj},
         year = 1988,
        month = dec,
       volume = {335},
        pages = {147},
          doi = {10.1086/166914},
       adsurl = {https://ui.adsabs.harvard.edu/abs/1988ApJ...335..147A}
}

@ARTICLE{Maraschi83,
       author = {{Maraschi}, L. and {Traversini}, R. and {Treves}, A.},
        title = "{A model for A0538-66: the fast flaring pulsar.}",
      journal = {\mnras},
         year = 1983,
        month = sep,
       volume = {204},
        pages = {1179-1184},
          doi = {10.1093/mnras/204.4.1179},
       adsurl = {https://ui.adsabs.harvard.edu/abs/1983MNRAS.204.1179M}
}

@ARTICLE{Ducci19b,
       author = {{Ducci}, L. and {Mereghetti}, S. and {Hryniewicz}, K. and {Santangelo}, A. and {Romano}, P.},
        title = "{Exploring the role of X-ray reprocessing and irradiation in the anomalous bright optical outbursts of A0538-66}",
      journal = {\aap},
         year = 2019,
        month = apr,
       volume = {624},
          eid = {A9},
        pages = {A9},
          doi = {10.1051/0004-6361/201834390},
archivePrefix = {arXiv},
       eprint = {1903.02918},
 primaryClass = {astro-ph.HE},
       adsurl = {https://ui.adsabs.harvard.edu/abs/2019A&A...624A...9D}
}

@ARTICLE{Medvedev10,
       author = {{Medvedev}, A. and {Fabrika}, S.},
        title = "{Evidence of supercritical disc funnel radiation in X-ray spectra of SS 433}",
      journal = {\mnras},
         year = 2010,
        month = feb,
       volume = {402},
       number = {1},
        pages = {479-491},
          doi = {10.1111/j.1365-2966.2009.15896.x},
archivePrefix = {arXiv},
       eprint = {0910.3186},
 primaryClass = {astro-ph.HE},
       adsurl = {https://ui.adsabs.harvard.edu/abs/2010MNRAS.402..479M}
}

@ARTICLE{Crook-Mansour26,
       author = {{Crook-Mansour}, Justine and {Fender}, Rob and {Andersson}, Alex and {Qiu}, Hao and {Hughes}, Andrew K. and {van den Eijnden}, Jakob and {Cowie}, Fraser J. and {Motta}, Sara and {Monageng}, Itumeleng and {Ducci}, Lorenzo and {Mereghetti}, Sandro and {Mathiba}, Andries and {Dobie}, Dougal and {Murphy}, Tara and {Kaplan}, David L. and {Carotenuto}, Francesco and {Charles}, Phil},
        title = "{Detection of an extremely luminous radio counterpart to the Be/X-ray binary A0538{\ensuremath{-}}66}",
      journal = {\mnras},
         year = 2026,
        month = may,
       volume = {548},
       number = {1},
          eid = {stag224},
        pages = {stag224},
          doi = {10.1093/mnras/stag224},
archivePrefix = {arXiv},
       eprint = {2601.22741},
 primaryClass = {astro-ph.HE},
       adsurl = {https://ui.adsabs.harvard.edu/abs/2026MNRAS.548ag224C}
}

@ARTICLE{Ducci25,
       author = {{Ducci}, Lorenzo and {Mereghetti}, Sandro},
        title = "{Sometimes They Come Back: Pulsations in A 0538─66 at 69 ms Rediscovered with NICER}",
      journal = {\apj},
         year = 2025,
        month = aug,
       volume = {989},
       number = {2},
          eid = {150},
        pages = {150},
          doi = {10.3847/1538-4357/adf338},
archivePrefix = {arXiv},
       eprint = {2507.16420},
 primaryClass = {astro-ph.HE},
       adsurl = {https://ui.adsabs.harvard.edu/abs/2025ApJ...989..150D}
}

@Inbook{Mushtukov23,
author="Mushtukov, Alexander
and Tsygankov, Sergey",
title="Accreting Strongly Magnetized Neutron Stars: X-ray Pulsars",
bookTitle="Handbook of X-ray and Gamma-ray Astrophysics",
year="2022",
publisher="Springer Nature Singapore",
address="Singapore",
pages="1--72",
isbn="978-981-16-4544-0",
doi="10.1007/978-981-16-4544-0_104-1",
url="https://doi.org/10.1007/978-981-16-4544-0_104-1"
}

@ARTICLE{Buccheri83,
       author = {{Buccheri}, R. and {Bennett}, K. and {Bignami}, G.~F. and {Bloemen}, J.~B.~G.~M. and {Boriakoff}, V. and {Caraveo}, P.~A. and {Hermsen}, W. and {Kanbach}, G. and {Manchester}, R.~N. and {Masnou}, J.~L. and {Mayer-Hasselwander}, H.~A. and {{\"O}zel}, M.~E. and {Paul}, J.~A. and {Sacco}, B. and {Scarsi}, L. and {Strong}, A.~W.},
        title = "{Search for pulsed {\ensuremath{\gamma}}-ray emission from radio pulsars in the COS-B data.}",
      journal = {\aap},
         year = 1983,
        month = dec,
       volume = {128},
        pages = {245-251},
       adsurl = {https://ui.adsabs.harvard.edu/abs/1983A&A...128..245B}
}

@ARTICLE{Ducci22,
       author = {{Ducci}, L. and {Mereghetti}, S. and {Santangelo}, A. and {Ji}, L. and {Carpano}, S. and {Covino}, S. and {Doroshenko}, V. and {Haberl}, F. and {Maitra}, C. and {Kreykenbohm}, I. and {Udalski}, A.},
        title = "{eROSITA detection of flares from the Be/X-ray binary A0538-66}",
      journal = {\aap},
         year = 2022,
        month = may,
       volume = {661},
          eid = {A22},
        pages = {A22},
          doi = {10.1051/0004-6361/202140867},
archivePrefix = {arXiv},
       eprint = {2106.14541},
 primaryClass = {astro-ph.HE},
       adsurl = {https://ui.adsabs.harvard.edu/abs/2022A&A...661A..22D}
}

@ARTICLE{Rajoelimanana17,
       author = {{Rajoelimanana}, A.~F. and {Charles}, P.~A. and {Meintjes}, P.~J. and {Townsend}, L.~J. and {Schurch}, M.~P.~E. and {Udalski}, A.},
        title = "{Orbital and superorbital monitoring of the Be/X-ray binary A0538-66: constraints on the system parameters}",
      journal = {\mnras},
         year = 2017,
        month = feb,
       volume = {464},
       number = {4},
        pages = {4133-4144},
          doi = {10.1093/mnras/stw2534},
archivePrefix = {arXiv},
       eprint = {1609.09811},
 primaryClass = {astro-ph.SR},
       adsurl = {https://ui.adsabs.harvard.edu/abs/2017MNRAS.464.4133R}
}

@ARTICLE{Nakajima26b,
       author = {{Nakajima}, M. and {Mihara}, T. and {Sugizaki}, M. and {Negoro}, H. and {Saito}, K. and {Yokoyama}, T. and {Wada}, M. and {Tamagawa}, T. and {Kawai}, N. and {Matsuoka}, M. and {Sakamoto}, T. and {Serino}, M. and {Sugita}, S. and {Kawakubo}, Y. and {Hiramatsu}, H. and {Yoshida}, A. and {Tsuboi}, Y. and {Sugai}, H. and {Nagashima}, N. and {Ishihara}, Y. and {Shidatsu}, M. and {Kang}, C. and {Nakamoto}, T. and {Uenishi}, M. and {Usuki}, T. and {Yatsuzuka}, S. and {Takahashi}, I. and {Yatsu}, Y. and {Nakahira}, S. and {Ueno}, S. and {Tomida}, H. and {Kurihara}, M. and {Ueda}, Y. and {Fujiwara}, K. and {Kobayashi}, S. and {Yamauchi}, M. and {Nishio}, M. and {Hiraizumi}, C. and {Yamaoka}, K. and {Iwakiri}, W. and {Kawamuro}, T. and {Yamada}, S. and {Ogawa}, S.},
        title = "{MAXI/GSC Observation of an X-ray Outburst from the Be/X-ray Binary Pulsar 1A 0535-66 (1A 0538-66)}",
      journal = {The Astronomer's Telegram},
         year = 2026,
        month = may,
       volume = {17815},
        pages = {1},
       adsurl = {https://ui.adsabs.harvard.edu/abs/2026ATel17815....1N}
}

@ARTICLE{Nakajima26a,
       author = {{Nakajima}, M. and {Negoro}, H. and {Mihara}, T. and {Sugizaki}, M. and {Takagi}, K. and {Nishio}, H. and {Tamagawa}, T. and {Kawai}, N. and {Matsuoka}, M. and {Sakamoto}, T. and {Serino}, M. and {Sugita}, S. and {Kawakubo}, Y. and {Hiramatsu}, H. and {Kondo}, Y. and {Iijima}, D. and {Yoshida}, A. and {Tsuboi}, Y. and {Sugai}, H. and {Nagashima}, N. and {Ishihara}, Y. and {Shidatsu}, M. and {Kang}, C. and {Nakamoto}, T. and {Uenishi}, M. and {Usuki}, T. and {Yatsuzuka}, S. and {Takahashi}, I. and {Yatsu}, Y. and {Nakahira}, S. and {Ueno}, S. and {Tomida}, H. and {Ogawa}, S. and {Kurihara}, M. and {Ueda}, Y. and {Fujiwara}, K. and {Kobayashi}, S. and {Yamauchi}, M. and {Nishio}, M. and {Hiraizumi}, C. and {Yamaoka}, K. and {Iwakiri}, W. and {Kawamuro}, T. and {Yamada}, S.},
        title = "{MAXI/GSC observation of a bright outburst from a Be/X-ray binary 1A 0538-66}",
      journal = {The Astronomer's Telegram},
         year = 2026,
        month = mar,
       volume = {17705},
        pages = {1},
       adsurl = {https://ui.adsabs.harvard.edu/abs/2026ATel17705....1N}
}

@ARTICLE{Reig26,
       author = {{Reig}, Pablo},
        title = "{Be/X-Ray Binaries: Phenomenology, Variability, and Accretion\raisebox{-0.5ex}\textasciitildeDynamics}",
      journal = {arXiv e-prints},
         year = 2026,
        month = jul,
          eid = {arXiv:2607.18980},
        pages = {arXiv:2607.18980},
archivePrefix = {arXiv},
       eprint = {2607.18980},
 primaryClass = {astro-ph.HE},
       adsurl = {https://ui.adsabs.harvard.edu/abs/2026arXiv260718980R}
}

@ARTICLE{Okazaki13,
       author = {{Okazaki}, Atsuo T. and {Hayasaki}, Kimitake and {Moritani}, Yuki},
        title = "{Origin of Two Types of X-Ray Outbursts in Be/X-Ray Binaries. I. Accretion Scenarios}",
      journal = {\pasj},
         year = 2013,
        month = apr,
       volume = {65},
          eid = {41},
        pages = {41},
          doi = {10.1093/pasj/65.2.41},
archivePrefix = {arXiv},
       eprint = {1211.5225},
 primaryClass = {astro-ph.HE},
       adsurl = {https://ui.adsabs.harvard.edu/abs/2013PASJ...65...41O}
}

@ARTICLE{Martin14,
       author = {{Martin}, Rebecca G. and {Nixon}, Chris and {Lubow}, Stephen H. and {Armitage}, Philip J. and {Price}, Daniel J. and {Do{\u{g}}an}, Suzan and {King}, Andrew},
        title = "{The Kozai-Lidov Mechanism in Hydrodynamical Disks}",
      journal = {\apjl},
         year = 2014,
        month = sep,
       volume = {792},
       number = {2},
          eid = {L33},
        pages = {L33},
          doi = {10.1088/2041-8205/792/2/L33},
archivePrefix = {arXiv},
       eprint = {1409.1226},
 primaryClass = {astro-ph.EP},
       adsurl = {https://ui.adsabs.harvard.edu/abs/2014ApJ...792L..33M}
}

@ARTICLE{Schulz20,
       author = {{Schulz}, N.~S. and {Kallman}, T.~E. and {Heinz}, S. and {Sell}, P. and {Jonker}, P. and {Brandt}, W.~N.},
        title = "{Origins of X-Ray Line Emissions in Circinus X-1 at Very Low X-Ray Flux}",
      journal = {\apj},
         year = 2020,
        month = mar,
       volume = {891},
       number = {2},
          eid = {150},
        pages = {150},
          doi = {10.3847/1538-4357/ab6dc8},
archivePrefix = {arXiv},
       eprint = {2001.05638},
 primaryClass = {astro-ph.HE},
       adsurl = {https://ui.adsabs.harvard.edu/abs/2020ApJ...891..150S}
}

@ARTICLE{Alcock01,
       author = {{Alcock}, C. and {Allsman}, R.~A. and {Alves}, D.~R. and {Axelrod}, T.~S. and {Becker}, A.~C. and {Bennett}, D.~P. and {Charles}, P.~A. and {Cook}, K.~H. and {Drake}, A.~J. and {Freeman}, K.~C. and {Geha}, M. and {Griest}, K. and {Lehner}, M.~J. and {Marshall}, S.~L. and {McGowan}, K.~E. and {Minniti}, D. and {Nelson}, C.~A. and {Peterson}, B.~A. and {Popowski}, P. and {Pratt}, M.~R. and {Quinn}, P.~J. and {Stubbs}, C.~W. and {Sutherland}, W. and {Tomaney}, A.~B. and {Vandehei}, T. and {Welch}, D.~L.},
        title = "{A 421-d activity cycle in the BeX recurrent transient A0538-66 from MACHO monitoring}",
      journal = {\mnras},
         year = 2001,
        month = mar,
       volume = {321},
       number = {4},
        pages = {678-684},
          doi = {10.1046/j.1365-8711.2001.04041.x},
archivePrefix = {arXiv},
       eprint = {astro-ph/0009415},
 primaryClass = {astro-ph},
       adsurl = {https://ui.adsabs.harvard.edu/abs/2001MNRAS.321..678A}
}

@ARTICLE{Ducci16,
       author = {{Ducci}, L. and {Covino}, S. and {Doroshenko}, V. and {Mereghetti}, S. and {Santangelo}, A. and {Sasaki}, M.},
        title = "{Optical and near-infrared photometric monitoring of the transient X-ray binary A0538-66 with REM}",
      journal = {\aap},
         year = 2016,
        month = nov,
       volume = {595},
          eid = {A103},
        pages = {A103},
          doi = {10.1051/0004-6361/201629236},
archivePrefix = {arXiv},
       eprint = {1610.01378},
 primaryClass = {astro-ph.HE},
       adsurl = {https://ui.adsabs.harvard.edu/abs/2016A&A...595A.103D}
}

@ARTICLE{White78,
       author = {{White}, N.~E. and {Carpenter}, G.~F.},
        title = "{The recurrent X-ray transient A 0538-66.}",
      journal = {\mnras},
         year = 1978,
        month = apr,
       volume = {183},
        pages = {11P-15},
          doi = {10.1093/mnras/183.1.11P},
       adsurl = {https://ui.adsabs.harvard.edu/abs/1978MNRAS.183P..11W}
}

@ARTICLE{Skinner82,
       author = {{Skinner}, G.~K. and {Bedford}, D.~K. and {Elsner}, R.~F. and {Leahy}, D. and {Weisskopf}, M.~C. and {Grindlay}, J.},
        title = "{Discovery of 69 ms periodic X-ray pulsations in A0538 - 66}",
      journal = {\nat},
         year = 1982,
        month = jun,
       volume = {297},
       number = {5867},
        pages = {568-570},
          doi = {10.1038/297568a0},
       adsurl = {https://ui.adsabs.harvard.edu/abs/1982Natur.297..568S}
}

@ARTICLE{Ducci25b,
       author = {{Ducci}, L. and {Mereghetti}, S. and {Bachetti}, M.},
        title = "{XMM-Newton detection of a bright outburst from the accreting pulsar 1A 0538-66}",
      journal = {The Astronomer's Telegram},
         year = 2025,
        month = oct,
       volume = {17455},
        pages = {1},
       adsurl = {https://ui.adsabs.harvard.edu/abs/2025ATel17455....1D}
}

@ARTICLE{Johnston99,
       author = {{Johnston}, Simon and {Manchester}, R.~N. and {McConnell}, D. and {Campbell-Wilson}, D.},
        title = "{Transient radio emission from the PSR B1259-63 system near periastron}",
      journal = {\mnras},
         year = 1999,
        month = jan,
       volume = {302},
       number = {2},
        pages = {277-287},
          doi = {10.1046/j.1365-8711.1999.02133.x},
       adsurl = {https://ui.adsabs.harvard.edu/abs/1999MNRAS.302..277J}
}

@INPROCEEDINGS{kaastraspex,
       author = {{Kaastra}, J.~S. and {Mewe}, R. and {Nieuwenhuijzen}, H.},
        title = "{SPEX: a new code for spectral analysis of X \& UV spectra.}",
    booktitle = {UV and X-ray Spectroscopy of Astrophysical and Laboratory Plasmas},
         year = 1996,
       editor = {{Yamashita}, K. and {Watanabe}, T.},
        month = jan,
        pages = {411-414},
       adsurl = {https://ui.adsabs.harvard.edu/abs/1996uxsa.conf..411K}
}

@ARTICLE{Cash1979,
       author = {{Cash}, W.},
        title = "{Parameter estimation in astronomy through application of the likelihood ratio.}",
      journal = {\apj},
         year = 1979,
        month = mar,
       volume = {228},
        pages = {939-947},
          doi = {10.1086/156922},
       adsurl = {https://ui.adsabs.harvard.edu/abs/1979ApJ...228..939C}
}

@ARTICLE{Kaastra2017,
       author = {{Kaastra}, J.~S.},
        title = "{On the use of C-stat in testing models for X-ray spectra}",
      journal = {\aap},
         year = 2017,
        month = sep,
       volume = {605},
          eid = {A51},
        pages = {A51},
          doi = {10.1051/0004-6361/201629319},
archivePrefix = {arXiv},
       eprint = {1707.09202},
 primaryClass = {astro-ph.HE},
       adsurl = {https://ui.adsabs.harvard.edu/abs/2017A&A...605A..51K}
}

@ARTICLE{Lodders2009,
       author = {{Lodders}, K. and {Palme}, H. and {Gail}, H.-P.},
        title = "{Abundances of the Elements in the Solar System}",
      journal = {Landolt B{\"o}rnstein},
         year = 2009,
        month = jan,
       volume = {4B},
        pages = {712},
          doi = {10.1007/978-3-540-88055-4_34},
archivePrefix = {arXiv},
       eprint = {0901.1149},
 primaryClass = {astro-ph.EP},
       adsurl = {https://ui.adsabs.harvard.edu/abs/2009LanB...4B..712L}
}

@ARTICLE{Barra2024,
       author = {{Barra}, F. and {Pinto}, C. and {Middleton}, M. and {Di Salvo}, T. and {Walton}, D.~J. and {G{\'u}rpide}, A. and {Roberts}, T.~P.},
        title = "{Examining the nature of the ultraluminous X-ray source Holmberg II X-1}",
      journal = {\aap},
         year = 2024,
        month = feb,
       volume = {682},
          eid = {A94},
        pages = {A94},
          doi = {10.1051/0004-6361/202348471},
archivePrefix = {arXiv},
       eprint = {2311.16243},
 primaryClass = {astro-ph.HE},
       adsurl = {https://ui.adsabs.harvard.edu/abs/2024A&A...682A..94B}
}

@ARTICLE{Pintore2017,
       author = {{Pintore}, F. and {Zampieri}, L. and {Stella}, L. and {Wolter}, A. and {Mereghetti}, S. and {Israel}, G.~L.},
        title = "{Pulsator-like Spectra from Ultraluminous X-Ray Sources and the Search for More Ultraluminous Pulsars}",
      journal = {\apj},
         year = 2017,
        month = feb,
       volume = {836},
       number = {1},
          eid = {113},
        pages = {113},
          doi = {10.3847/1538-4357/836/1/113},
archivePrefix = {arXiv},
       eprint = {1701.03595},
 primaryClass = {astro-ph.HE},
       adsurl = {https://ui.adsabs.harvard.edu/abs/2017ApJ...836..113P}
}

@ARTICLE{Kosec2021,
       author = {{Kosec}, P. and {Pinto}, C. and {Reynolds}, C.~S. and {Guainazzi}, M. and {Kara}, E. and {Walton}, D.~J. and {Fabian}, A.~C. and {Parker}, M.~L. and {Valtchanov}, I.},
        title = "{Ionized emission and absorption in a large sample of ultraluminous X-ray sources}",
      journal = {\mnras},
         year = 2021,
        month = dec,
       volume = {508},
       number = {3},
        pages = {3569-3588},
          doi = {10.1093/mnras/stab2856},
archivePrefix = {arXiv},
       eprint = {2109.14683},
 primaryClass = {astro-ph.HE},
       adsurl = {https://ui.adsabs.harvard.edu/abs/2021MNRAS.508.3569K}
}

@ARTICLE{Pinto2020b,
       author = {{Pinto}, C. and {Walton}, D.~J. and {Kara}, E. and {Parker}, M.~L. and {Soria}, R. and {Kosec}, P. and {Middleton}, M.~J. and {Alston}, W.~N. and {Fabian}, A.~C. and {Guainazzi}, M. and {Roberts}, T.~P. and {Fuerst}, F. and {Earnshaw}, H.~P. and {Sathyaprakash}, R. and {Barret}, D.},
        title = "{XMM-Newton campaign on ultraluminous X-ray source NGC 1313 X-1: wind versus state variability}",
      journal = {\mnras},
         year = 2020,
        month = mar,
       volume = {492},
       number = {4},
        pages = {4646-4665},
          doi = {10.1093/mnras/staa118},
archivePrefix = {arXiv},
       eprint = {1911.09568},
 primaryClass = {astro-ph.HE},
       adsurl = {https://ui.adsabs.harvard.edu/abs/2020MNRAS.492.4646P}
}

@Inbook{Pinto2023,
author="Pinto, Ciro
and Walton, Dominic J.",
title="Ultra-Luminous X-Ray Sources: Extreme Accretion and Feedback",
bookTitle="High-Resolution X-ray Spectroscopy: Instrumentation, Data Analysis, and Science",
year="2023",
publisher="Springer Nature Singapore",
address="Singapore",
pages="345--391",
isbn="978-981-99-4409-5",
doi="10.1007/978-981-99-4409-5_12",
url="https://doi.org/10.1007/978-981-99-4409-5_12"
}

@ARTICLE{Pinto2026a,
       author = {{Pinto}, C. and {Del Santo}, M. and {D'A{\`\i}}, A. and {Pintore}, F. and {Russell}, T. and {Parra}, M. and {Ferreira}, J. and {Petrucci}, P.-O. and {Fukumura}, K. and {Marino}, A. and {Mu{\~n}oz-Darias}, T. and {Rodr{\'\i}guez Castillo}, G.~A. and {Segreto}, A.},
        title = "{A high-resolution X-ray view of the ultra-fast outflow in MAXI J1810{\ensuremath{-}}222}",
      journal = {\aap},
         year = 2026,
        month = apr,
       volume = {708},
          eid = {A332},
        pages = {A332},
          doi = {10.1051/0004-6361/202558315},
       adsurl = {https://ui.adsabs.harvard.edu/abs/2026A&A...708A.332P}
}

@ARTICLE{Pinto2026b,
       author = {{Pinto}, C. and {Caserta}, S. and {Barra}, F. and {Xu}, Y. and {Barret}, D. and {Kosec}, P. and {La Palombara}, N. and {Marino}, A. and {Pintore}, F. and {Riggio}, A. and {Roberts}, T.~P. and {Salvaggio}, C. and {Sidoli}, L. and {Soria}, R. and {Walton}, D.~J.},
        title = "{XMM-Newton multi-year campaign on NGC 55 ULX-1: Resolving the wind and its variability with RGS}",
      journal = {\aap},
         year = 2026,
        month = feb,
       volume = {706},
          eid = {A266},
        pages = {A266},
          doi = {10.1051/0004-6361/202556631},
archivePrefix = {arXiv},
       eprint = {2512.16522},
 primaryClass = {astro-ph.HE},
       adsurl = {https://ui.adsabs.harvard.edu/abs/2026A&A...706A.266P}
}

@ARTICLE{Nakajima25,
       author = {{Nakajima}, M. and {Sugizaki}, M. and {Mihara}, T. and {Negoro}, H. and {Takagi}, K. and {Takahashi}, H. and {Nishio}, H. and {Tamagawa}, T. and {Kawai}, N. and {Matsuoka}, M. and {Sakamoto}, T. and {Serino}, M. and {Sugita}, S. and {Kawakubo}, Y. and {Hiramatsu}, H. and {Kondo}, Y. and {Yoshida}, A. and {Tsuboi}, Y. and {Sugai}, H. and {Nagashima}, N. and {Ishihara}, Y. and {Shidatsu}, M. and {Kang}, C. and {Nakamoto}, T. and {Uenishi}, M. and {Usuki}, T. and {Yatsuzuka}, S. and {Takahashi}, I. and {Yatsu}, Y. and {Nakahira}, S. and {Ueno}, S. and {Tomida}, H. and {Ishikawa}, M. and {Ogawa}, S. and {Kurihara}, M. and {Ueda}, Y. and {Fujiwara}, K. and {Yamauchi}, M. and {Nishio}, M. and {Hiraizumi}, C. and {Yamaoka}, K. and {Iwakiri}, W. and {Kawamuro}, T. and {Yamada}, S.},
        title = "{MAXI observation of a bright outburst from a Be/X-ray binary 1A 0538-66 around September 30, 2025.}",
      journal = {The Astronomer's Telegram},
         year = 2025,
        month = nov,
       volume = {17467},
        pages = {1},
       adsurl = {https://ui.adsabs.harvard.edu/abs/2025ATel17467....1N}
}

@ARTICLE{Charles83,
       author = {{Charles}, P.~A. and {Booth}, L. and {Densham}, R.~H. and {Bath}, G.~T. and {Thorstensen}, J.~R. and {Howarth}, I.~D. and {Willis}, A.~J. and {Skinner}, G.~K. and {Olszewski}, E.},
        title = "{Extreme variability in the Be-type, periodic recurrent X-ray transient AO538-66 : a highly eccentric interacting binary.}",
      journal = {\mnras},
         year = 1983,
        month = feb,
       volume = {202},
        pages = {657-682},
          doi = {10.1093/mnras/202.3.657},
       adsurl = {https://ui.adsabs.harvard.edu/abs/1983MNRAS.202..657C}
}

@ARTICLE{Skinner80,
       author = {{Skinner}, G.~K.},
        title = "{Observations of optical flares in the recurrent X-ray transient A0538-66}",
      journal = {\nat},
         year = 1980,
        month = nov,
       volume = {288},
       number = {5787},
        pages = {141-143},
          doi = {10.1038/288141a0},
       adsurl = {https://ui.adsabs.harvard.edu/abs/1980Natur.288..141S}
}

@ARTICLE{Densham83,
       author = {{Densham}, R.~H. and {Charles}, P.~A. and {Menzies}, J.~W. and {van der Klis}, M. and {van Paradijs}, J.},
        title = "{Four outburst cycles of A 0538-66: evidence for a rapidly evolving envelope around the primary.}",
      journal = {\mnras},
         year = 1983,
        month = dec,
       volume = {205},
        pages = {1117-1133},
          doi = {10.1093/mnras/205.4.1117},
       adsurl = {https://ui.adsabs.harvard.edu/abs/1983MNRAS.205.1117D}
}

@ARTICLE{Mushtukov19,
       author = {{Mushtukov}, Alexander A. and {Ingram}, Adam and {Middleton}, Matthew and {Nagirner}, Dmitrij I. and {van der Klis}, Michiel},
        title = "{Timing properties of ULX pulsars: optically thick envelopes and outflows}",
      journal = {\mnras},
         year = 2019,
        month = mar,
       volume = {484},
       number = {1},
        pages = {687-697},
          doi = {10.1093/mnras/sty3525},
archivePrefix = {arXiv},
       eprint = {1811.02049},
 primaryClass = {astro-ph.HE},
       adsurl = {https://ui.adsabs.harvard.edu/abs/2019MNRAS.484..687M}
}

@ARTICLE{Middleton21,
       author = {{Middleton}, M.~J. and {Walton}, D.~J. and {Alston}, W. and {Dauser}, T. and {Eikenberry}, S. and {Jiang}, Y.-F. and {Fabian}, A.~C. and {Fuerst}, F. and {Brightman}, M. and {Marshall}, H. and {Parker}, M. and {Pinto}, C. and {Harrison}, F.~A. and {Bachetti}, M. and {Altamirano}, D. and {Bird}, A.~J. and {Perez}, G. and {Miller-Jones}, J. and {Charles}, P. and {Boggs}, S. and {Christensen}, F. and {Craig}, W. and {Forster}, K. and {Grefenstette}, B. and {Hailey}, C. and {Madsen}, K. and {Stern}, D. and {Zhang}, W.},
        title = "{NuSTAR reveals the hidden nature of SS433}",
      journal = {\mnras},
         year = 2021,
        month = sep,
       volume = {506},
       number = {1},
        pages = {1045-1058},
          doi = {10.1093/mnras/stab1280},
archivePrefix = {arXiv},
       eprint = {1810.10518},
 primaryClass = {astro-ph.HE},
       adsurl = {https://ui.adsabs.harvard.edu/abs/2021MNRAS.506.1045M}
}

@ARTICLE{Veledina24,
       author = {{Veledina}, Alexandra and {Muleri}, Fabio and {Poutanen}, Juri and {Podgorn{\'y}}, Jakub and {Dov{\v{c}}iak}, Michal and {Capitanio}, Fiamma and {Churazov}, Eugene and {De Rosa}, Alessandra and {Di Marco}, Alessandro and {Forsblom}, Sofia V. and {Kaaret}, Philip and {Krawczynski}, Henric and {La Monaca}, Fabio and {Loktev}, Vladislav and {Lutovinov}, Alexander A. and {Molkov}, Sergey V. and {Mushtukov}, Alexander A. and {Ratheesh}, Ajay and {Rodriguez Cavero}, Nicole and {Steiner}, James F. and {Sunyaev}, Rashid A. and {Tsygankov}, Sergey S. and {Weisskopf}, Martin C. and {Zdziarski}, Andrzej A. and {Bianchi}, Stefano and {Bright}, Joe S. and {Bursov}, Nikolaj and {Costa}, Enrico and {Egron}, Elise and {Garcia}, Javier A. and {Green}, David A. and {Gurwell}, Mark and {Ingram}, Adam and {Kajava}, Jari J.~E. and {Kale}, Ruta and {Kraus}, Alex and {Malyshev}, Denys and {Marin}, Fr{\'e}d{\'e}ric and {Matt}, Giorgio and {McCollough}, Michael and {Mereminskiy}, Ilya A. and {Nizhelsky}, Nikolaj and {Piano}, Giovanni and {Pilia}, Maura and {Pittori}, Carlotta and {Rao}, Ramprasad and {Righini}, Simona and {Soffitta}, Paolo and {Shevchenko}, Anton and {Svoboda}, Jiri and {Tombesi}, Francesco and {Trushkin}, Sergei A. and {Tsybulev}, Peter and {Ursini}, Francesco and {Wu}, Kinwah and {Agudo}, Iv{\'a}n and {Antonelli}, Lucio A. and {Bachetti}, Matteo and {Baldini}, Luca and {Baumgartner}, Wayne H. and {Bellazzini}, Ronaldo and {Bongiorno}, Stephen D. and {Bonino}, Raffaella and {Brez}, Alessandro and {Bucciantini}, Niccol{\`o} and {Castellano}, Simone and {Cavazzuti}, Elisabetta and {Chen}, Chien-Ting and {Ciprini}, Stefano and {Del Monte}, Ettore and {Di Gesu}, Laura and {Di Lalla}, Niccol{\`o} and {Donnarumma}, Immacolata and {Doroshenko}, Victor and {Ehlert}, Steven R. and {Enoto}, Teruaki and {Evangelista}, Yuri and {Fabiani}, Sergio and {Ferrazzoli}, Riccardo and {Gunji}, Shuichi and {Hayashida}, Kiyoshi and {Heyl}, Jeremy and {Iwakiri}, Wataru and {Jorstad}, Svetlana G. and {Karas}, Vladimir and {Kislat}, Fabian and {Kitaguchi}, Takao and {Kolodziejczak}, Jeffery J. and {Latronico}, Luca and {Liodakis}, Ioannis and {Maldera}, Simone and {Manfreda}, Alberto and {Marinucci}, Andrea and {Marscher}, Alan P. and {Marshall}, Herman L. and {Massaro}, Francesco and {Mitsuishi}, Ikuyuki and {Mizuno}, Tsunefumi and {Negro}, Michela and {Ng}, Chi-Yung and {O'Dell}, Stephen L. and {Omodei}, Nicola and {Oppedisano}, Chiara and {Papitto}, Alessandro and {Pavlov}, George G. and {Peirson}, Abel L. and {Perri}, Matteo and {Pesce-Rollins}, Melissa and {Petrucci}, Pierre-Olivier and {Possenti}, Andrea and {Puccetti}, Simonetta and {Ramsey}, Brian D. and {Rankin}, John and {Roberts}, Oliver and {Romani}, Roger W. and {Sgr{\`o}}, Carmelo and {Slane}, Patrick and {Spandre}, Gloria and {Swartz}, Doug and {Tamagawa}, Toru and {Tavecchio}, Fabrizio and {Taverna}, Roberto and {Tawara}, Yuzuru and {Tennant}, Allyn F. and {Thomas}, Nicholas E. and {Trois}, Alessio and {Turolla}, Roberto and {Vink}, Jacco and {Xie}, Fei and {Zane}, Silvia},
        title = "{Cygnus X-3 revealed as a Galactic ultraluminous X-ray source by IXPE}",
      journal = {Nature Astronomy},
         year = 2024,
        month = aug,
       volume = {8},
        pages = {1031-1046},
          doi = {10.1038/s41550-024-02294-9},
archivePrefix = {arXiv},
       eprint = {2303.01174},
 primaryClass = {astro-ph.HE},
       adsurl = {https://ui.adsabs.harvard.edu/abs/2024NatAs...8.1031V}
}

@ARTICLE{Vasilopoulos21,
       author = {{Vasilopoulos}, Georgios and {Koliopanos}, Filippos and {Haberl}, Frank and {Treiber}, Helena and {Brightman}, Murray and {Earnshaw}, Hannah P. and {G{\'u}rpide}, Andr{\'e}s},
        title = "{Chandra Probes the X-Ray Variability of M51 ULX-7: Evidence of Propeller Transition and X-Ray Dips on Orbital Periods}",
      journal = {\apj},
         year = 2021,
        month = mar,
       volume = {909},
       number = {1},
          eid = {50},
        pages = {50},
          doi = {10.3847/1538-4357/abda49},
archivePrefix = {arXiv},
       eprint = {2102.07996},
 primaryClass = {astro-ph.HE},
       adsurl = {https://ui.adsabs.harvard.edu/abs/2021ApJ...909...50V}
}

@ARTICLE{Fuerst21,
       author = {{F{\"u}rst}, F. and {Walton}, D.~J. and {Heida}, M. and {Bachetti}, M. and {Pinto}, C. and {Middleton}, M.~J. and {Brightman}, M. and {Earnshaw}, H.~P. and {Barret}, D. and {Fabian}, A.~C. and {Kretschmar}, P. and {Pottschmidt}, K. and {Ptak}, A. and {Roberts}, T. and {Stern}, D. and {Webb}, N. and {Wilms}, J.},
        title = "{Long-term pulse period evolution of the ultra-luminous X-ray pulsar NGC 7793 P13}",
      journal = {\aap},
         year = 2021,
        month = jul,
       volume = {651},
          eid = {A75},
        pages = {A75},
          doi = {10.1051/0004-6361/202140625},
       adsurl = {https://ui.adsabs.harvard.edu/abs/2021A&A...651A..75F}
}

@ARTICLE{Karmakar25,
       author = {{Karmakar}, Sricheta and {Mushtukov}, Alexander A. and {Middleton}, Matthew},
        title = "{Angle-dependent hardening of the reprocessed spectra in ULXs powered by accretion on to neutron stars}",
      journal = {\mnras},
         year = 2025,
        month = oct,
       volume = {543},
       number = {2},
        pages = {1447-1455},
          doi = {10.1093/mnras/staf1569},
archivePrefix = {arXiv},
       eprint = {2509.11682},
 primaryClass = {astro-ph.HE},
       adsurl = {https://ui.adsabs.harvard.edu/abs/2025MNRAS.543.1447K}
}

@ARTICLE{Huppenkothen19,
       author = {{Huppenkothen}, Daniela and {Bachetti}, Matteo and {Stevens}, Abigail L. and {Migliari}, Simone and {Balm}, Paul and {Hammad}, Omar and {Khan}, Usman Mahmood and {Mishra}, Himanshu and {Rashid}, Haroon and {Sharma}, Swapnil and {Martinez Ribeiro}, Evandro and {Valles Blanco}, Ricardo},
        title = "{Stingray: A Modern Python Library for Spectral Timing}",
      journal = {\apj},
         year = 2019,
        month = aug,
       volume = {881},
       number = {1},
          eid = {39},
        pages = {39},
          doi = {10.3847/1538-4357/ab258d},
archivePrefix = {arXiv},
       eprint = {1901.07681},
 primaryClass = {astro-ph.IM},
       adsurl = {https://ui.adsabs.harvard.edu/abs/2019ApJ...881...39H}
}

@ARTICLE{Bachetti24,
       author = {{Bachetti}, Matteo and {Huppenkothen}, Daniela and {Stevens}, Abigail and {Swinbank}, John and {Mastroserio}, Guglielmo and {Lucchini}, Matteo and {Lai}, Eleonora and {Buchner}, Johannes and {Desai}, Amogh and {Joshi}, Gaurav and {Pisanu}, Francesco and {Pisupati}, Sri and {Sharma}, Swapnil and {Tripathi}, Mihir and {Vats}, Dhruv},
        title = "{Stingray 2: A fast and modern Python library for spectral timing}",
      journal = {The Journal of Open Source Software},
         year = 2024,
        month = oct,
       volume = {9},
       number = {102},
          eid = {7389},
        pages = {7389},
          doi = {10.21105/joss.07389},
       adsurl = {https://ui.adsabs.harvard.edu/abs/2024JOSS....9.7389B}
}

@INPROCEEDINGS{Zerbi04,
       author = {{Zerbi}, Filippo M. and {Chincarini}, Guido and {Ghisellini}, Gabriele and {Rodono}, Marcello and {Tosti}, Gino and {Antonelli}, Lucio A. and {Conconi}, Paolo and {Covino}, Stefano and {Cutispoto}, Giuseppe and {Molinari}, Emilio and {Nicastro}, Luciano and {Palazzi}, Eliana},
        title = "{The REM telescope: a robotic multiwavelength facility}",
    booktitle = {Ground-based Instrumentation for Astronomy},
         year = 2004,
       editor = {{Moorwood}, Alan F.~M. and {Iye}, Masanori},
       series = {Society of Photo-Optical Instrumentation Engineers (SPIE) Conference Series},
       volume = {5492},
        month = sep,
        pages = {1590-1601},
          doi = {10.1117/12.551145},
       adsurl = {https://ui.adsabs.harvard.edu/abs/2004SPIE.5492.1590Z}
}

@INPROCEEDINGS{Covino04,
       author = {{Covino}, Stefano and {Stefanon}, Mauro and {Sciuto}, Giorgio and {Fernandez-Soto}, Alberto and {Tosti}, Gino and {Zerbi}, Filippo M. and {Chincarini}, Guido and {Antonelli}, Lucio A. and {Conconi}, Paolo and {Cutispoto}, Giuseppe and {Molinari}, Emilio and {Nicastro}, Luciano and {Rodono}, Marcello},
        title = "{REM: a fully robotic telescope for GRB observations}",
    booktitle = {Ground-based Instrumentation for Astronomy},
         year = 2004,
       editor = {{Moorwood}, Alan F.~M. and {Iye}, Masanori},
       series = {Society of Photo-Optical Instrumentation Engineers (SPIE) Conference Series},
       volume = {5492},
        month = sep,
        pages = {1613-1622},
          doi = {10.1117/12.551532},
       adsurl = {https://ui.adsabs.harvard.edu/abs/2004SPIE.5492.1613C}
}

@dataset{2020yCat.1350....0G,
       author = {{Gaia Collaboration}},
        title = "{VizieR Online Data Catalog: Gaia EDR3 (Gaia Collaboration, 2020)}",
 howpublished = {VizieR On-line Data Catalog: I/350.  Originally published in: 2021A\&A...649A...1G},
         year = 2020,
        month = nov,
          eid = {I/350},
          doi = {10.26093/cds/vizier.1350},
       adsurl = {https://ui.adsabs.harvard.edu/abs/2020yCat.1350....0G}
}

@software{2015ascl.soft02021B,
       author = {{Bachetti}, Matteo},
        title = "{MaLTPyNT: Quick look timing analysis for NuSTAR data}",
 howpublished = {Astrophysics Source Code Library, record ascl:1502.021},
         year = 2015,
        month = feb,
          eid = {ascl:1502.021},
archivePrefix = {ascl},
       eprint = {1502.021},
       adsurl = {https://ui.adsabs.harvard.edu/abs/2015ascl.soft02021B}
}

@ARTICLE{Kosec2018b,
       author = {{Kosec}, P. and {Pinto}, C. and {Fabian}, A.~C. and {Walton}, D.~J.},
        title = "{Searching for outflows in ultraluminous X-ray sources through high-resolution X-ray spectroscopy}",
      journal = {\mnras},
         year = 2018,
        month = feb,
       volume = {473},
       number = {4},
        pages = {5680-5697},
          doi = {10.1093/mnras/stx2695},
archivePrefix = {arXiv},
       eprint = {1710.06438},
 primaryClass = {astro-ph.HE},
       adsurl = {https://ui.adsabs.harvard.edu/abs/2018MNRAS.473.5680K}
}

@ARTICLE{Morii16,
       author = {{Morii}, Mikio and {Yamaoka}, Hitoshi and {Mihara}, Tatehiro and {Matsuoka}, Masaru and {Kawai}, Nobuyuki},
        title = "{Search for soft X-ray flashes at the fireball phase of classical/recurrent novae using MAXI/GSC data}",
      journal = {\pasj},
         year = 2016,
        month = jun,
       volume = {68},
          eid = {S11},
        pages = {S11},
          doi = {10.1093/pasj/psw007},
archivePrefix = {arXiv},
       eprint = {1601.04394},
 primaryClass = {astro-ph.HE},
       adsurl = {https://ui.adsabs.harvard.edu/abs/2016PASJ...68S..11M}
}

@ARTICLE{Corbet97,
       author = {{Corbet}, R.~H.~D. and {Charles}, P.~A. and {Southwell}, K.~A. and {Smale}, A.~P.},
        title = "{A Low-Amplitude X-Ray and Optical Outburst from the Periodic Transient A0538-66: Accretion onto a Magnetosphere?}",
      journal = {\apj},
         year = 1997,
        month = feb,
       volume = {476},
       number = {2},
        pages = {833-841},
          doi = {10.1086/303644},
       adsurl = {https://ui.adsabs.harvard.edu/abs/1997ApJ...476..833C}
}

@ARTICLE{Mukai93,
       author = {{Mukai}, K.},
        title = "{PIMMS and Viewing: proposal preparation tools}",
      journal = {Legacy},
         year = 1993,
        month = may,
       volume = {3},
        pages = {21-31},
       adsurl = {https://ui.adsabs.harvard.edu/abs/1993Legac...3...21M}
}

@ARTICLE{Dai06,
       author = {{Dai}, Hai-Lang and {Liu}, Xi-Wei and {Li}, Xiang-Dong},
        title = "{Exploration of the P$_{s}$-P$_{orb}$ Relation for Wind-fed X-Ray Pulsars}",
      journal = {\apj},
         year = 2006,
        month = dec,
       volume = {653},
       number = {2},
        pages = {1410-1416},
          doi = {10.1086/508735},
archivePrefix = {arXiv},
       eprint = {astro-ph/0608621},
 primaryClass = {astro-ph},
       adsurl = {https://ui.adsabs.harvard.edu/abs/2006ApJ...653.1410D}
}

@ARTICLE{Mavromatakis93,
       author = {{Mavromatakis}, F. and {Haberl}, F.},
        title = "{Two outbursts from A 0538-66 in the ROSAT all-sky survey.}",
      journal = {\aap},
         year = 1993,
        month = jul,
       volume = {274},
        pages = {304-308},
       adsurl = {https://ui.adsabs.harvard.edu/abs/1993A&A...274..304M}
}

@ARTICLE{Campana97,
       author = {{Campana}, S.},
        title = "{ROSAT observations of A0538-66 during quiescence.}",
      journal = {\aap},
         year = 1997,
        month = apr,
       volume = {320},
        pages = {840-844},
          doi = {10.48550/arXiv.astro-ph/9610239},
archivePrefix = {arXiv},
       eprint = {astro-ph/9610239},
 primaryClass = {astro-ph},
       adsurl = {https://ui.adsabs.harvard.edu/abs/1997A&A...320..840C}
}

@ARTICLE{Campana02,
       author = {{Campana}, S. and {Stella}, L. and {Israel}, G.~L. and {Moretti}, A. and {Parmar}, A.~N. and {Orlandini}, M.},
        title = "{The Quiescent X-Ray Emission of Three Transient X-Ray Pulsars}",
      journal = {\apj},
         year = 2002,
        month = nov,
       volume = {580},
       number = {1},
        pages = {389-393},
          doi = {10.1086/343074},
archivePrefix = {arXiv},
       eprint = {astro-ph/0207422},
 primaryClass = {astro-ph},
       adsurl = {https://ui.adsabs.harvard.edu/abs/2002ApJ...580..389C}
}

@ARTICLE{Nakajima26c,
       author = {{Nakajima}, M. and {Hiramatsu}, H. and {Negoro}, H. and {Saito}, K. and {Yokoyama}, T. and {Wada}, M. and {Mihara}, T. and {Tamagawa}, T. and {Kawai}, N. and {Matsuoka}, M. and {Sakamoto}, T. and {Serino}, M. and {Sugita}, S. and {Kawakubo}, Y. and {Yoshida}, A. and {Tsuboi}, Y. and {Ishihara}, Y. and {Ogaki}, H. and {Shidatsu}, M. and {Kang}, C. and {Uenishi}, M. and {Usuki}, T. and {Yatsuzuka}, S. and {Aoe}, H. and {Takahashi}, I. and {Yatsu}, Y. and {Nakahira}, S. and {Ueno}, S. and {Tomida}, H. and {Kurihara}, M. and {Ueda}, Y. and {Fujiwara}, K. and {Kobayashi}, S. and {Yamauchi}, M. and {Nishio}, M. and {Hiraizumi}, C. and {Yamaoka}, K. and {Sugizaki}, M. and {Iwakiri}, W. and {Kawamuro}, T. and {Yamada}, S. and {Ogawa}, S.},
        title = "{MAXI/GSC observation of a major X-ray outburst from the Be/X-ray Binary Pulsar 1A 0535-66 (1A 0538-66)}",
      journal = {The Astronomer's Telegram},
         year = 2026,
        month = aug,
       volume = {18020},
        pages = {1},
       adsurl = {https://ui.adsabs.harvard.edu/abs/2026ATel18020....1N}
}

@ARTICLE{Vasilopoulos20,
       author = {{Vasilopoulos}, G. and {Ray}, P.~S. and {Gendreau}, K.~C. and {Jenke}, P.~A. and {Jaisawal}, G.~K. and {Wilson-Hodge}, C.~A. and {Strohmayer}, T.~E. and {Altamirano}, D. and {Iwakiri}, W.~B. and {Wolff}, M.~T. and {Guillot}, S. and {Malacaria}, C. and {Stevens}, A.~L.},
        title = "{The 2019 super-Eddington outburst of RX J0209.6-7427: detection of pulsations and constraints on the magnetic field strength}",
      journal = {\mnras},
         year = 2020,
        month = jun,
       volume = {494},
       number = {4},
        pages = {5350-5359},
          doi = {10.1093/mnras/staa991},
archivePrefix = {arXiv},
       eprint = {2004.03022},
 primaryClass = {astro-ph.HE},
       adsurl = {https://ui.adsabs.harvard.edu/abs/2020MNRAS.494.5350V}
}

@ARTICLE{Wilson-Hodge18,
       author = {{Wilson-Hodge}, Colleen A. and {Malacaria}, Christian and {Jenke}, Peter A. and {Jaisawal}, Gaurava K. and {Kerr}, Matthew and {Wolff}, Michael T. and {Arzoumanian}, Zaven and {Chakrabarty}, Deepto and {Doty}, John P. and {Gendreau}, Keith C. and {Guillot}, Sebastien and {Ho}, Wynn C.~G. and {LaMarr}, Beverly and {Markwardt}, Craig B. and {{\"O}zel}, Feryal and {Prigozhin}, Gregory Y. and {Ray}, Paul S. and {Ramos-Lerate}, Mercedes and {Remillard}, Ronald A. and {Strohmayer}, Tod E. and {Vezie}, Michael L. and {Wood}, Kent S. and {NICER Science Team}},
        title = "{NICER and Fermi GBM Observations of the First Galactic Ultraluminous X-Ray Pulsar Swift J0243.6+6124}",
      journal = {\apj},
         year = 2018,
        month = aug,
       volume = {863},
       number = {1},
          eid = {9},
        pages = {9},
          doi = {10.3847/1538-4357/aace60},
archivePrefix = {arXiv},
       eprint = {1806.10094},
 primaryClass = {astro-ph.HE},
       adsurl = {https://ui.adsabs.harvard.edu/abs/2018ApJ...863....9W}
}

@ARTICLE{Matsuoka09,
       author = {{Matsuoka}, Masaru and {Kawasaki}, Kazuyoshi and {Ueno}, Shiro and {Tomida}, Hiroshi and {Kohama}, Mitsuhiro and {Suzuki}, Motoko and {Adachi}, Yasuki and {Ishikawa}, Masaki and {Mihara}, Tatehiro and {Sugizaki}, Mutsumi and {Isobe}, Naoki and {Nakagawa}, Yujin and {Tsunemi}, Hiroshi and {Miyata}, Emi and {Kawai}, Nobuyuki and {Kataoka}, Jun and {Morii}, Mikio and {Yoshida}, Atsumasa and {Negoro}, Hitoshi and {Nakajima}, Motoki and {Ueda}, Yoshihiro and {Chujo}, Hirotaka and {Yamaoka}, Kazutaka and {Yamazaki}, Osamu and {Nakahira}, Satoshi and {You}, Tetsuya and {Ishiwata}, Ryoji and {Miyoshi}, Sho and {Eguchi}, Satoshi and {Hiroi}, Kazuo and {Katayama}, Haruyoshi and {Ebisawa}, Ken},
        title = "{The MAXI Mission on the ISS: Science and Instruments for Monitoring All-Sky X-Ray Images}",
      journal = {\pasj},
         year = 2009,
        month = oct,
       volume = {61},
        pages = {999},
          doi = {10.1093/pasj/61.5.999},
archivePrefix = {arXiv},
       eprint = {0906.0631},
 primaryClass = {astro-ph.IM},
       adsurl = {https://ui.adsabs.harvard.edu/abs/2009PASJ...61..999M}
}

@ARTICLE{Mihara11,
       author = {{Mihara}, Tatehiro and {Nakajima}, Motoki and {Sugizaki}, Mutsumi and {Serino}, Motoko and {Matsuoka}, Masaru and {Kohama}, Mitsuhiro and {Kawasaki}, Kazuyoshi and {Tomida}, Hiroshi and {Ueno}, Shiro and {Kawai}, Nobuyuki and {Kataoka}, Jun and {Morii}, Mikio and {Yoshida}, Atsumasa and {Yamaoka}, Kazutaka and {Nakahira}, Satoshi and {Negoro}, Hitoshi and {Isobe}, Naoki and {Yamauchi}, Makoto and {Sakurai}, Ikuya},
        title = "{Gas Slit Camera (GSC) onboard MAXI on ISS}",
      journal = {\pasj},
         year = 2011,
        month = nov,
       volume = {63},
        pages = {S623-S634},
          doi = {10.1093/pasj/63.sp3.S623},
archivePrefix = {arXiv},
       eprint = {1103.4224},
 primaryClass = {astro-ph.IM},
       adsurl = {https://ui.adsabs.harvard.edu/abs/2011PASJ...63S.623M}
}

@ARTICLE{Pottschmidt26,
       author = {{Pottschmidt}, Katja and {Kretschmar}, Peter and {Sokolova-Lapa}, Ekaterina and {Ambrosi}, Elena and {Ballhausen}, Ralf and {Becker}, Peter A. and {Berger}, Katrin and {Brumback}, McKinley C. and {Coley}, Joel B. and {Corbet}, Robin H.~D. and {D'A{\`\i}}, Antonino and {DeCesar}, Megan E. and {Ferrigno}, Carlo and {F{\"u}rst}, Felix and {Islam}, Nazma and {Kreykenbohm}, Ingo and {Madurga-Favieres}, Vicente and {Pradhan}, Pragati and {Stierhof}, Jakob and {Thalhammer}, Philipp and {West}, Brent F. and {Wolff}, Michael T. and {Zainab}, Aafia and {Zalot}, Nicolas and {Malacaria}, Christian and {Rothschild}, Richard E. and {Wilms}, J{\"o}rn and {Wood}, Kent S.},
        title = "{Pulse profiles of accreting neutron stars - A review of analysis methods, observations, and theoretical models}",
      journal = {arXiv e-prints},
         year = 2026,
        month = aug,
          eid = {arXiv:2608.20876},
        pages = {arXiv:2608.20876},
          doi = {10.48550/arXiv.2608.20876},
archivePrefix = {arXiv},
       eprint = {2608.20876},
 primaryClass = {astro-ph.HE},
       adsurl = {https://ui.adsabs.harvard.edu/abs/2026arXiv260820876P}
}

\end{document}